\documentclass[12pt,a4paper]{article}
\usepackage{amssymb}
\usepackage{amsmath}
\usepackage{amsfonts}
\usepackage{geometry}
\usepackage{setspace}

\newtheorem{theorem}{Theorem}
\newtheorem{assumption}{Assumption}

\newtheorem{corollary}{Corollary}

\newtheorem{lemma}{Lemma}

\input{tcilatex}
\begin{document}

\title{A further look at Modified ML estimation of the panel AR(1) model
with fixed effects and arbitrary initial conditions.}
\author{Hugo Kruiniger\thanks{%
Address: hugo.kruiniger@durham.ac.uk; Dept.\ of Economics, 1 Mill Hill Lane,
Durham DH1 3HY, England. } \\
Durham University}
\date{This version: 4 January 2026}
\maketitle

\begin{center}
\textbf{Abstract}
\end{center}

In this paper we consider two generalizations of Lancaster's (\textit{Review
of Economic Studies}, 2002) Modified Maximum Likelihood estimator (MMLE) for
the panel AR(1) model with fixed effects, arbitrary initial conditions and
strictly exogenous covariates when the time dimension of the panel, $T$, is
fixed. When the autoregressive parameter $\rho =1,$ the limiting modified
profile log-likelihood function for this model has a stationary point of
inflection and $\rho $ is first-order underidentified but second-order
identified. We show that, unlike the Random Effects and Transformed MLEs for
this type of model, the generalized MMLEs are uniquely defined in finite
samples w.p.1. for any value of $\left\vert \rho \right\vert \leq 1$. When $%
\rho =1,$ the rate of convergence of the MMLEs is $N^{1/4},$ where $N$ is
the cross-sectional dimension of the panel. We derive the limiting
distributions of the MMLEs when $\rho =1$. They are generally asymmetric. We
also show that Quasi LM tests that are based on the modified profile
log-likelihood function and use its expected rather than observed Hessian
for hypotheses that include a restriction on $\rho $, and confidence sets
that are based on inverting these tests have correct asymptotic size in a
uniform sense when $\left\vert \rho \right\vert \leq 1$. Finally, we
investigate the finite sample properties of the MMLEs and the QLM test in a
Monte Carlo study. \bigskip \vspace{0.1in}

\noindent JEL\ classification: C11, C13, C23.\bigskip \vspace{0.1in}


\bigskip

\baselineskip=18.8pt

\renewcommand{\baselinestretch}{1.5}

\setcounter{page}{0} \thispagestyle{empty}\newpage

\section{Introduction\protect\vspace{-0.1in}}

In this paper we consider generalized Modified ML (cf. Neyman and Scott,
1948) estimators and identification robust inference methods for panel AR(1)
models with fixed effects (FE), arbitrary initial conditions and strictly
exogenous covariates when $T$ is fixed.

It is well known that the FE ML estimator for the autoregressive parameter $%
\rho $ that is equal to the LSDV estimator is inconsistent when $T$ is
fixed, cf. Nickell (1981).\footnote{%
FE estimators only use data in differences.} To obtain a consistent FE\
estimator for $\rho $ (or for $\theta _{0}=(\rho $ $\sigma ^{2}$ $\beta
^{\prime })^{\prime }$, where $\sigma ^{2}$ is the error variance and $\beta 
$ is the vector of coefficients of the covariates) based on the likelihood
function for the model, Lancaster (2002) proposed a Bayesian approach that
involves reparametrizing the fixed effects and integrating the new effects
from the likelihood function using a uniform prior density. He defined his
estimator for $\rho $ (or for $\theta _{0}$) as a local rather than a global
maximizer of the resulting marginal (or joint) posterior density because
this posterior density is improper and has a "global maximum" at $r=-\infty $
or $r=\infty $ for any sample size, cf. Dhaene and Jochmans (2016). Bun and
Carree (2005) took a different route and proposed a bias-corrected LSDV
estimator for $\theta _{0}$ with the correction based on formulae for the
asymptotic biases of the LSDV estimators for $\rho $ and $\beta $. However,
a version of their estimator is equal to Lancaster's estimator for $\theta
_{0}$, cf. Dhaene and Jochmans (2016), and both of them can be viewed as a
Modified ML estimator (MMLE). Bun and Carree (2005) investigated the finite
sample properties of their estimator using various Monte Carlo experiments.
They reported non-convergence of their estimator in about 40\% of the
replications in some experiments where $N=100,$ $T=6$ and $\rho =0.8.$ The
possible non-existence of the MMLE is also related to the fact that the
posterior density is improper. Specifically, when $\rho =1,$ the limiting
modified profile log-likelihood function of $r$ has a stationary point of
inflection at $r=1$, cf. Ahn and Thomas (2023). Dhaene and Jochmans (2016)
addressed the non-existence problem by generalizing the MMLE for $\theta
_{0} $: they defined their Adjusted Likelihood estimator for $\rho $ as the
minimizer of the 2-norm of the gradient of modified profile log-likelihood
function of $r$ subject to some constraints: two bounds on $r$ that depend
on the LSDV estimate for $\rho $ and were imposed to ensure uniqueness of
the estimator asymptotically, and a condition for a maximum. $\sigma ^{2}$
and $\beta $ are\linebreak estimated by solving the likelihood equations for 
$\sigma ^{2}$ and $\beta $ and replacing $r$ by $\widehat{\rho }$ ($\widehat{%
\rho }_{ADJ}$).

In this paper we discuss two generalized MMLEs for $\theta _{0}$ that exist
with probability approaching one (w.p.a.1) as $N$ increases for any $%
\left\vert \rho \right\vert \leq 1$.\footnote{%
Note that w.p.a.1. means with probability approaching one, i.e., w.p.1
asymptotically.} The first generalized MMLE minimizes a quadratic form in
the modified profile score vector for $\theta _{0}$ subject to $r\in \lbrack
-1,\infty )$ and a condition for a maximum, while the second one minimizes
the 2-norm of the modified profile score for $\rho $ on $[-1,\infty )$
subject to a condition for a maximum. The former MMLE depends on a weight
matrix, while the latter MMLE\ depends on different constraints on $r$ than
the generalized MMLE of Dhaene and Jochmans (2016) does.

While the likelihood functions of the Transformed MLE of Hsiao et al. (2002)
and the Random Effects (RE) MLE of Chamberlain (1980) and Anderson and Hsiao
(1982) may each have two local maxima, see Bun et al. (2017), we show that
the modified profile likelihood function of $r$ has at most one local
maximum on the interval $[-1,\infty ),$ and that if $\left\vert \rho
\right\vert \leq 1,$ then both types of generalized MMLEs are uniquely
defined w.p.1. and consistent. If the latter function has a local maximum on 
$[-1,\infty )$, then all the generalized MMLEs will be equal to each other.
However, if that function has no local maximum on $[-1,\infty )$, then these
estimators are different from each other and the value of the first type of
generalized MMLE will depend on the choice of the weight matrix.

We also derive the limiting distributions of two generalized MMLEs. Similar
to the cases of the REMLE and the Transformed MLE, which we will hereafter
refer to as the FEMLE, if $\rho =1,$ then $\rho $ is only second-order
identified by their objective functions and as a result the rate of
convergence of the MMLEs for $\rho $ is $N^{1/4}$, cf. Ahn and Thomas (2023)
and Kruiniger (2013). Our analysis for $\rho =1$ is closely related to
Sargan (1983) for instrumental variable and ML estimators and also to
Rotnitzky et al. (2000) for MLEs when a parameter is only second-order
identified, although there are some important differences. We view the MMLEs
as GMM estimators in order to derive their limiting distributions when $\rho
=1$. Using an appropriate reparametrization of the modified profile
likelihood, we find that if $\rho =1$ and the data are i.i.d. and normal,
then the limiting distributions of the MMLEs are generally asymmetric unlike
those of the RE- and FEMLE and other MLEs for parameters that are only
second-order identified.

We also discuss inference methods related to the modified profile likelihood
function. Wald tests, some versions of (Quasi) LM tests, and (Quasi) LR
tests that are used for testing hypotheses involving $\rho $ and are based
on the (reparametrized) modified\ profile likelihood function do not
uniformly converge to their fixed parameter first-order limiting
distributions when $\rho $ is close or equal to one, cf. Rotnitzky et al.
(2000) and Bottai (2003). As a consequence these tests do not asymptotically
have correct size in a uniform sense when $\left\vert \rho \right\vert \leq
1 $. Similarly to Kruiniger (2025a) in the case of (Quasi) LM tests related
to the RE- and the FE(Q)MLE, we show that (Q)LM test statistics that are
based on the modified profile log-likelihood function and use its \textit{%
expected} rather than \textit{observed} Hessian for hypotheses that include
a restriction on $\rho $, and confidence sets that are based on inverting
these tests have correct asymptotic size in a uniform sense when $\left\vert
\rho \right\vert \leq 1$.

Monte Carlo results confirm that the QLM\ tests have correct size and show
that when the data are i.i.d. and normal and $\left\vert \rho \right\vert <1$%
, the MMLEs for $\rho $ can have a significantly smaller RMSE than the
asymptotically efficient REMLE in panels as large as $T=9$ and $N=500$. When
the data are not i.i.d. and normal, it is generally not possible to rank the
Quasi MMLEs, the RE- and the FEQMLE in terms of asymptotic efficiency.

Both types of generalized MMLEs are also useful for estimating other models
with parameters that may correspond to stationary points of inflection of
the profile likelihood function. Examples of such models are the sample
selection model and the stochastic production frontier model for a
cross-section of units that are discussed in Lee and Chesher (1986) and
models with skew-normal distributions, see e.g. Hallin and Ley (2014).

Dhaene and Jochmans (2016) discuss several alternative approaches to
constructing modified (profile) objective functions for the nonstationary
panel AR(1) model that yield estimators similar to Lancaster's MMLE. They
have shown that their Adjusted Likelihood estimator for the nonstationary
panel AR(1) model is uniquely defined asymptotically. However, they have not
proven uniqueness of their estimator in finite samples nor have\linebreak
they derived its limiting distribution when $\rho =1$.\thinspace
Furthermore, our paper is the first paper\linebreak that provides MMLE-based
inference methods that have correct uniform asymptotic size.

Hahn and Kuersteiner (2002) modified the LSDV\ estimator to remove bias up
to order $O(T^{-1}).$ Other FE estimators for dynamic panel models include
the first-difference (FD) instrumental variable estimator of Anderson and
Hsiao (1982), the FE GMM estimators of Kruiniger (2001), the Maximum
Invariant Likelihood estimator of Moreira (2009), the FDMLE of Kruiniger
(2008) and the Panel Fully Aggregated Estimator of Han et al.\linebreak
(2014). The latter two estimators rely on covariance stationarity of the
data when $\left\vert \rho \right\vert <1.$

Dovonon and Hall (2018) present a limiting distribution theory for GMM
estimators when first-order identification fails but second-order
identification holds. However, as shown in Kruiniger (2025b), their theory,
specifically their Theorem 1(c), depends on a condition, namely $\Pr (%
\mathbb{R}
_{1}=0)=0$, that is only satisfied in the overidentified case and therefore
cannot be used to derive the results in this paper. Kruiniger (2025b)
completes the theory in Dovonon and Hall (2018) by adding a limiting
distribution theory for the exactly identified case.

The paper is organised as follows. Section 2 presents the panel AR(1) model
and the assumptions. Section 3 discusses existence, uniqueness and
consistency of the generalized MMLEs as well as their asymptotic
distributions. Section 4 discusses inference methods that have correct
asymptotic size in a uniform sense. Section 5 studies the finite sample
properties of the MMLEs and a (Q)LM test. Finally, section 6 offers some
concluding remarks. Derivations and proofs can be found in the appendix.%
\vspace{-0.13in}

\section{The panel AR(1) model\protect\vspace{-0.07in}}

We consider ML-type estimators for the panel AR(1) model with $K$ strictly
exogenous covariates $x_{i,t,k},$ $k=1,...,K:\vspace{-0.12in}$ 
\begin{equation}
y_{i,t}=\rho y_{i,t-1}+x_{i,t}^{\prime }\beta +\alpha _{i}+\varepsilon _{i,t}%
\text{ with }\beta =(1-\rho )\check{\beta}\text{ and }\alpha _{i}=(1-\rho
)\mu _{i},\vspace{-0.04in}  \label{mdl2}
\end{equation}%
for $i=1,...,N$ and $t=1,...,T,$ where $x_{i,t}^{\prime }$ is the $t-th$ row
of the $T\times K$ matrix $X_{i},$ $\alpha _{i}$ is a fixed effect and $%
\varepsilon _{i,t}$ is an error term. We can also allow for time effects in
the model.

Let $y_{i}=(y_{i,1}$ $...$ $y_{i,T})^{\prime },$ $y_{i,-1}=(y_{i,0}$ $...$ $%
y_{i,T-1})^{\prime },$ $\varepsilon _{i}=(\varepsilon _{i,1}$ $...$ $%
\varepsilon _{i,T})^{\prime }$ and $\overline{x}_{i}^{\prime }=T^{-1}\iota
^{\prime }X_{i}$, with $\iota $ equal to a $T-$vector of ones. If we let $%
v_{i}=(\rho -1)y_{i,0}+\alpha _{i}+\overline{x}_{i}^{\prime }\beta $ for $%
i=1,...,N,$ then the model in (\ref{mdl2}) can also be written as $%
y_{i}-y_{i,0}\iota =\rho (y_{i,-1}-y_{i,0}\iota )+QX_{i}\beta +v_{i}\iota
+\varepsilon _{i}$ for $i=1,...,N$, where $Q=I_{T}-T^{-1}\iota \iota
^{\prime }$ and $I_{T}$ is an identity matrix with dimension $T,$ cf.
Lancaster (2002). We make the following assumption:

\begin{assumption}
The variable $y_{i,t}$ is generated by (\ref{mdl2}) with (i) $T\geq 2$; (ii) 
$-1\leq \rho \leq 1$;\newline
(iii) $\{(\varepsilon _{i}^{\prime },v_{i},(vech(QX_{i}))^{\prime })^{\prime
}\}_{i=1}^{N}$ is a sequence of $i.i.d.$ random vectors with $E(v_{i})=0,$%
\linebreak $\qquad Var(v_{i})=\sigma _{v}^{2}<\infty $ and $E(X_{i}^{\prime
}QX_{i})$ is a finite and positive definite matrix; and\newline
(iv) $\varepsilon _{i}\perp (v_{i},(vech(QX_{i}))^{\prime })^{\prime },$ $%
E(\varepsilon _{i})=0$ and $Var(\varepsilon _{i})=\sigma ^{2}I_{T}<\infty ,$ 
$i=1,...,N$.
\end{assumption}

Thus we assume cross-sectional independence, strict exogeneity of the
regressors in first-differences, homoskedasticity and no multicollinearity.
On the other hand, we allow for ARCH and non-normality of the error terms,
the $\varepsilon _{i,t}.$

We require that $T\geq 2$ and $\rho \geq -1$ for identification. In
economics the assumption $\rho \geq -1$ can reasonably be expected to hold
when the covariates are strictly exogenous. The restrictive parametrization $%
\alpha _{i}=(1-\rho )\mu _{i}$ and $\beta =(1-\rho )\check{\beta}$ prevents
the fixed effects and the means of the individual regressors from turning
into trends at $\rho =1$ and thereby avoids a discontinuity in the data
generating process at $\rho =1$. These restrictions and the restriction $%
\rho \leq 1$ are only imposed on the DGP but not in estimation.

We are interested in consistent estimation of the common parameters $\rho ,$ 
$\sigma ^{2}$ and $\beta $ under large $N$, fixed $T$ asymptotics. We will
treat the individual effects as nuisance parameters. We will work with a
Gaussian homoskedastic (quasi-)likelihood but we note that consistency of
the MMLEs (for $\rho $ and $\beta $) does not depend on normality or
cross-sectional homoskedasticity of the errors.

\section{Modified ML estimation of the panel AR(1) model}

Conditional on $y_{i,0}$ and $X_{i},$ $i=1,...,N$ and normalized by $N$, the
Gaussian FE log-likelihood function for the model in (\ref{mdl2}) is, up to
an additive constant, given by:\vspace{-0.05in}%
\begin{equation}
-\frac{T}{2}\log s^{2}-\frac{1}{2s^{2}}\frac{1}{N}%
\sum_{i=1}^{N}(y_{i}-ry_{i,-1}-X_{i}b-a_{i}\iota )^{\prime
}(y_{i}-ry_{i,-1}-X_{i}b-a_{i}\iota ).  \label{lolik}
\end{equation}%
To obtain a consistent FE estimator for $\theta _{0}$ based on (\ref{lolik}%
), Lancaster (2002) proposed a Bayesian approach that involves using a
reparametrization of the fixed effects, which aims to achieve information
orthogonality (but fails to do so when covariates are present), and
integrating the new effects from the likelihood function using a uniform
prior density. He defines his estimator for $\theta _{0}$ as a local maximum
of the joint posterior density. Letting $\theta =(r$ $s^{2}$ $b)^{\prime }$,
his joint posterior log-density for the model in (\ref{mdl2}), normalized by 
$N$, which can be interpreted as a (normalized) modified profile
log-likelihood function, is given by:\vspace{-0.05in}%
\begin{align}
\widetilde{l}_{N}(\theta )& =\widetilde{l}_{N}(r,s^{2},b)=(T-1)\xi
(r)+l_{N}(\theta )\quad \text{where }  \label{mdlf} \\
\xi (r)& =\frac{1}{T(T-1)}\sum_{t=1}^{T-1}\frac{(T-t)}{t}r^{t}\quad \text{and%
}  \notag \\
l_{N}(\theta )& =-\frac{T-1}{2}\log s^{2}-\frac{1}{2s^{2}}\frac{1}{N}%
\sum_{i=1}^{N}(y_{i}-ry_{i,-1}-X_{i}b)^{\prime }Q(y_{i}-ry_{i,-1}-X_{i}b). 
\notag
\end{align}%
and the corresponding modified profile likelihood equations are given by:%
\vspace{-0.05in}%
\begin{eqnarray}
\Psi _{\rho }(\theta ) &=&(T-1)\xi ^{\prime }(r)+\frac{1}{s^{2}}\frac{1}{N}%
\sum_{i=1}^{N}(y_{i}-ry_{i,-1}-X_{i}b)^{\prime }Qy_{i,-1}=0,
\label{modlikeqs} \\
\Psi _{\sigma ^{2}}(\theta ) &=&-\frac{T-1}{2s^{2}}+\frac{1}{2s^{4}}\frac{1}{%
N}\sum_{i=1}^{N}(y_{i}-ry_{i,-1}-X_{i}b)^{\prime
}Q(y_{i}-ry_{i,-1}-X_{i}b)=0,  \notag \\
\Psi _{\beta }(\theta ) &=&\frac{1}{s^{2}}\frac{1}{N}\sum_{i=1}^{N}X_{i}^{%
\prime }Q(y_{i}-ry_{i,-1}-X_{i}b)=0.  \notag
\end{eqnarray}%
Note that the joint posterior density is not proper.

Let $\widehat{\theta }_{LAN}$ denote Lancaster's estimator for $\theta _{0}$
and let $\Theta _{N}$ be the set of roots of $\frac{\partial \widetilde{l}%
_{N}}{\partial \theta }=0$ corresponding to local maxima of $\widetilde{l}%
_{N}$ on $\Omega $ which is an open subset of $%
\mathbb{R}
\times 
\mathbb{R}
^{+}\times 
\mathbb{R}
^{K}.$ Thus $\widehat{\theta }_{LAN}\in \Theta _{N}$ unless $\Theta _{N}$ is
empty, in which case (we will say that) $\widehat{\theta }_{LAN}$ does not
exist. In that case Lancaster effectively puts $\widehat{\theta }_{LAN}=%
\mathbf{0,}$ see his consistency proof. This `trick' ensures that $\widehat{%
\theta }_{LAN}$ always exists so that one can consider whether $\widehat{%
\theta }_{LAN}$ is a consistent estimator for $\theta _{0}.$ Note that none
of the roots of $\frac{\partial \widetilde{l}_{N}}{\partial \theta }=0$
correspond to the global maxima that can occur at $r=\infty $ and, if $T$ is
odd, at $r=-\infty .$

Lancaster showed that $\widetilde{l}_{N}(\theta )$ converges uniformly in
probability to a nonstochastic differentiable function of $\theta ,$ say $%
\widetilde{l}(\theta ),$ and that $\frac{\partial \widetilde{l}(\theta )}{%
\partial \theta }|_{\theta _{0}}=0.$ Next we derive necessary and sufficient
conditions for negative definiteness of the Hessian of $\widetilde{l}(\theta
)$ at $\theta _{0},$ viz.:\vspace{-0.1in}%
\begin{equation}
MH=\left( 
\begin{array}{ccc}
(T-1)\xi ^{\prime \prime }(\rho )-tr(\Phi ^{\prime }Q\Phi )-\frac{\Sigma
_{zqz}}{\sigma ^{2}} & \frac{(T-1)\xi ^{\prime }(\rho )}{\sigma ^{2}} & -%
\frac{\Sigma _{xqz}^{\prime }}{\sigma ^{2}} \\ 
\frac{(T-1)\xi ^{\prime }(\rho )}{\sigma ^{2}} & -\frac{T-1}{2\sigma ^{4}} & 
0 \\ 
-\frac{\Sigma _{xqz}}{\sigma ^{2}} & 0 & -\frac{\Sigma _{xqx}}{\sigma ^{2}}%
\end{array}%
\right) ,\vspace{-0.1in}  \label{mh}
\end{equation}%
where $\Sigma _{zqz}=$ p$\lim_{N\rightarrow \infty }N^{-1}\sum_{i=1}^{N}%
\widetilde{Z}_{i}Q\widetilde{Z}_{i},$ $\Sigma _{xqx}=$ p$\lim_{N\rightarrow
\infty }N^{-1}\sum_{i=1}^{N}X_{i}^{\prime }QX_{i}$ and $\Sigma _{xqz}=$ p$%
\lim_{N\rightarrow \infty }N^{-1}\sum_{i=1}^{N}X_{i}^{\prime }Q\widetilde{Z}%
_{i}$ with $\widetilde{Z}_{i}=\varphi v_{i}+\Phi QX_{i}\beta ,$\vspace{-0.1in%
}%
\begin{equation}
\Phi =\Phi (\rho )=\left( 
\begin{array}{cccccc}
0 & . & . & 0 & 0 & 0 \\ 
1 & 0 &  &  & 0 & 0 \\ 
\rho & 1 & 0 &  &  & 0 \\ 
. & \rho & 1 & 0 &  & . \\ 
. &  & \rho & 1 & 0 & . \\ 
\rho ^{T-2} & . & . & \rho & 1 & 0%
\end{array}%
\right) \text{ and }\varphi =\varphi (\rho )=\left( 
\begin{array}{c}
1 \\ 
\rho \\ 
\rho ^{2} \\ 
\vdots \\ 
\rho ^{T-2} \\ 
\rho ^{T-1}%
\end{array}%
\right) .\vspace{-0.1in}
\end{equation}

It follows from lemma 4.1 in Dhaene and Jochmans (2016) that if $T=2$ and $%
\Sigma _{zqz}>0$ (so that $\rho \neq 1$) or if $T>2$ and $\rho \neq 1$, then 
$MH$ is negative definite so that $\widetilde{l}(\theta )$ has a local
maximum at $\theta _{0}$.\footnote{%
Their lemma 4.1 implies that $\xi ^{\prime \prime }(\rho )-(T-1)^{-1}tr(\Phi
^{\prime }Q\Phi )+2(\xi ^{\prime }(\rho ))^{2}\leq 0$ with equality if and
only if $T=2$ or $\rho =1$.} Kruiniger (2001) had already shown that if $%
\rho =1$ and $T\geq 2,$ then $MH$ is singular. Moreover, Ahn and Thomas
(2023) have shown that $\widetilde{l}(\theta )$ actually has a stationary
point of inflection when $\rho =1$ rather than a local maximum. This
property is related to the fact that the posterior density is not proper.
Later on, in the context of Theorem 1 below, we will show that if $\rho =1,$ 
$\widetilde{l}_{N}$ may not have any local maximum on $\widetilde{\Omega }%
=[-1,\infty )\times (0,\infty )\times 
\mathbb{R}
^{K}$ asymptotically, so that $\widehat{\theta }_{LAN}$ is inconsistent.%
\footnote{%
Lancaster's model is $y_{i}=\rho y_{i,-1}+X_{i}\beta +\alpha _{i}\iota
+\varepsilon _{i}$ without the restrictions $\beta =(1-\rho )\check{\beta}$
and $\alpha _{i}=(1-\rho )\mu _{i}.$ Therefore, if $\rho =1$ and $\beta \neq
0,$ then the probability limit of the Hessian of \textit{his} modified
log-likelihood function at $\theta _{0}$ is still negative definite and his
estimator is consistent. However, if $\rho =1,$ $\beta =0$ and $\alpha
_{i}=0 $ for $i=1,...,N,$ then his estimator is inconsistent.} $\widehat{%
\theta }_{LAN}$ has two more drawbacks. Firstly, $\widetilde{l}_{N}(\theta )$
may not have any local maximum in small samples, in which case $\widehat{%
\theta }_{LAN}$ does not exist. This may happen when $\rho $ is close or
equal to unity. Secondly, Lancaster did not rule out that $\widetilde{l}%
_{N}(\theta )$ and $\widetilde{l}(\theta )$ have multiple local maxima on $%
\Omega $ and he did not explain how to find the consistent estimator if that
were the case.\vspace{-0.14in}

\subsection{Generalized Modified ML estimators}

We will now introduce two generalizations of $\widehat{\theta }_{LAN}$. We
have assumed that $\left\vert \rho \right\vert \leq 1.$ Under this
assumption we will be able to show below that $\widetilde{l}_{N}(\theta )$
can have one local maximum on $\widetilde{\Omega }$ at most. To ensure that
the MMLE for $\theta _{0}$ is also defined in most cases where $\Theta
_{N}\cap \widetilde{\Omega }=\varnothing ,$ we will generalize its
definition as follows:\vspace{-0.09in}%
\begin{equation}
\widehat{\theta }_{W}=\arg \min_{\theta \in \widetilde{\Omega }}\left( \frac{%
\partial \widetilde{l}_{N}(\theta )}{\partial \theta }\right) ^{\prime
}W_{N}\left( \frac{\partial \widetilde{l}_{N}(\theta )}{\partial \theta }%
\right) \text{ s.t. }x^{\prime }\left( \frac{\partial ^{2}\widetilde{l}%
_{N}(\theta )}{\partial \theta \partial \theta ^{\prime }}\right) x\leq 0%
\text{ }\forall x\in 
\mathbb{R}
^{2+K},  \label{mml}
\end{equation}%
where $W_{N}$ is a positive definite (PD) symmetric weight matrix and plim$%
_{N\rightarrow \infty }W_{N}=W$ where $W$ is PD. Thus our MMLE is defined as
the minimizer of a quadratic form in the modified profile score vector, $%
\frac{\partial \widetilde{l}_{N}}{\partial \theta },$ subject to the Hessian
of $\widetilde{l}_{N}$ being negative semi-definite. If $\widetilde{l}%
_{N}(\theta )$ has a local maximum, then our MMLE for $\theta _{0}$ does not
depend on $W_{N}$ and is equal to $\widehat{\theta }_{LAN}$. Theorem 1 below
asserts that $\widehat{\theta }_{W}$ exists w.p.a.1, is uniquely defined
(given $W_{N}$) w.p.1 and is consistent for any $\theta _{0}\in \widetilde{%
\Omega }$.

Note that among the likelihood equations in (\ref{modlikeqs}) only the one
for $r$ is modified. Hence, when solving $\Psi _{\beta }(\theta )=0$ for $b$
we obtain the unique solution $\widehat{\beta }(r)=(\sum_{i=1}^{N}X_{i}^{%
\prime }QX_{i})^{-1}\times $ $\sum_{i=1}^{N}X_{i}^{\prime
}Q(y_{i}-ry_{i,-1}) $ and when solving $\Psi _{\sigma ^{2}}(\theta )=0$ for $%
s^{2}$ we obtain the unique solution $\widehat{\sigma }%
^{2}(r,b)=(T-1)^{-1}N^{-1}\sum_{i=1}^{N}(y_{i}-ry_{i,-1}-X_{i}b)^{\prime
}Q(y_{i}-ry_{i,-1}-X_{i}b).$ Let $\widehat{\theta }(r)=(r,\widehat{\sigma }%
^{2}(r,\widehat{\beta }(r)),\widehat{\beta }(r))^{\prime },$ then the
(normalized) modified profile log-likelihood function of $r$, $\widetilde{l}%
_{N}^{c}(r),$ is defined by the equality $\widetilde{l}_{N}^{c}(r)=%
\widetilde{l}_{N}(\widehat{\theta }(r)),$ i.e., $\widetilde{l}_{N}^{c}(r)=%
\widetilde{l}_{N}(r,\widehat{\sigma }^{2}(r,\widehat{\beta }(r)),\widehat{%
\beta }(r)).$

An alternative generalized MMLE for $\theta _{0}$, which is based on $%
\widetilde{l}_{N}^{c}(r)$, is given by $\widehat{\theta }_{C}$ with\vspace{%
-0.18in} \footnote{%
One can also define a class of MMLEs where only $s^{2}$ is profiled out but
not $b$.}\linebreak 
\begin{gather}
\widehat{\rho }_{C}=\arg \min_{r\in \lbrack -1,\infty )}\left( \frac{%
\partial \widetilde{l}_{N}^{c}(r)}{\partial r}\right) ^{2}\text{ s.}\bigskip 
\text{t. }\frac{\partial ^{2}\widetilde{l}_{N}^{c}(r)}{\partial r^{2}}\leq
0,\quad  \label{mml2} \\
\widehat{\beta }_{C}=\widehat{\beta }(\widehat{\rho }_{C})\quad \text{%
and\medskip }\quad \widehat{\sigma }_{C}^{2}=\widehat{\sigma }^{2}(\widehat{%
\rho }_{C},\widehat{\beta }_{C}).  \notag
\end{gather}%
The Adjusted Likelihood estimator of Dhaene and Jochmans (2016), viz. $%
\widehat{\theta }_{ADJ}$, is defined similarly to $\widehat{\theta }_{C}$: $%
\widehat{\rho }_{ADJ}=\arg \min_{r\in \mathcal{E}}\left( \frac{\partial 
\widetilde{l}_{N}^{c}(r)}{\partial r}\right) ^{2}$ s.t. $\frac{\partial ^{2}%
\widetilde{l}_{N}^{c}(r)}{\partial r^{2}}\leq 0$ with $\mathcal{E=\{}r:$ $%
\mathcal{(}r\mathcal{-}\widehat{\rho }_{LSDV})^{2}(-\frac{\partial
^{2}l_{N}^{c}(r)}{\partial r^{2}}|_{\widehat{\rho }_{LSDV}})\leq 1\}$ where $%
l_{N}^{c}(r)=l_{N}(r,\widehat{\sigma }^{2}(r,\widehat{\beta }(r)),\widehat{%
\beta }(r))$. $\widehat{\beta }_{ADJ}=\widehat{\beta }(\widehat{\rho }%
_{ADJ}) $ and $\widehat{\sigma }_{ADJ}^{2}=\widehat{\sigma }^{2}(\widehat{%
\rho }_{ADJ},\widehat{\beta }_{ADJ}).$ However, they have only shown that $%
\widehat{\theta }_{ADJ}$ is uniquely defined asymptotically. On the other
hand, Theorem 1 below states that when $\left\vert \rho \right\vert \leq 1,$ 
$\widehat{\theta }_{C}$ is uniquely defined in finite samples. It is unclear
how rare the event $\widehat{\rho }_{C}\notin \mathcal{E}$ is in finite
samples when $\left\vert \rho \right\vert \leq 1$, but if $\widehat{\rho }%
_{C}\notin \mathcal{E}$, then $\widehat{\rho }_{C}$ is likely to be a better
estimate than $\widehat{\theta }_{ADJ}$ because either $\left\vert \frac{%
\partial \widetilde{l}_{N}^{c}(r)}{\partial r}|_{\widehat{\rho }%
_{C}}\right\vert <\left\vert \frac{\partial \widetilde{l}_{N}^{c}(r)}{%
\partial r}|_{\widehat{\rho }_{ADJ}}\right\vert $ or $\widehat{\theta }%
_{ADJ}<-1.$\vspace{-0.05in}

There exists no $W_{N}$ such that the $\widehat{\theta }_{W}$ estimator
always equals the $\widehat{\theta }_{C}$ estimator: if $\frac{\partial 
\widetilde{l}_{N}^{c}(r)}{\partial r}|_{\widehat{\rho }_{C}}=0,$ then $\frac{%
\partial \widetilde{l}_{N}(\theta )}{\partial \theta }|_{\widehat{\theta }%
_{W}}=0$\ and both estimates of $\theta $ are equal, but if $\frac{\partial 
\widetilde{l}_{N}^{c}(r)}{\partial r}|_{\widehat{\rho }_{C}}$ $\neq 0,$ then 
$\frac{\partial \widetilde{l}_{N}(\theta )}{\partial \theta }|_{\widehat{%
\theta }_{W}}\neq 0$\ and the two estimates of $\theta $ are unequal
although the value of $\widehat{\theta }_{W}$ will be close to that of $%
\widehat{\theta }_{C}$ for $W_{N}$ that give relatively little weight to $%
\frac{\partial \widetilde{l}_{N}(\theta )}{\partial r}.$ We also consider a
variation on $\widehat{\theta }_{W}$ with the first element of $\frac{%
\partial \widetilde{l}_{N}(\theta )}{\partial \theta }$ replaced by $\frac{%
\partial \widetilde{l}_{N}^{c}(r)}{\partial r}.$ We call this MMLE $\widehat{%
\theta }_{F}.$\linebreak If $W_{N}=diag(\infty ,\underline{W}_{N,2,2})$ and
the elements of $\underline{W}_{N,2,2}$ are finite, then $\widehat{\theta }%
_{F}=\widehat{\theta }_{C}.$

In the appendix we show that $\widetilde{l}_{N}^{c}(r)$ converges uniformly
in probability to a nonstochastic differentiable function of $r,$ say $%
\widetilde{l}^{c}(r),$ that $\frac{\partial \widetilde{l}^{c}(r)}{\partial r}%
|_{\rho }=0$ and that $\frac{\partial ^{2}\widetilde{l}^{c}(r)}{\partial
r^{2}}|_{\rho }\leq 0,$ with equality holding if $\rho =1$ or if $T=2$ and $%
\sigma _{v}^{2}=\beta =0$ (i.e., $\Sigma _{zqz}=0$). Thus, similar to $%
\widetilde{l}(\theta ),$ $\widetilde{l}^{c}(r)$ has a local maximum at $\rho 
$ when $\rho \neq 1$ and, in case $T=2$, $\Sigma _{zqz}>0$. In the appendix
we also show that $\widetilde{l}^{c}(r)$ has a stationary point of
inflection at $\rho $ when $\rho =1$. To simplify the exposition we assume
in the remainder of this paper that if $T=2$ and $\rho \neq 1,$ then either $%
\sigma _{v}^{2}>0$ or $\beta \neq 0$ so that $\Sigma _{zqz}>0.$

Note that $\widehat{\theta }_{C}$ would only fail to exist in the extremely
unlikely case that $\frac{\partial ^{2}\widetilde{l}_{N}^{c}(r)}{\partial
r^{2}}>0$ on the entire interval $[-1,\infty ).$ Similarly, $\widehat{\theta 
}_{W}$ and $\widehat{\theta }_{F}$ would only fail to exist in the extremely
unlikely case that for no $\theta \in \widetilde{\Omega },$ $x^{\prime
}\left( \frac{\partial ^{2}\widetilde{l}_{N}(\theta )}{\partial \theta
\partial \theta ^{\prime }}\right) x\leq 0$ $\forall x\in 
\mathbb{R}
^{2+K}.$ \footnote{%
One could ensure that $\widehat{\theta }_{W},$ $\widehat{\theta }_{F}$ and $%
\widehat{\theta }_{C}$ are always defined by replacing them by $\widehat{%
\theta }(\widehat{\rho }_{ML}+\frac{3}{T+1})$ in these improbable cases,
where $-\frac{3}{T+1}$ is the asymptotic bias of $\widehat{\rho }_{ML}$ when 
$\rho =1$.\ The rationale for this proposed solution is that the
non-existence problem most likely only occurs (if ever) when the sample size
is very small and $\rho $ is close or equal to unity.} The second-order
conditions $\frac{\partial ^{2}\widetilde{l}_{N}^{c}(r)}{\partial r^{2}}\leq
0$ and $x^{\prime }\left( \frac{\partial ^{2}\widetilde{l}_{N}(\theta )}{%
\partial \theta \partial \theta ^{\prime }}\right) x\leq 0$ $\forall x\in 
\mathbb{R}
^{2+K}$ are a crucial part of the definitions of $\widehat{\theta }_{C},$ $%
\widehat{\theta }_{W}$ and $\widehat{\theta }_{F}$ because $\widetilde{l}%
_{N}^{c}(r)$ and $\widetilde{l}_{N}(r)$ may attain a minimum on $[-1,\infty
) $ and $\widetilde{\Omega }$, respectively, see lemma 1 in the appendix.

The next theorem asserts uniqueness and consistency of $\widehat{\theta }%
_{W},$ $\widehat{\theta }_{F}$ and $\widehat{\theta }_{C}$:\vspace{-0.12in}

\begin{theorem}
Let Assumption 1 hold. Then the Modified MLEs $\widehat{\theta }_{W},$ $%
\widehat{\theta }_{F}$ and $\widehat{\theta }_{C}$ for $\theta _{0}$ are
uniquely defined w.p.1 when they exist, exist w.p.a.1 and are consistent.%
\vspace{-0.06in}
\end{theorem}

If $-1\leq \rho <1,$ $\lim_{N\rightarrow \infty }\Pr (\Theta _{N}\cap 
\widetilde{\Omega }=\varnothing )=0$, i.e., $\widehat{\theta }_{LAN}$ exists
w.p.a.1. In this case $\widehat{\theta }_{LAN}$ is also unique w.p.1. (if it
exists) and consistent. However, if $\rho =1,$ $\lim_{N\rightarrow \infty
}\Pr (\Theta _{N}\cap \widetilde{\Omega }=\varnothing )>0$ by lemma 4 in the
appendix (and $\theta _{0}\neq \mathbf{0}$), i.e., $\widehat{\theta }_{LAN}$
may not exist even asymptotically, which implies that $\widehat{\theta }%
_{LAN}$ is inconsistent.

When $-1\leq \rho <1,$ the first-order, fixed parameter asymptotic
distributions of $\widehat{\theta }_{W},$ $\widehat{\theta }_{F},$ $\widehat{%
\theta }_{C}$ and $\widehat{\theta }_{LAN}$ are the same and given by (cf.
Kruiniger, 2001):$\vspace{-0.14in}$ 
\begin{equation}
\sqrt{N}\left( \widehat{\theta }-\theta _{0}\right) \overset{d}{\rightarrow }%
N\left( 0,\left( MH\right) ^{-1}MIM\left( MH\right) ^{-1}\right) ,\vspace*{%
0.01in}  \label{asydist}
\end{equation}%
where $MH$ is given in (\ref{mh}) and under normality of the $\varepsilon
_{i}$ $MIM$ (Modified Information Ma-\linebreak trix) equals:\footnote{%
To derive (\ref{mim}) we have used that if $\varepsilon
_{i}|(v_{i},QX_{i})\sim N(0,\sigma ^{2}I_{T}),$ then for any constant $%
T\times T$ matrices $M_{1}$ and $M_{2},$ $E(\varepsilon _{i}^{\prime
}M_{1}\varepsilon _{i}\varepsilon _{i}^{\prime }M_{2}\varepsilon
_{i})=\sigma ^{4}(tr(M_{1})tr(M_{2})+tr(M_{1}M_{2}+M_{1}^{\prime }M_{2}))$.} 
\begin{equation}
MIM=\left( 
\begin{array}{ccc}
tr(Q\Phi Q\Phi )+\frac{\sigma ^{2}tr(\Phi ^{\prime }Q\Phi )+\Sigma _{zqz}}{%
\sigma ^{2}} & -\frac{(T-1)\xi ^{\prime }(\rho )}{\sigma ^{2}} & \frac{%
\Sigma _{xqz}^{\prime }}{\sigma ^{2}} \\ 
-\frac{(T-1)\xi ^{\prime }(\rho )}{\sigma ^{2}} & \frac{T-1}{2\sigma ^{4}} & 
0 \\ 
\frac{\Sigma _{xqz}}{\sigma ^{2}} & 0 & \frac{\Sigma _{xqx}}{\sigma ^{2}}%
\end{array}%
\right) .\hspace{-0.85in}  \label{mim}
\end{equation}

It can easily be checked that $tr(Q\Phi Q\Phi )\neq -(T-1)\xi ^{\prime
\prime }(\rho )$ and hence $MH\neq -MIM$.

If $T=2$, $\widehat{\rho }_{LAN}$ is equal to the FEMLE for $\rho $ that has
been proposed by Hsiao et al. (2002), henceforth $\widehat{\rho }_{FEML}$,
but if $T>2,$ the data are i.i.d. and normal and $\left\vert \rho
\right\vert <1,$ $\widehat{\rho }_{LAN}$ is asymptotically less efficient
than $\widehat{\rho }_{FEML}$, see Ahn and Thomas (2023); when the data are
not i.i.d and normal, $\widehat{\rho }_{LAN}$ may be asymptotically more
efficient than $\widehat{\rho }_{FEML}$.

If $\rho =1,$ $\det (MIM)\neq 0$ but $\frac{\partial ^{2}\widetilde{l}^{c}(r)%
}{\partial r^{2}}|_{\rho }=0$ and $\det (MH)=0.$ Thus $\rho $ and $\theta $
are first- order underidentified when $\rho =1$. Although we cannot directly
apply the results of Rot-\linebreak nitzky et al. (2000), who developed an
asymptotic theory for MLEs when the information matrix is singular, to $%
\widehat{\theta }_{W},$ $\widehat{\theta }_{F}$ and $\widehat{\theta }_{C}$
when $\rho =1$, because they are \textit{Modified}\linebreak MLEs and $\det
(MIM)\neq 0$, arguments similar to theirs suggest that these MMLEs have a
slower than $\sqrt{N}$ rate of convergence and that their limiting
distributions are non-standard. When deriving the limiting distributions of $%
\widehat{\theta }_{C}$ and $\widehat{\theta }_{F}$ for $\rho =1$ below, we
will view the MMLEs as GMM estimators.\footnote{%
The limiting distribution of $\widehat{\theta }_{W}$ for $\rho =1$ can be
derived using results in Kruiniger (2025b).} If $\rho $ is close to 1, $\det
(MH)$\ and $\frac{\partial ^{2}\widetilde{l}^{c}(r)}{\partial r^{2}}|_{\rho
} $ are close to zero and the MMLEs will have a "weak moment conditions"
problem. \vspace{-0.1in}

\subsection{The limiting distributions of $\protect\widehat{\protect\theta }%
_{C}$ and $\protect\widehat{\protect\theta }_{F}$ when $\protect\rho =1$}

W.p.a.1 $\widehat{\rho }_{C}$ is a solution of the first-order condition
(f.o.c.) $G_{N}^{c}(r)\equiv \frac{\partial ^{2}\widetilde{l}_{N}^{c}(r)}{%
\partial r^{2}}\frac{\partial \widetilde{l}_{N}^{c}(r)}{\partial r}=0.$
Using a Taylor expansion of $G_{N}^{c}(\widehat{\rho }_{C})$ around $r=1,$
we show in the appendix that when $\rho =1,$ $N^{1/4}(\widehat{\rho }%
_{C}-1)=O_{p}(1),$ i.e., the rate of convergence of $\widehat{\rho }_{C}$ is
at least $N^{1/4}$. This quartic root rate of convergence reflects the fact
that $\frac{\partial ^{2}\widetilde{l}^{c}(1)}{\partial r^{2}}=0$ and $\frac{%
\partial ^{3}\widetilde{l}^{c}(1)}{\partial r^{3}}=\frac{T(T-1)(T+1)}{12}%
\neq 0$, which means that $\rho $ is second-order identified when $\rho =1$,
and is in line with results in Sargan (1983), Rotnitzky et al. (2000), Ahn
and Thomas (2023), Madsen (2009), Dovonon and Renault (2013) and Kruiniger
(2013) who also study estimation when a parameter is only second-order
identified. Note that this rate is faster than the $N^{1/6}$\thinspace -rate
of the MLEs of the parameters that correspond to the inflection point of the
likelihood functions of the sample selection model and the stochastic
production frontier model for a cross-section that are discussed in Lee and
Chesher (1986) and the models with skew-normal distributions that are
discussed in Hallin and Ley (2014).

Next we discuss the derivation of the limiting distribution of $\widehat{%
\theta }_{C}$ when $\rho =1.$ Let $M_{N}^{c}(r)=N\left( \frac{\partial 
\widetilde{l}_{N}^{c}(r)}{\partial r}\right) ^{2}.$ Analogously to Sargan
(1983) and Rotnitzky et al. (2000) consider the following Taylor expansion
of $M_{N}^{c}(r)$ around $r=1\vspace{-0.1in}:$%
\begin{equation}
M_{N}^{c}(r)=M_{N}^{c}(1)+\sum_{j=1}^{4}\frac{1}{j!}\frac{\partial
^{j}M_{N}^{c}(1)}{\partial r^{j}}(r-1)^{j}+P_{3,N}(N^{1/4}(r-1)),\vspace{%
-0.1in}  \label{tobj}
\end{equation}%
where $P_{3,N}(N^{1/4}(r-1))$ is a polynomial in $N^{1/4}(r-1)$ with
coefficients that are $o_{p}(1)$. Let $\widehat{\rho }=\widehat{\rho }_{C}.$
Substituting $\widehat{\rho }$ for $r$ in (\ref{tobj}) we obtain%
\begin{eqnarray}
M_{N}^{c}(\widehat{\rho }) &=&N\left( \frac{\partial \widetilde{l}_{N}^{c}(1)%
}{\partial r}\right) ^{2}+\frac{\partial ^{3}\widetilde{l}_{N}^{c}(1)}{%
\partial r^{3}}N^{1/2}\frac{\partial \widetilde{l}_{N}^{c}(1)}{\partial r}%
N^{1/2}(\widehat{\rho }-1)^{2}+  \label{mnc} \\
&&\frac{1}{4}\left( \frac{\partial ^{3}\widetilde{l}_{N}^{c}(1)}{\partial
r^{3}}\right) ^{2}N(\widehat{\rho }-1)^{4}+R_{1,N}^{c}(N^{1/4}(\widehat{\rho 
}-1)),  \notag
\end{eqnarray}%
where $R_{1,N}^{c}(N^{1/4}(\widehat{\rho }-1))=o_{p}(1).$

Let $Z_{1,N}=\left( -\frac{1}{2}\frac{\partial ^{3}\widetilde{l}_{N}^{c}(1)}{%
\partial r^{3}}\right) ^{-1}N^{1/2}\left( \frac{\partial \widetilde{l}%
_{N}^{c}(1)}{\partial r}\right) .$ In the proof of Theorem 2 we show that $%
Z_{1,N}=O_{p}(1)$ and that there exists a sequence $\{U_{N}\}$ with $%
U_{N}=O_{p}(N^{-1/2})$ such that if $Z_{1,N}+U_{N}>0,$ then $M_{N}^{c}(r)$\
has two local minima attained at values $\widetilde{\rho }$ such that $%
N^{1/2}(\widetilde{\rho }-1)^{2}=Z_{1,N}+o_{p}(1),$ whereas if $%
Z_{1,N}+U_{N}<0,$ then $M_{N}^{c}(r)$\ has one local minimum attained at $r=%
\widehat{\rho }$ with $N^{1/2}(\widehat{\rho }-1)^{2}=o_{p}(1).$
Furthermore, when $Z_{1,N}+U_{N}>0,$ the sign of $N^{1/4}(\widehat{\rho }-1)$
is determined by the remainder $R_{1,N}^{c}(N^{1/4}(\widehat{\rho }-1))$.

To obtain the limiting distribution of $\widehat{\theta }_{C}$ when $\rho =1$
we use the following new parametrization (indicated by the subscript $n$),
cf. Kruiniger (2013): $\theta _{n}=(r_{n},s_{n}^{2},b_{n}^{\prime })^{\prime
}$ where $r_{n}=r,$ $s_{n}^{2}=s^{2}/r$ and $b_{n}=b.$ Noting that we can
express the elements of $\theta $ as functions of the elements of $\theta
_{n},$ viz. $\theta =\theta (\theta
_{n})=(r_{n},s_{n}^{2}r_{n},b_{n}^{\prime })^{\prime },$ the reparameterized
modified profile log-likelihood function is given by $\widetilde{l}%
_{n,N}(\theta _{n})=\widetilde{l}_{N}(\theta (\theta _{n})).$ Similarly to
Lancaster (2002), it can be shown that $\widetilde{l}_{n,N}(\theta _{n})$
converges uniformly in probability to a nonstochastic continuous function of 
$\theta _{n},$ i.e. $\widetilde{l}_{n}(\theta _{n})=\widetilde{l}(\theta
(\theta _{n})).$ The reparametrization is such that the elements of the
first row and the first column of the Hessian of $\widetilde{l}_{n}(\theta
_{n})$ at $\theta _{0,n}=(\rho _{n},\sigma _{n}^{2},\beta _{n}^{\prime
})^{\prime }=\theta _{\ast }\equiv (1,\sigma ^{2},0^{\prime })^{\prime }$
are equal to zero. Note that if $\rho =1$, then $\theta _{0}=\theta
_{0,n}=\theta _{\ast }$ for some $\sigma ^{2}$.

We also need to introduce some additional notation. Let $\widehat{\theta }=%
\widehat{\theta }_{C}$ and $\widehat{\theta }_{n}=\widehat{\theta }_{n,C}=$%
\linebreak $(\widehat{\rho }_{C},$ $\widehat{\sigma }_{n,C}^{2},$ $\widehat{%
\beta }_{C}^{\prime })^{\prime }$ with $\widehat{\sigma }_{n,C}^{2}=\widehat{%
\sigma }_{C}^{2}/\widehat{\rho }_{C}.$ Furthermore, let $Z_{2,N}=N^{1/2}(%
\widehat{\sigma }^{2}(1,\widehat{\beta }(1))-\sigma ^{2})$, $Z_{3,N}=N^{1/2}(%
\widehat{\beta }-\beta )$ and $Z_{N}=(Z_{1,N},Z_{2,N},Z_{3,N}^{\prime
})^{\prime }.$ Then we have the following results:\vspace{-0.1in}\pagebreak

\begin{theorem}
Let Assumption 1 hold, $\varepsilon _{i}\sim N(0,\sigma ^{2}I),$ $i=1,...,N,$
and $\rho =1.$ Then $\vspace{0.08in}$\newline
(i) $Z_{N}\overset{d}{\rightarrow }Z=(Z_{1},Z_{2},Z_{3}^{\prime })^{\prime
}\sim N(0,\Sigma _{Z}),$ where $E(Z_{1}Z_{2})=0,$ $E(Z_{1}Z_{3})=0,$ $%
E(Z_{2}Z_{3})=0,$ \newline
$Var(Z_{1})=48T^{-2}((T-1)(T+1))^{-1},$ $Var(Z_{2})=2\sigma ^{4}(T-1)^{-1}$
and $Var(Z_{3})=\sigma ^{2}(\Sigma _{xqx})^{-1};\vspace{0.08in}$ (ii)
letting $K_{+}=\sigma ^{2}(T+1)/6$ and $B^{c}=\mathbf{1}(R^{c}>0)$ with the
r.v. $R^{c}$ defined in (\ref{rc})$,\vspace{0.09in}\newline
\left[ 
\begin{array}{c}
N^{1/4}(\widehat{\rho }_{C}-1) \\ 
N^{1/2}(\widehat{\sigma }_{n,C}^{2}-\sigma ^{2}) \\ 
N^{1/2}\widehat{\beta }_{C}%
\end{array}%
\right] \overset{d}{\rightarrow }\left[ 
\begin{array}{c}
(-1)^{B^{c}}Z_{1}^{1/2} \\ 
Z_{2}+K_{+}Z_{1} \\ 
Z_{3}%
\end{array}%
\right] \mathbf{1}\{Z_{1}>0\}+\left[ 
\begin{array}{c}
0 \\ 
Z_{2} \\ 
Z_{3}%
\end{array}%
\right] \mathbf{1}\{Z_{1}\leq 0\}.$
\end{theorem}

\noindent C\textit{omments: }In the proof of Theorem 2 we show that the sign
of $N^{1/4}(\widehat{\rho }_{C}-1)$ depends on $\frac{\partial ^{5}%
\widetilde{l}_{N}^{c}(1)}{\partial r^{5}}$, whereas it follows from
Kruiniger (2013) and corollary 1 in Rotnitzky et al. (2000) that the sign of 
$N^{1/4}(\widehat{\rho }_{FEML}-1)$ only depends on the second and third
derivatives of the FE\ log-likelihood. The latter is generally true for MLEs
of parameters that are only second-order identified, cf. Rotnitzky et al.
(2000);

Relaxing the assumption of normality of the $\varepsilon _{i}$ affects $%
\Sigma _{Z}$ and the conditional distribution of $B^{c}$ given $Z$ but
otherwise does not change Theorem 2;

The limiting distribution of $\widehat{\rho }_{C}$ is asymmetric unlike that
of $\widehat{\rho }_{FEML}$ and other MLEs of parameters that are only
second-order identified, cf. Rotnitzky et al. (2000);

From $\widehat{\theta }_{C}=\theta (\widehat{\theta }_{n,C})$ we have $%
\widehat{\sigma }_{C}^{2}=\widehat{\sigma }_{n,C}^{2}\widehat{\rho }_{C}.$
Hence the rate of convergence of $\widehat{\sigma }_{C}^{2}$ is also $%
N^{1/4} $ and $N^{1/4}(\widehat{\sigma }_{C}^{2}-\sigma ^{2})=N^{1/4}(%
\widehat{\rho }_{C}-1)\sigma ^{2}+o_{p}(1)$;

Finally, the following result implies the sign of the asymptotic bias of $%
\widehat{\rho }_{C}$\ and $\widehat{\sigma }_{C}^{2}$:

\begin{corollary}
Let Assumption 1 hold, $\varepsilon _{i}\sim N(0,\sigma ^{2}I),$ $i=1,...,N,$
and $\rho =1.$ Then if $T\geq 4,$ $E((-1)^{B^{c}}Z_{1}^{1/2}|Z_{1}>0)>0$
whereas if $T=2$ or $T=3,$ $E((-1)^{B^{c}}Z_{1}^{1/2}|Z_{1}>0)<0.$
\end{corollary}

We now consider the minimum rate of convergence of $\widehat{\rho }=\widehat{%
\rho }_{F}$ and the limiting distribution of $\widehat{\theta }_{F}$ when $%
\rho =1$. Details of the derivations of these properties of $\widehat{\rho }%
_{F}$ and $\widehat{\theta }_{F}$ are given in the appendix. There we show
that $N^{1/4}(\widehat{\rho }-1)=O_{p}(1),$ cf. Lemma 5.

Let $\Psi _{N,n}(\theta _{n})=(\frac{\partial \widetilde{l}_{N}^{c}(r)}{%
\partial r},s_{n}^{2}r\frac{\partial \widetilde{l}_{n,N}(\theta _{n})}{%
\partial s_{n}^{2}},s_{n}^{2}r\frac{\partial \widetilde{l}_{n,N}(\theta _{n})%
}{\partial b^{\prime }})^{\prime },$ $\widehat{\underline{\omega }}_{n}=((%
\widehat{\sigma }_{n,F}^{2}-\sigma ^{2}),\widehat{\beta }_{F}^{\prime
})^{\prime }$ and $\underline{w}_{n}=(s_{n}^{2},b^{\prime })^{\prime }$.
Then we have the following results:

\begin{theorem}
Let Assumption 1 hold, $\varepsilon _{i}\sim N(0,\sigma ^{2}I),$ $i=1,...,N,$
$\rho =1,$ and let $W_{N}$ be a PD matrix. Then $\vspace{0.08in}\vspace{%
0.08in}\newline
\left[ 
\begin{array}{c}
N^{1/4}(\widehat{\rho }_{F}-1) \\ 
N^{1/2}\widehat{\underline{\omega }}_{n}%
\end{array}%
\right] \overset{d}{\rightarrow }\left[ 
\begin{array}{c}
(-1)^{B}Z_{1}^{1/2} \\ 
\underline{\omega }_{+}%
\end{array}%
\right] \mathbf{1}\{Z_{1}>0\}+\left[ 
\begin{array}{c}
0 \\ 
\underline{\omega }_{+}+K_{-}Z_{1}%
\end{array}%
\right] \mathbf{1}\{Z_{1}\leq 0\},\vspace{0.06in}\vspace{0.08in}\newline
$where $(Z_{1},$ $\underline{\omega }_{+}^{\prime })^{\prime }\sim
N(0,\Sigma _{\omega }),$ $B=\mathbf{1}(R>0)$ and the r.v. $R,$ the matrix $%
\Sigma _{\omega }$ and the constant vector $K_{-}$ are implicitly defined in
the proof.
\end{theorem}

\noindent C\textit{omments: }In the proof of Theorem 3 we see that the sign
of $N^{1/4}(\widehat{\rho }_{F}-1)$ depends on $\frac{\partial ^{5}%
\widetilde{l}_{N}^{c}(1)}{\partial r^{5}}$. The order of that derivative is
the same as what Kruiniger (2013) found for Quasi MLEs of second-order
identified parameters but different from what Rotnitzky et al. (2000) found
for MLEs;

Relaxing the assumption of normality of the $\varepsilon _{i}$ affects $%
\Sigma _{\omega }$ and the conditional distributions of $B$ and $R$ given $%
(Z_{1},$ $\underline{\omega }_{+}^{\prime })^{\prime }$ but otherwise does
not fundamentally change the results in Theorem 3;

Like $\widehat{\rho }_{C}$ and $\widehat{\sigma }_{C}^{2},$ when $\rho =1,$ $%
\widehat{\rho }_{F}$ and $\widehat{\sigma }_{F}^{2}$ converge at a rate of
at least $N^{1/4}$ to $\rho $ and $\sigma ^{2}$, whereas $\widehat{\beta }%
_{F}$ converges at a rate of $N^{1/2}$ to $\beta $ just like $\widehat{\beta 
}_{C}$;

For any $W,$ $(\widehat{\rho }_{F}-1)^{2}$ is first-order asymptotically
equivalent to $(\widehat{\rho }_{C}-1)^{2}$ and\ hence the RMSEs of $%
\widehat{\rho }_{F}$ and $\widehat{\rho }_{C}$ are asymptotically the same.
However, the limiting distribution of $B$ and hence that of $N^{1/4}(%
\widehat{\rho }_{F}-1)$ depends on $W.$ The limiting distributions of $%
\widehat{\sigma }_{F}^{2}$ and $\widehat{\beta }_{F}$ also depend on $W$ and
are different from those of $\widehat{\sigma }_{C}^{2}$ and $\widehat{\beta }%
_{C}$ unless $W_{N}=diag(W_{N,1,1},\underline{W}_{N,2,2})$ where $W_{N,1,1}$
is a scalar. In the latter case $\underline{\omega }%
_{+}+K_{-}Z_{1}=(Z_{2},Z_{3}^{\prime })^{\prime }\ $and $K_{-}=(-K_{+},0)^{%
\prime }.$ If in addition $W_{N,1,1}=\infty $ while the elements of $%
\underline{W}_{N,2,2}$ are finite, then the limiting distributions of $%
N^{1/4}(\widehat{\rho }_{F}-1)$ and $N^{1/4}(\widehat{\rho }_{C}-1)$ are
also the same;

It can be expected that the MMLEs also have non-standard asymptotic
properties close to the singularity point, $\theta _{\ast }$. Rotnitzky et
al. (2000) informally discuss a richness of possibilities for the MLEs close
to the singularity point and one can expect several possibilities for the
MMLEs too. To save space we don't explore them here. Nonetheless they are a
warning of the care needed in conducting inference close to $\theta _{\ast }$%
. Finally, we note that the local-to-unity asymptotic behaviour of various
GMM estimators for the panel AR(1) model discussed in Kruiniger (2009) is
unrelated to second-order identification.\vspace{-0.15in}

\section{Modified likelihood based inference}

Wald tests, some versions of (Quasi) LM tests, and (Quasi) LR tests that are
used for testing hypotheses involving a parameter that is only second-order
identified do not have correct asymptotic size in a uniform sense, cf.
Bottai (2003), who explains this in the setting of one-dimensional
parametric models. Generalizing the LM-type testing approach in Bottai
(2003) that has correct uniform asymptotic size in that setting to a
multiple parameter setting, Kruiniger (2025a) has shown that (Quasi) LM
tests that are related to the RE- and the FE(Q)MLE for the panel AR(1) model
and standardised by using (a sandwich formula involving) the \textit{expected%
} rather than the \textit{observed} average Hessian have correct uniform
asymptotic size when $\left\vert \rho \right\vert \leq 1$. However, the
situation is somewhat different in the case of the Quasi LM (QLM) tests that
are based on the (normalized) reparametrized modified profile log-likelihood
function $\widetilde{l}_{n,N}(\theta _{n})$ and used for testing hypotheses
that include a restriction on $\rho $. Firstly, in this case the singularity
point, $\theta _{\ast }$, corresponds to a stationary point of inflection of 
$\widetilde{l}(\theta )$ and $\widetilde{l}_{n}(\theta _{n})$ rather than a
maximum. Secondly, although $\det (MH)=0$ when $\theta _{0}=\theta _{\ast }$
for any $\sigma ^{2},$ $\det (MIM)\neq 0$ in this case.$\linebreak $As a
result in finite samples $\widetilde{l}_{n,N}(\theta _{n})$ may not even
have a local maximum when $\rho $ is close$\linebreak $to one. Nevertheless,
the expected average Hessian of $\widetilde{l}_{n,N}(\theta _{n})$ at $%
\underline{\mathcal{\theta }}_{n}=\underline{\mathcal{\breve{\theta}}}_{n},$
viz. $\overline{H}(\underline{\mathcal{\breve{\theta}}}_{n})\equiv E_{%
\underline{\mathcal{\breve{\theta}}}_{n}}(\frac{\partial ^{2}\widetilde{l}%
_{n,N}(\theta _{n})}{\partial \theta _{n}\partial \theta _{n}^{\prime }}|_{%
\underline{\mathcal{\breve{\theta}}}_{n}})$, where $\underline{\mathcal{%
\theta }}_{n}=(\theta _{n}^{\prime }$ $s_{v,n}^{2})^{\prime }$ with $%
s_{v,n}^{2}=s_{v}^{2}/s^{2}-(1-r)$, is negative definite and hence
nonsingular for any value of $\underline{\mathcal{\breve{\theta}}}_{n}$ that
differs from the singularity point $\underline{\mathcal{\theta }}_{\ast
}=\linebreak (\theta _{\ast }^{\prime }$ $0)^{\prime }$.\footnote{%
Note that $\overline{H}(\underline{\mathcal{\breve{\theta}}}_{n})$ depends
on $\underline{\mathcal{\breve{\theta}}}_{n}=(\check{\theta}_{n}^{\prime }$ $%
\check{s}_{v,n}^{2})^{\prime }$, whereas the observed Hessian $\partial ^{2}%
\widetilde{l}_{N,n}(\theta _{n})/\partial \theta _{n}\partial \theta
_{n}^{\prime }|_{\check{\theta}_{n}}$ only depends on $\check{\theta}_{n}$.} 
\footnote{%
This reparametrization is the same as the one used in Kruiniger (2013) for
the FE(Q)MLE.} Note that the values of the elements of $\overline{H}(%
\underline{\mathcal{\breve{\theta}}}_{n})$ do not depend on the true
distribution of the data.\ We will now introduce the QLM test-statistic $QLM(%
\widetilde{\underline{\mathcal{\theta }}}_{n})$ for testing $H_{0}:$ $A%
\mathcal{\theta }_{0,n}=a$, which includes a restriction on $\rho $, (i.e., $%
\rho =a_{1},$) where $\widetilde{\underline{\mathcal{\theta }}}_{n}$ is a
restricted estimate of $\underline{\mathcal{\theta }}_{0,n}$ such that $A%
\widetilde{\mathcal{\theta }}_{n}=a,$ $A$ is a $J\times \dim (\mathcal{%
\theta })$ constant matrix of rank $J,$ and $a\ $is a constant vector.
Define the average information matrix $\overline{\mathcal{J}}\mathcal{(%
\mathcal{\breve{\theta}}}_{n}\mathcal{)}=N^{-1}\tsum_{i=1}^{N}\mathcal{J}_{i}%
\mathcal{(\mathcal{\breve{\theta}}}_{n}\mathcal{)}$ with $\mathcal{J}_{i}%
\mathcal{(\breve{\theta}}_{n}\mathcal{)}=\left( \frac{\partial \widetilde{l}%
_{n,i}(\theta _{n})}{\partial \theta _{n}}|_{\mathcal{\breve{\theta}}%
_{n}}\right) \left( \frac{\partial \widetilde{l}_{n,i}(\theta _{n})}{%
\partial \theta _{n}^{\prime }}|_{\mathcal{\breve{\theta}}_{n}}\right) ,$
where $\widetilde{l}_{n,i}(\theta _{n})$ is the contribution to the modified
profile log-likelihood function $N\times \widetilde{l}_{n,N}(\theta _{n})$
by individual $i$. If $\widetilde{\mathcal{\theta }}_{n}\neq \mathcal{\theta 
}_{\ast }$ for all $\sigma ^{2}>0$ or $J=1,$ then $QLM(\widetilde{\underline{%
\mathcal{\theta }}}_{n})$ is given by:\vspace{-0.1in}%
\begin{eqnarray}
QLM(\widetilde{\underline{\mathcal{\theta }}}_{n}) &=&N\times \frac{\partial 
\widetilde{l}_{n,N}(\widetilde{\mathcal{\theta }}_{n})}{\partial \theta _{n}}%
\overline{H}^{-1}(\widetilde{\underline{\mathcal{\theta }}}_{n})A^{\prime
}\times  \label{qlm} \\
&&(A\overline{H}^{-1}(\widetilde{\underline{\mathcal{\theta }}}_{n})%
\overline{\mathcal{J}}(\widetilde{\mathcal{\theta }}_{n})\overline{H}^{-1}(%
\widetilde{\underline{\mathcal{\theta }}}_{n})A^{\prime })^{-1}A\overline{H}%
^{-1}(\widetilde{\underline{\mathcal{\theta }}}_{n})\frac{\partial 
\widetilde{l}_{n,N}(\widetilde{\mathcal{\theta }}_{n})}{\partial \theta _{n}}%
.  \notag
\end{eqnarray}%
The parameter $\sigma _{v,n}^{2}$ can be estimated by the restricted
FE(Q)MLE, cf. Kruiniger (2025a).

If $H_{0}$ is true and $\widetilde{\mathcal{\theta }}_{n}\neq \mathcal{%
\theta }_{\ast }$ (for all $\sigma ^{2}>0$) or $J=1$, then $QLM(\widetilde{%
\underline{\mathcal{\theta }}}_{n})\sim \chi ^{2}(J).$ To test $H_{0}:$ $%
\rho =a$ for some known value of $a\in (-1,1]$, one can use $QLM(\widetilde{%
\underline{\mathcal{\theta }}}_{n})$ in (\ref{qlm}) with $A=(1$ $\mathbf{0}%
^{\prime })$ and $\frac{\partial \widetilde{l}_{n,N}(\widetilde{\mathcal{%
\theta }}_{n})}{\partial \theta _{n}}=A^{\prime }\frac{\partial \widetilde{l}%
_{n,N}(\widetilde{\mathcal{\theta }}_{n})}{\partial r}$. Note that the value
of $QLM(\widetilde{\underline{\mathcal{\theta }}}_{n})$ remains the same
when $\overline{H}^{-1}(\widetilde{\underline{\mathcal{\theta }}}_{n})$ is
replaced by $adj(\overline{H}(\widetilde{\underline{\mathcal{\theta }}}%
_{n})) $. Furthermore, $\overline{\mathcal{J}}(\widetilde{\mathcal{\theta }}%
_{n})$ and p$\lim_{N\rightarrow \infty }\overline{\mathcal{J}}(\widetilde{%
\mathcal{\theta }}_{n})$ are positive definite.

If $\widetilde{\mathcal{\theta }}_{n}=\mathcal{\theta }_{\ast }$ for some $%
\sigma ^{2}>0,$ $rk(\overline{H}(\widetilde{\underline{\mathcal{\theta }}}%
_{n}))=\dim (\overline{H}(\widetilde{\underline{\mathcal{\theta }}}_{n}))-1$
and hence $rk(adj(\overline{H}(\widetilde{\underline{\mathcal{\theta }}}%
_{n})))=1.$ Furthermore, $\overline{H}(\widetilde{\underline{\mathcal{\theta 
}}}_{n})_{i,j}=0$ iff $i=1$ and/or $j=1,$ $adj(\overline{H}(\widetilde{%
\underline{\mathcal{\theta }}}_{n}))_{i,j}\neq 0$ iff $i=j=1,$ and $adj(%
\overline{H}(\widetilde{\underline{\mathcal{\theta }}}_{n}))(1$ $\mathbf{0}%
^{\prime })^{\prime }\propto (1$ $\mathbf{0}^{\prime })^{\prime }.$ Thus, if 
$\widetilde{\mathcal{\theta }}_{n}=\mathcal{\theta }_{\ast }$ (for some $%
\sigma ^{2}>0$), the QLM test statistic in (\ref{qlm}) for testing $H_{0}:$ $%
\rho =a$ equals $N\times (\frac{\partial \widetilde{l}_{n,N}(\mathcal{\theta 
}_{n})}{\partial r}|_{\theta _{\ast }})^{2}/(N^{-1}\tsum_{i=1}^{N}(\frac{%
\partial \widetilde{l}_{n,i}(\mathcal{\theta }_{n})}{\partial r}|_{\theta
_{\ast }})^{2}).$\linebreak As p$\lim_{N\rightarrow \infty
}N^{-1}\tsum_{i=1}^{N}(\frac{\partial \widetilde{l}_{n,i}(\mathcal{\theta }%
_{n})}{\partial r}|_{\theta _{\ast }})^{2}>0,$ it follows that one can still
use (\ref{qlm}) for testing $H_{0}:$ $\rho =a$ when $\widetilde{\mathcal{%
\theta }}_{n}=\mathcal{\theta }_{\ast }$ (for some $\sigma ^{2}>0$).
However, if $\widetilde{\mathcal{\theta }}_{n}=\mathcal{\theta }_{\ast }$
(for some $\sigma ^{2}>0$) and $J\geq 2,$ then p$\lim_{N\rightarrow \infty
}\det (A\,adj(\overline{H}(\widetilde{\underline{\mathcal{\theta }}}_{n}))%
\overline{\mathcal{J}}(\widetilde{\mathcal{\theta }}_{n})adj(\overline{H}(%
\widetilde{\underline{\mathcal{\theta }}}_{n}))A^{\prime })=0\mathbf{.}$ To
test $H_{0}:$ $A\mathcal{\theta }_{0,n}=a$ when $\widetilde{\mathcal{\theta }%
}_{n}=\mathcal{\theta }_{\ast }$ (for some $\sigma ^{2}>0$) and $J\geq 2,$
one can use the following QLM test statistic, cf. Bottai (2003) and
Kruiniger (2025a):\vspace{-0.1in} 
\begin{eqnarray}
QLM(\widetilde{\underline{\mathcal{\theta }}}_{n}) &=&N\times \widetilde{S}%
_{N}^{\prime }(\widetilde{\mathcal{\theta }}_{n})\widetilde{\mathcal{H}}%
^{-1}(\widetilde{\underline{\mathcal{\theta }}}_{n})\widetilde{A}^{\prime
}\times  \label{qlm1} \\
&&(\widetilde{A}\widetilde{\mathcal{H}}^{-1}(\widetilde{\underline{\mathcal{%
\theta }}}_{n})\widetilde{\mathcal{J}}(\widetilde{\mathcal{\theta }}_{n})%
\widetilde{\mathcal{H}}^{-1}(\widetilde{\underline{\mathcal{\theta }}}_{n})%
\widetilde{A}^{\prime })^{-1}\widetilde{A}\widetilde{\mathcal{H}}^{-1}(%
\widetilde{\underline{\mathcal{\theta }}}_{n})\widetilde{S}_{N}(\widetilde{%
\mathcal{\theta }}_{n}),  \notag
\end{eqnarray}%
$\vspace{-0.08in}$with$\vspace{-0.08in}$%
\begin{eqnarray*}
\widetilde{S}_{N}(\widetilde{\mathcal{\theta }}_{n})
&=&N^{-1}\sum\nolimits_{i=1}^{N}S_{i},\text{\quad }\widetilde{\mathcal{J}}(%
\widetilde{\mathcal{\theta }}_{n})=N^{-1}\sum%
\nolimits_{i=1}^{N}(S_{i}S_{i}^{\prime }),\text{\quad }\widetilde{A}%
=A,\medskip \\
S_{i} &=&(S_{i,1},S_{i,2}^{\prime })^{\prime },\text{\quad }S_{i,1}=\frac{1}{%
2}\frac{\partial ^{2}\widetilde{l}_{n,i}}{\partial r_{n}^{2}}|_{\widetilde{%
\mathcal{\theta }}_{n}},\text{\quad }S_{i,2}=\frac{\partial \widetilde{l}%
_{n,i}}{\partial d_{n}}|_{\widetilde{\mathcal{\theta }}_{n}},\vspace{-0.14in}
\\
\widetilde{\mathcal{H}}_{1,1} &=&\frac{2}{4!}E_{\widetilde{\underline{%
\mathcal{\theta }}}_{n}}(\frac{\partial ^{4}\widetilde{l}_{n,N}}{\partial
r_{n}^{4}}|_{\widetilde{\underline{\mathcal{\theta }}}_{n}}),\text{\quad }%
\widetilde{\mathcal{H}}_{1,2}^{\prime }=\widetilde{\mathcal{H}}%
_{2,1}=\medskip \frac{1}{2!}E_{\widetilde{\underline{\mathcal{\theta }}}%
_{n}}(\frac{\partial ^{3}\widetilde{l}_{n,N}}{\partial r_{n}^{2}\partial
d_{n}}|_{\widetilde{\underline{\mathcal{\theta }}}_{n}}), \\
\widetilde{\mathcal{H}}_{2,2} &=&\frac{2}{2!}E_{\widetilde{\underline{%
\mathcal{\theta }}}_{n}}(\frac{\partial ^{2}\widetilde{l}_{n,N}}{\partial
d_{n}\partial d_{n}^{\prime }}|_{\widetilde{\underline{\mathcal{\theta }}}%
_{n}}),\text{\quad }\widetilde{\mathcal{H}}(\widetilde{\underline{\mathcal{%
\theta }}}_{n})=\left[ 
\begin{array}{cc}
\widetilde{\mathcal{H}}_{1,1} & \widetilde{\mathcal{H}}_{1,2} \\ 
\widetilde{\mathcal{H}}_{2,1} & \widetilde{\mathcal{H}}_{2,2}%
\end{array}%
\right] ,
\end{eqnarray*}%
where we have partitioned $\theta _{n}$ as $\theta _{n}=(r_{n},d_{n}^{\prime
})^{\prime }$ and used $\widetilde{l}_{n,N}$ and $\widetilde{l}_{n,i}$ as
short for $\widetilde{l}_{n,N}(\theta _{n})$ and $\widetilde{l}_{n,i}(\theta
_{n})$. Note that $\frac{\partial \widetilde{l}_{n,i}(\widetilde{\mathcal{%
\theta }}_{n})}{\partial r}$ and $\overline{H}(\widetilde{\underline{%
\mathcal{\theta }}}_{n})$ in (\ref{qlm}) have been replaced by $S_{i,1}$ and 
$\widetilde{\mathcal{H}}(\widetilde{\underline{\mathcal{\theta }}}_{n})$ in (%
\ref{qlm1}). Furthermore, to derive the QLM test-statistic $QLM(\widetilde{%
\underline{\mathcal{\theta }}}_{n})$ given in (\ref{qlm1}), the hypothesis $%
\rho =1$ had to be reformulated as $(\rho -1)^{2}=0$. As a result $%
\widetilde{A}_{1,1}$ is equal to $\frac{1}{2}\frac{\partial ^{2}(\rho -1)^{2}%
}{\partial \rho ^{2}}=1$. If $H_{0}$ is true, $\widetilde{\mathcal{\theta }}%
_{n}=\mathcal{\theta }_{\ast }$ (for some $\sigma ^{2}>0$) and $J\geq 2$,
then $QLM(\widetilde{\underline{\mathcal{\theta }}}_{n})\sim \chi ^{2}(J).$

\begin{theorem}
Under regularity conditions (A1)-(A7) given in the appendix, the Quasi LM
test for $H_{0}:$ $A\mathcal{\theta }_{0,n}=a,$ where $A_{1,.}=(1$ $\mathbf{0%
}^{\prime }),$ that is based on (\ref{qlm}) if $\widetilde{\mathcal{\theta }}%
_{n}\neq \mathcal{\theta }_{\ast }$ for all $\sigma ^{2}>0,$ and on (\ref%
{qlm1}) if $\widetilde{\mathcal{\theta }}_{n}=\mathcal{\theta }_{\ast }$ for
some $\sigma ^{2}>0,$ has correct asymptotic size in a uniform sense.
\end{theorem}

Confidence sets (CSs) that are obtained by inverting the tests based on (\ref%
{qlm}) and (\ref{qlm1}) have correct asymptotic size in a uniform sense.
Other tests (and CSs) for $\rho $ that have correct asymptotic size include
(CSs based on) the GMM LM test(-statistic)s of Newey and West (1987) that
exploit the moments conditions of the System GMM and the nonlinear
Ahn-Schmidt (AS) GMM estimator, respectively, see Kruiniger (2009) for the
System version and Bun and Kleibergen (2022) for the AS version of the test,
and identication-robust test(-statistics)s such as the GMM AR test of Stock
and Wright (2000) and the KLM and GMM-CLR tests of Kleibergen (2005) that
exploit System and AS moments conditions, cf. Bun and Kleibergen (2022).
Kruiniger (2025a) has shown that if the data are i.i.d. and normal, then the
QLM test for testing an hypothesis about $\rho $ that is based on the FEMLE
and uses the expected Hessian\ shares the optimal power properties of the
KLM\ test in a worst case scenario. To test $H_{0}:$ $\rho =1$ one could
also use a Wald test based on\linebreak $\sqrt{N}(\widehat{\rho }%
_{C}-1)^{2}. $ Under $H_{0}\ \sqrt{N}(\widehat{\rho }_{C}-1)^{2}\overset{d}{%
\rightarrow }Z_{1}\mathbf{1}\{Z_{1}>0\},$ cf. Theorem 2. Recall that $%
Z_{1,N}=\left( -\frac{1}{2}\frac{\partial ^{3}\widetilde{l}_{N}^{c}(1)}{%
\partial r^{3}}\right) ^{-1}N^{1/2}\left( \frac{\partial \widetilde{l}%
_{N}^{c}(1)}{\partial r}\right) \overset{d}{\rightarrow }Z_{1},$ with p$%
\lim\nolimits_{N\rightarrow \infty }\frac{\partial ^{3}\widetilde{l}%
_{N}^{c}(1)}{\partial r^{3}}=\frac{\partial ^{3}\widetilde{l}^{c}(1)}{%
\partial r^{3}}=\frac{T(T-1)(T+1)}{12}$ and $\frac{\partial \widetilde{l}%
_{N}^{c}(1)}{\partial r}$ given in (\ref{lcn}). When the data are
heterogeneous and/or non-normal, one can bootstrap the distribution of $%
N^{1/2}\left( \frac{\partial \widetilde{l}_{N}^{c}(1)}{\partial r}\right) $
or estimate the averages of the second and the fourth moments of the $%
\varepsilon _{i,t}$ by using that under $H_{0}$\ $\varepsilon
_{i}=y_{i}-y_{i,-1}$ for $i=1,...,N.$ To test $H_{0}:$ $\rho =1$ one could
also use any other panel unit root test, e.g. the test of Harris and
Tzavalis (1999) that is based on the bias-corrected LSDV estimator for $\rho
,$ i.e., $\widehat{\rho }_{ML}+\frac{3}{T+1}$, where $-\frac{3}{T+1}$ is the
asymptotic bias of $\widehat{\rho }_{ML}$ when $\rho =1.$ The rate of
convergence of $\widehat{\rho }_{ML}$ is $N^{1/2}$ which is faster than $%
N^{1/4},$ the rate of $\widehat{\rho }_{C}.$ Hence if $N$ is large enough
inference based on $\widehat{\rho }_{ML}$ is better in terms of power and
size. Finally, to test a hypothesis that only involves $\beta $, one can use
a Wald test based on $\widehat{\beta }_{C}$.\vspace{-0.15in}

\section{The finite sample performance of the Modified ML\protect\linebreak %
estimators and the Quasi LM test}

In this section we compare through Monte Carlo simulations the finite sample
properties of three estimators in various panel AR(1) models without
covariates: $\widehat{\rho }_{C}$; the REMLE for $\rho $ that has been
proposed by both Chamberlain (1980) and Anderson and Hsiao (1982),
henceforth $\widehat{\rho }_{REML}$; and the FEMLE for $\rho $ (i.e., $%
\widehat{\rho }_{FEML}$) that has been proposed by Hsiao et al. (2002). We
study how the properties of these estimators are affected if we change (1)
the distributions of the $v_{i}=y_{i,0}-\mu _{i}$ or (2) the ratio of the
variances of the error components, i.e. $\sigma _{\mu }^{2}/\sigma ^{2}$. We
conducted the simulation experiments for $(T,N)=(4,100),$ $(9,100),$ $%
(4,500) $ or $(9,500)$ and $\rho =0.5,$ $0.8,$ $0.9,$ $0.95,$ $0.98$ or $1.$

In all simulation experiments the error components have been drawn from
normal distributions with zero means. We assumed that $\sigma _{\mu }^{2}=0,$
$1$ or $25.$ For the $\varepsilon _{i,t}$ we assumed homoskedasticity and no
autocorrelation: $E(\varepsilon _{i}\varepsilon _{i}^{\prime })=\sigma ^{2}I$
with $\sigma ^{2}=1.$

In order to assess how the assumptions with respect to $y_{i,0}-\mu _{i}$, $%
i=1,...,N,$ affect the properties of the estimators, we conducted two
different sets of experiments, which are identified by a capital: in one
set, labeled NS, the initial observations are non-stationary, i.e., $%
y_{i,0}-\mu _{i}=0$, $i=1,...,N,$ whereas in the other set, labeled S, the
initial observations are drawn from stationary distributions when $%
\left\vert \rho \right\vert <1$, i.e., $(y_{i,0}-\mu _{i})\sim N(0,\sigma
_{i,0}^{2}/(1-\rho ^{2}))$ with $\sigma _{i,0}^{2}=\sigma ^{2},$ although $%
y_{i,0}-\mu _{i}=0$, $i=1,...,N,$ when $\rho =1$.

Note that all four estimators suffer from a weak moment conditions problem
when $\rho $ is close to one, cf. Kruiniger (2013).

In the cases of the RE- and FEMLE $(1-\rho )\mu _{i}+\varepsilon _{i}$ is
decomposed as $(1-\rho )\pi y_{i,0}-(1-\rho )v_{i}+\varepsilon _{i}=(1-\rho
)\pi y_{i,0}+u_{i}$ with $\pi =1$ for the FE case. In the experiments we
imposed homoskedasticity on their likelihood functions and added the
restrictions $\sigma ^{2}>0$ and $(T-1)(1-\rho )^{2}\sigma _{v}^{2}+\sigma
^{2}>0$ to ensure that the estimates of $E(u_{i}u_{i}^{\prime })$ were PD.

We allowed for time effects by subtracting cross-sectional averages from the
data.

To speed up the computations, we computed $\widehat{\rho }_{C}$ by
maximizing $\widetilde{l}_{N}(\theta )$ subject to\thinspace $-1\leq r\leq
1.4$ rather than $-1\leq r<\infty .$ (We also tried using $-1\leq r\leq 2,$
but never found an internal local maximum between $1.4$ and $2$.) If no
internal local maximum was found, we computed $\widehat{\rho }_{C}$ by
solving (\ref{mml2}) s.t.\thinspace $-1\leq r\leq 1.4$ using grid search.

Tables 1-6 report the simulation results in terms of the biases and root
mean squared errors (RMSEs) of the estimators and the relative frequencies
that $\widehat{\rho }_{LAN}$ did not exist (NM). The tables differ with
respect to the dimensions of the panel and the assumptions made about the $%
y_{i,0}-\mu _{i}$, $i=1,...,N$. Inspection of the\thinspace results
leads\thinspace to\thinspace the\thinspace following conclusions:\thinspace 
\footnote{%
Dhaene and Jochmans (2016) report simulations results on the finite sample
properties of their Adjusted Likelihood estimator ($\widehat{\rho }_{ADJ}$),
the bias corrected LSDV estimator of Hahn and Kuersteiner (2002) ($\widehat{%
\rho }_{HK}$) and the 1-step\ GMM estimator of Arellano and Bond (1991) ($%
\widehat{\rho }_{AB}$). Some of their simulation experiments are equal to
some of our experiments. The results for these experiments show that $%
\widehat{\rho }_{ADJ}$ and $\widehat{\rho }_{C}$ are very similar and that $%
\widehat{\rho }_{HK}$ has a large bias when $T$ is small. $\widehat{\rho }%
_{AB}$ has poor properties when $\rho $ is close to 1 due to weak
instruments.} \vspace{-0.04in}

\begin{enumerate}
\item In almost all experiments (the exception is design NS with $N=100$ and 
$\rho =.0.5$) $\widehat{\rho }_{REML}$ is superior in terms of RMSE for
`smaller' values of $\rho $ (i.e., values closer to 0), $\widehat{\rho }%
_{FEML}$ is superior for `larger' values of $\rho $ (i.e., values closer to
1), while $\widehat{\rho }_{C}$ is superior on an interval of `intermediate'
values of $\rho $, which includes $\rho =0.8$ when $T=4$ and $N=100,$ and $%
\rho =0.9$ when $T=4$ and $N=500.$ In most experiments $\widehat{\rho }%
_{REML}$ is superior when $\rho =0.5,$ while $\widehat{\rho }_{FEML}$ is
superior when $\rho $ is near/equals $1.$\linebreak\ When $\rho $ is near $%
1, $ the bias of $\widehat{\rho }_{C}$ is larger than the biases of $%
\widehat{\rho }_{FEML}$ and $\widehat{\rho }_{REML}$.

\item When $T$ or $N$ increases, the values of the bounds of the interval
for $\rho $ on which $\widehat{\rho }_{C}$ is superior increase. When $T=9$
and $N=500,$ $\widehat{\rho }_{C}$ is superior around $\rho =0.95.$
Furthermore, when $\rho =0.50$ and $T=9$ or $N=500,$ $\widehat{\rho }_{FEML}$
is often the most efficient estimator after $\widehat{\rho }_{REML}$.

\item When $\sigma _{\mu }^{2}/\sigma ^{2}$ increases, the RMSE of $\widehat{%
\rho }_{REML}$ increases and hence the value of the lowerbound of the
interval of values of $\rho $ on which $\widehat{\rho }_{C}$ is superior
decreases.

\item When $Var(y_{i,0}-\mu _{i})/\sigma ^{2}$ decreases, the bias and the
RMSE of $\widehat{\rho }_{C}$ and the RMSE of $\widehat{\rho }_{REML}$
increase and the value of the upperbound of the interval of values of $\rho $
on which $\widehat{\rho }_{C}$ is superior decreases.

\item When $T=4$ and $N=100,$ $NM>0.35$ for $\rho \geq 0.8;$ when $T=4$ and $%
N=500,$ $NM>0.29$ for $\rho \geq 0.8;$ when $T=9$ and $N=100,$ $NM>0.35$ for 
$\rho \geq 0.9;$ and when $T=9$ and $N=500,$ $NM>0.25$ for $\rho \geq 0.9.$
Generally, the higher the value of $\rho ,$ the higher the value of $NM$.
When $\rho =1,$ $NM\approx 0.50$ for all panels considered, which supports
the idea that even asymptotically $\widehat{\rho }_{LAN}$ may not exist when 
$\rho =1$. If the value of $Var(y_{i,0}-\mu _{i})/\sigma ^{2}$ decreases,
the value of $NM$ increases. Under design NS, when $T=4$, $N=100$ and $\rho
=0.5,$ we still have $NM>0.3.$
\end{enumerate}

We have also investigated the size and power properties of the modified
likelihood based QLM-test for testing $H_{0}:$ $\rho =a$, that is, $QLM(%
\mathcal{\rho }).$ To this end, we conducted three types of Monte Carlo
experiments. The designs of two of them, labelled S-Normal and NS-Normal,
were similar to designs S and NS described above. The designs of the third
kind of experiments, labelled S-ChiSq., were also similar to S with one
difference: the $\varepsilon _{i,t}$ were i.i.d. $(\chi ^{2}(1)-1)/\sqrt{2}$
instead of i.i.d. $N(0,1)$ so that $(y_{i,0}-\mu _{i})\sim (\chi ^{2}(1)-1)/%
\sqrt{2(1-\rho ^{2})}$ instead of $N(0,1/(1-\rho ^{2})).$ In all experiments 
$\mu _{i}\sim N(0,1).$ We used various true values for $\rho $ including $%
0.5,$ $0.9,$ $0.95$ and $0.99$. The results for the power of $QLM(\mathcal{%
\rho })$ were based on testing $H_{0}:$ $\rho =0.8$. In all experiments $T=9$
and $N\in \{100,500\}$.

$QLM(\mathcal{\rho })$ depends on $\mathcal{H}(\widetilde{\underline{%
\mathcal{\theta }}}_{n})$, i.e., an estimate of the expected Hessian that is
based on the restricted estimate $\widetilde{\underline{\mathcal{\theta }}}%
_{n}$. One of the parameters in $\mathcal{H}(\underline{\mathcal{\theta }}%
_{0,n})$ is $\sigma _{v,n}^{2}$. However, the latter is not estimated by a
MMLE. Instead we used the restricted FE(Q)MLE for $\sigma _{v,n}^{2}$.

Tables 7 and 8 report the simulation results for the size and the power of $%
QLM(\mathcal{\rho })$, respectively.\footnote{%
Dhaene and Jochmans (2016) report simulations results on the finite sample
properties of confidence intervals for $\rho $ based on $\widehat{\rho }%
_{ADJ}$, $\widehat{\rho }_{HK}$ and $\widehat{\rho }_{AB},$ respectively,
and their first-order asymptotic standard errors. Their results show that
none of these intervals have correct size when $\rho $ is close to one, with
the size distortions being particularly large for the confidence intervals
based on $\widehat{\rho }_{HK}$ and $\widehat{\rho }_{AB}$.} Table 7 shows
that the empirical size of the test is very close to the nominal size of 5\%
in all experiments, including those where $\rho $ is close to one. Finally,
table 8 shows that the power properties of $QLM(\mathcal{\rho })$ do not
change much across the three types of experiments and also that its power is
still high when (true) $\rho =0.99$ despite weak identification in that case.%
\vspace{-0.1in}

\section{Concluding remarks}

Alvarez and Arellano (2022) and Juodis (2013) have extended the MMLE of
Lancaster to panel AR(1) models that allow for time-series
heteroskedasticity. Their estimators suffer from the same problems as
Lancaster's MMLE, namely a weak moment conditions problem if the parameter
values are close to the unit root \textit{and} time-series homoskedasticity,
cf. Alvarez and Arellano (2022) and Kruiniger (2013); the related problem of
possible non-existence; and the possibility of non-uniqueness of local
maxima of the modified profile likelihood function. The non-existence
problem can be solved by generalizing their estimators in a similar way as
Lancaster's estimator has been generalized to (\ref{mml}) or (\ref{mml2}).
However, it is unclear whether the modified profile likelihood function has
at most one local maximum even when the parameter space for $\rho $ is
restricted to $[-1,1]$.\footnote{%
The modified profile likelihood equation for $\rho $ is a polynomial in $r$.
If the model has no covariates, then the coefficients of this polynomial are
functions of $\rho ,$ $\sigma _{v}^{2}$ and $T$ variance parameters instead
of one.} If uniqueness would not hold, then one could select a local maximum
that is (plausible and) closest to the value of $\widehat{\rho }_{FEML}$ (or 
$\widehat{\rho }_{REML}$), which is a consistent estimator, as the MMLE.%
\footnote{%
Note that this method of selecting the MMLE is also \textquotedblleft
sensible\textquotedblright\ in finite samples.}

Alvarez and Arellano (2022) and Dhaene and Jochmans (2016) have also
extended the MMLE of Lancaster to panel AR(p) models, while Juodis (2013)
has also extended the MMLE of Lancaster to panel VARX(1) models. Comments
similar to those made in the previous paragraph apply to these extensions.
The MMLEs discussed in section 3 are inconsistent for models with endogenous
or predetermined covariates. However, in some cases these models can be
replaced by VAR models.

It seems reasonable to expect that the aforementioned extensions of the
MMLEs to more general models may also outperform the RE- and FEMLEs for
those models in panels of realistic dimensions for some parts of the
parameter space.\ However, a comprehensive Monte Carlo study of their finite
sample properties is left for future research.

Finally, we note that Bester and Hansen (2007) and Arellano and Bonhomme
(2009) have proposed priors that result in first-order unbiased Bayesian
estimators for $\rho $ in a version of model (\ref{mdl2}) that does not
include the $K$ exogenous covariates. \pagebreak

\appendix 

\section{Proofs and derivations$\protect\vspace{-0.04in}$}

\textbf{\noindent The asymptotic bias of the LSDV estimator for }$\rho ,$ $%
\widehat{\rho }_{ML}$\textbf{:}

The LSDV\ estimators for $\rho $ and $\beta ,$ $\widehat{\rho }_{ML}$ and $%
\widehat{\beta }_{ML},$ satisfy the profile likelihood equations for $\rho $
and $\beta :$ 
\begin{eqnarray}
\sum_{i=1}^{N}y_{i,-1}^{\prime }Q(y_{i}-\widehat{\rho }_{ML}y_{i,-1}-X_{i}%
\widehat{\beta }_{ML}) &=&0\text{ and }  \label{likeqs} \\
\sum_{i=1}^{N}X_{i}^{\prime }Q(y_{i}-\widehat{\rho }_{ML}y_{i,-1}-X_{i}%
\widehat{\beta }_{ML}) &=&0.  \notag
\end{eqnarray}%
Let $r_{xy_{-1}}^{2}=(\sum_{i=1}^{N}y_{i-1}^{\prime
}Qy_{i,-1})^{-1}\sum_{i=1}^{N}(y_{i-1}^{\prime
}QX_{i})(\sum_{i=1}^{N}X_{i}^{\prime
}QX_{i})^{-1}\sum_{i=1}^{N}(X_{i}^{\prime }Qy_{i,-1})$, $%
s_{y}^{2}=(T-1)^{-1}N^{-1}\tsum\nolimits_{i=1}^{N}y_{i,-1}^{\prime
}Qy_{i,-1} $ and $\rho _{ML}=$ p$\lim_{N\rightarrow \infty }\widehat{\rho }%
_{ML}.$ Using that $y_{i,-1}-\mu _{i}\iota -\iota \overline{x}_{i}^{\prime }%
\check{\beta}=\varphi v_{i}+\Phi QX_{i}\beta +\Phi \varepsilon _{i}=%
\widetilde{Z}_{i}+\Phi \varepsilon _{i}$ and $Q\iota =0$, it can be shown
that the asymptotic bias of $\widehat{\rho }_{ML}$ is given by (cf. e.g. Bun
and Carree, 2005):%
\begin{equation}
\rho _{ML}-\rho =-\frac{\sigma ^{2}h(\rho )}{(1-\rho _{xy_{-1}}^{2})\sigma
_{y}^{2}},  \label{abia}
\end{equation}%
where $\rho _{xy_{-1}}^{2}=$ p$\lim_{N\rightarrow \infty }r_{xy_{-1}}^{2},$ $%
\sigma _{y}^{2}=$ p$\lim_{N\rightarrow \infty }s_{y}^{2}$ and $h(\rho
)=-(T-1)^{-1}tr(Q\Phi )=\linebreak \frac{1}{T(T-1)}\sum_{t=1}^{T-1}(T-t)\rho
^{t-1}=\xi ^{\prime }(\rho ).$ Note that $h(\rho )=\frac{T-1-T\rho +\rho ^{T}%
}{T(T-1)(1-\rho )^{2}},$ when $\rho \neq 1,$ and $h(1)=\frac{1}{2}.$
Assumption 1 implies that $\sigma _{y}^{2}=\frac{\sigma ^{2}}{T-1}tr(\Phi
^{\prime }Q\Phi )+\frac{1}{T-1}E(\widetilde{Z}_{i}Q\widetilde{Z}_{i})$ and $%
\frac{\sigma ^{2}}{T-1}tr(\Phi ^{\prime }Q\Phi )>0$ and hence $\sigma
_{y}^{2}>0.$ We also have $\rho _{xy_{-1}}^{2}<1.$ Furthermore, if $%
\left\vert \rho \right\vert \leq 1,$ $h(\rho )>0$ and hence $\rho _{ML}-\rho
<0$ (cf. e.g. Bun and Carree, 2005).

It can also be shown that if $\rho =1,$ then $\rho _{ML}-\rho =-\frac{3}{T+1}
$. Note that $E(\widetilde{Z}_{i}Q\widetilde{Z}_{i})=\sigma _{v}^{2}\varphi
^{\prime }Q\varphi +2E(v_{i}\varphi ^{\prime }Q\Phi QX_{i})\beta +\beta
^{\prime }E(X_{i}^{\prime }Q\Phi Q\Phi QX_{i})\beta .$ Let $f(\rho )=\frac{1%
}{T-1}tr(\Phi ^{\prime }Q\Phi )$ and $g(\rho )=\frac{1}{T-1}\varphi ^{\prime
}Q\varphi $. Below we show that $f(1)=\frac{1}{6}(T+1).$ Furthermore, $%
g(1)=0 $ and when $\rho =1,$ we also have $\beta =\rho _{xy_{-1}}^{2}=0.$ We
conclude that when $\rho =1,$ then $E(\widetilde{Z}_{i}Q\widetilde{Z}%
_{i})=0, $ $\sigma _{y}^{2}=\frac{\sigma ^{2}}{6}(T+1)\ $and $\rho
_{ML}-\rho =-\frac{3}{T+1}$ (cf. Harris and Tzavalis, 1999).$\bigskip $

\textbf{\noindent }Proof of the claim that $f(1)=\frac{1}{6}(T+1):$

We have $f(\rho )=(T-1)^{-1}tr(\Phi ^{\prime }Q\Phi )=(T-1)^{-1}(tr\Phi
^{\prime }\Phi -T^{-1}\iota ^{\prime }\Phi \Phi ^{\prime }\iota
)=(T-1)^{-1}\times \linebreak (\sum_{t=0}^{T-2}\sum_{s=0}^{t}\rho
^{2s}-T^{-1}\sum_{t=0}^{T-2}(\sum_{s=0}^{t}\rho ^{s})^{2}).$ It follows that 
$f(1)=(T-1)^{-1}(\sum_{t=0}^{T-2}(t+1)-T^{-1}%
\sum_{t=1}^{T-1}t^{2})=(T-1)^{-1}(\frac{1}{2}(T-1)T-\frac{1}{6}(T-1)(2T-1))=%
\frac{1}{6}(T+1).\quad \square $\pagebreak

\textbf{\noindent Some results related to }$\widetilde{l}_{N}^{c}(r)$ 
\textbf{and }$\frac{\partial \widetilde{l}_{N}^{c}(r)}{\partial r}:\medskip $

By the envelope theorem we have $\frac{\partial \widetilde{l}_{N}^{c}(r)}{%
\partial r}=\Psi _{\rho }(r,\widehat{\sigma }^{2}(r,\widehat{\beta }(r)),%
\widehat{\beta }(r))$, i.e.,%
\begin{equation}
\frac{\partial \widetilde{l}_{N}^{c}(r)}{\partial r}=(T-1)\xi ^{\prime }(r)+%
\widehat{\sigma }^{-2}(r,\widehat{\beta }(r))N^{-1}%
\sum_{i=1}^{N}(y_{i}-ry_{i,-1}-X_{i}\widehat{\beta }(r))^{\prime }Qy_{i,-1}.
\label{foc1}
\end{equation}

Let $\widehat{\sigma }_{ML}^{2}=(T-1)^{-1}N^{-1}\sum_{i=1}^{N}[(y_{i}-%
\widehat{\rho }_{ML}y_{i,-1}-X_{i}\widehat{\beta }_{ML})^{\prime }Q(y_{i}-%
\widehat{\rho }_{ML}y_{i,-1}-X_{i}\widehat{\beta }_{ML})].$ Next we show
that the first-order condition for a local maximum of $\widetilde{l}%
_{N}^{c}(r)$ can be written as 
\begin{equation}
\frac{\partial \widetilde{l}_{N}^{c}(r)}{\partial r}=(T-1)\xi ^{\prime }(r)-%
\frac{(T-1)(r-\widehat{\rho }_{ML})}{\widehat{\sigma }%
_{ML}^{2}/(s_{y}^{2}(1-r_{xy_{-1}}^{2}))+(r-\widehat{\rho }_{ML})^{2}}=0.
\label{foc1b}
\end{equation}

\textbf{\noindent }Derivation of (\ref{foc1b}): Using $\sum_{i=1}^{N}X_{i}^{%
\prime }Q(y_{i}-\widehat{\rho }_{ML}y_{i,-1}-X_{i}\widehat{\beta }_{ML})=0$
from (\ref{likeqs}) and $\sum_{i=1}^{N}X_{i}^{\prime
}Q(y_{i}-ry_{i,-1}-X_{i}b)=0$ from (\ref{modlikeqs}), we obtain%
\begin{equation}
\widehat{\beta }_{ML}-b=(\sum_{i=1}^{N}X_{i}^{\prime
}QX_{i})^{-1}\sum_{i=1}^{N}(X_{i}^{\prime }Qy_{i,-1})(r-\widehat{\rho }%
_{ML}).  \label{bml}
\end{equation}%
Next, using $\sum_{i=1}^{N}y_{i,-1}^{\prime }Q(y_{i}-\widehat{\rho }%
_{ML}y_{i,-1}-X_{i}\widehat{\beta }_{ML})=0$ from (\ref{likeqs}), we obtain%
\begin{multline*}
\sum_{i=1}^{N}y_{i,-1}^{\prime
}Q(y_{i}-ry_{i,-1}-X_{i}b)=\sum_{i=1}^{N}y_{i,-1}^{\prime }Q(y_{i,-1}(%
\widehat{\rho }_{ML}-r)+X_{i}(\widehat{\beta }_{ML}-b))= \\
(\widehat{\rho }_{ML}-r)(\sum_{i=1}^{N}(y_{i-1}^{\prime
}Qy_{i,-1})-\sum_{i=1}^{N}(y_{i-1}^{\prime
}QX_{i})(\sum_{i=1}^{N}X_{i}^{\prime
}QX_{i})^{-1}\sum_{i=1}^{N}(X_{i}^{\prime }Qy_{i,-1})).
\end{multline*}%
Hence\ 
\begin{equation}
(T-1)^{-1}N^{-1}\sum_{i=1}^{N}(y_{i}-ry_{i,-1}-X_{i}\widehat{\beta }%
(r))^{\prime }Qy_{i,-1}=(\widehat{\rho }_{ML}-r)s_{y}^{2}(1-r_{xy_{-1}}^{2}).
\label{tusstap}
\end{equation}%
Using $\sum_{i=1}^{N}y_{i,-1}^{\prime }Q(y_{i}-\widehat{\rho }%
_{ML}y_{i,-1}-X_{i}\widehat{\beta }_{ML})=0$ and $\sum_{i=1}^{N}X_{i}^{%
\prime }Q(y_{i}-\widehat{\rho }_{ML}y_{i,-1}-X_{i}\widehat{\beta }_{ML})=0$
from (\ref{likeqs}), we obtain%
\begin{equation*}
\sum_{i=1}^{N}[(y_{i}-ry_{i,-1}-X_{i}b)^{\prime }Q(y_{i}-ry_{i,-1}-X_{i}b)]=
\end{equation*}%
\begin{multline*}
\sum_{i=1}^{N}[(y_{i}-\widehat{\rho }_{ML}y_{i,-1}-X_{i}\widehat{\beta }%
_{ML})^{\prime }Q(y_{i}-\widehat{\rho }_{ML}y_{i,-1}-X_{i}\widehat{\beta }%
_{ML})+ \\
((\widehat{\rho }_{ML}-r)y_{i,-1}+X_{i}(\widehat{\beta }_{ML}-b))^{\prime
}Q(y_{i,-1}(\widehat{\rho }_{ML}-r)+X_{i}(\widehat{\beta }_{ML}-b))].
\end{multline*}%
In addition, by using (\ref{bml}) once more, we obtain%
\begin{multline*}
\sum_{i=1}^{N}[((\widehat{\rho }_{ML}-r)y_{i,-1}+X_{i}(\widehat{\beta }%
_{ML}-b))^{\prime }Q(y_{i,-1}(\widehat{\rho }_{ML}-r)+X_{i}(\widehat{\beta }%
_{ML}-b))]= \\
(\widehat{\rho }_{ML}-r)^{2}[\sum_{i=1}^{N}y_{i,-1}^{\prime
}Qy_{i,-1}-\sum_{i=1}^{N}(y_{i-1}^{\prime
}QX_{i})(\sum_{i=1}^{N}X_{i}^{\prime
}QX_{i})^{-1}\sum_{i=1}^{N}(X_{i}^{\prime }Qy_{i,-1})].
\end{multline*}%
Hence 
\begin{gather}
\widehat{\sigma }^{2}(r,\widehat{\beta }(r))=(T-1)^{-1}N^{-1}%
\sum_{i=1}^{N}(y_{i}-ry_{i,-1}-X_{i}\widehat{\beta }(r))^{\prime
}Q(y_{i}-ry_{i,-1}-X_{i}\widehat{\beta }(r))=  \notag \\
\widehat{\sigma }_{ML}^{2}+(\widehat{\rho }%
_{ML}-r)^{2}s_{y}^{2}(1-r_{xy_{-1}}^{2}).  \label{appsig}
\end{gather}%
Finally, combining (\ref{foc1}) with (\ref{tusstap}) and (\ref{appsig})
yields (\ref{foc1b}).$\bigskip $

Next we show that $\sigma _{ML}^{2}=$ p$\lim_{N\rightarrow \infty }\widehat{%
\sigma }_{ML}^{2}>0$.$\medskip $

\textbf{\noindent }Proof of the claim that $\sigma _{ML}^{2}>0:$

Using $Q(y_{i}-\widehat{\rho }_{ML}y_{i,-1}-X_{i}\widehat{\beta }%
_{ML})=Q(\varepsilon _{i}+(\rho -\widehat{\rho }_{ML})y_{i,-1}+X_{i}(\beta -%
\widehat{\beta }_{ML}))$ and $Qy_{i,-1}=Q(\widetilde{Z}_{i}+\Phi \varepsilon
_{i}),$ where $\widetilde{Z}_{i}=\varphi v_{i}+\Phi QX_{i}\beta ,$ we obtain 
$\widehat{\sigma }_{ML}^{2}=(T-1)^{-1}N^{-1}\sum_{i=1}^{N}[(\varepsilon
_{i}+(\rho -\widehat{\rho }_{ML})(\widetilde{Z}_{i}+\Phi \varepsilon
_{i})+X_{i}(\beta -\widehat{\beta }_{ML}))^{\prime }Q(\varepsilon _{i}+(\rho
-\widehat{\rho }_{ML})(\widetilde{Z}_{i}+\Phi \varepsilon _{i})+X_{i}(\beta -%
\widehat{\beta }_{ML}))].$

Assumption 1 implies that $\varepsilon _{i}|(v_{i},QX_{i})\sim i.i.d.\mathit{%
N}(0,\sigma ^{2}I_{T}),$ $i=1,...,N,$ with $\sigma ^{2}>0.$ It follows that $%
\sigma _{ML}^{2}=$ p$\lim_{N\rightarrow \infty }\widehat{\sigma }%
_{ML}^{2}\geq $ p$\lim_{N\rightarrow \infty
}(T-1)^{-1}N^{-1}\sum_{i=1}^{N}[(\varepsilon _{i}+(\rho -\widehat{\rho }%
_{ML})\Phi \varepsilon _{i})^{\prime }Q\times $ $(\varepsilon _{i}+(\rho -%
\widehat{\rho }_{ML})\Phi \varepsilon _{i})]=\sigma
^{2}(T-1)^{-1}tr((I+(\rho -\rho _{ML})\Phi )^{\prime }Q(I+(\rho -\rho
_{ML})\Phi ))>0.\quad \square \bigskip $

\textbf{\noindent Proof of the claim that }$\widetilde{l}_{N}^{c}(r)$ 
\textbf{converges uniformly in probability to} $\widetilde{l}^{c}(r):$

We have $\widetilde{l}_{N}^{c}(r)=\widetilde{l}_{N}(r,\widehat{\sigma }%
^{2}(r,\widehat{\beta }(r)),\widehat{\beta }(r))=(T-1)\xi (r)-\frac{T-1}{2}%
\log (\widehat{\sigma }^{2}(r,\widehat{\beta }(r)))-\frac{T-1}{2}$ and from (%
\ref{appsig}), $\widehat{\sigma }^{2}(r,\widehat{\beta }(r))=\widehat{\sigma 
}_{ML}^{2}+(\widehat{\rho }_{ML}-r)^{2}s_{y}^{2}(1-r_{xy_{-1}}^{2}).$ Note
that $-\log (\widehat{\sigma }^{2}(r,\widehat{\beta }(r)))$ is a concave
function of $r.$ Then it follows from pointwise convergence of $\log (%
\widehat{\sigma }^{2}(r,\widehat{\beta }(r)))$ to the function $\log (\sigma
^{2}(r))\equiv \log (\sigma _{ML}^{2}+(\rho _{ML}-r)^{2}\sigma
_{y}^{2}(1-\rho _{xy_{-1}}^{2}))$ that p$\lim_{N\rightarrow \infty
}\sup_{r\in \lbrack -1,\infty )}\left\vert \widetilde{l}_{N}^{c}(r)-\right. $
$\left. \widetilde{l}^{c}(r)\right\vert =0,$ see e.g. Newey and McFadden
(1994, section 2.6).$\quad \square \bigskip $\pagebreak

\textbf{\noindent Some results related to }$\widetilde{l}^{c}(r),$ $\frac{%
\partial \widetilde{l}^{c}(r)}{\partial r}|_{\rho }=0$ \textbf{and} $\frac{%
\partial ^{2}\widetilde{l}^{c}(r)}{\partial r^{2}}|_{\rho }:\medskip $

The first-order condition for a local maximum of $\widetilde{l}^{c}(r)$ can
be written as:%
\begin{equation}
\frac{\partial \widetilde{l}^{c}(r)}{\partial r}=(T-1)\xi ^{\prime }(r)-%
\frac{(T-1)(r-\rho _{ML})}{\sigma _{ML}^{2}/(\sigma _{y}^{2}(1-\rho
_{xy_{-1}}^{2}))+(r-\rho _{ML})^{2}}=0.  \label{foc2b}
\end{equation}%
The second-order condition for a local maximum of $\widetilde{l}^{c}(r)$ is
given by:%
\begin{equation*}
\frac{\partial ^{2}\widetilde{l}^{c}(r)}{\partial r^{2}}=(T-1)\xi ^{\prime
\prime }(r)-\frac{(T-1)(\sigma _{ML}^{2}/(\sigma _{y}^{2}(1-\rho
_{xy_{-1}}^{2}))-(r-\rho _{ML})^{2})}{(\sigma _{ML}^{2}/(\sigma
_{y}^{2}(1-\rho _{xy_{-1}}^{2}))+(r-\rho _{ML})^{2})^{2}}<0.
\end{equation*}

Below we show that%
\begin{equation}
\sigma _{ML}^{2}/(\sigma _{y}^{2}(1-\rho _{xy_{-1}}^{2}))=-\left( \frac{%
\sigma ^{2}\xi ^{\prime }(\rho )}{\sigma _{y}^{2}(1-\rho _{xy_{-1}}^{2})}%
\right) ^{2}+\sigma ^{2}/(\sigma _{y}^{2}(1-\rho _{xy_{-1}}^{2})).
\label{sml}
\end{equation}%
Then it is easily verified that $\frac{\partial \widetilde{l}^{c}(r)}{%
\partial r}|_{\rho }=0$ and%
\begin{equation*}
\frac{\partial ^{2}\widetilde{l}^{c}(r)}{\partial r^{2}}|_{\rho }=(T-1)\xi
^{\prime \prime }(\rho )+(T-1)(2\left( \xi ^{\prime }(\rho )\right)
^{2}-\sigma _{y}^{2}(1-\rho _{xy_{-1}}^{2})/\sigma ^{2}).
\end{equation*}%
Note that $(T-1)\sigma _{y}^{2}=\sigma ^{2}tr(\Phi ^{\prime }Q\Phi )+E(%
\widetilde{Z}_{i}Q\widetilde{Z}_{i})$. Let $\sigma _{x}^{2}=$ p$%
\lim_{N\rightarrow \infty }\frac{1}{(T-1)N}\tsum\nolimits_{i=1}^{N}X_{i}^{%
\prime }QX_{i}$ and $\sigma _{xy_{-1}}^{2}=$ p$\lim_{N\rightarrow \infty }%
\frac{1}{(T-1)N}\tsum\nolimits_{i=1}^{N}X_{i}^{\prime }Qy_{i,-1}.$ Using $%
Qy_{i,-1}=Q(\widetilde{Z}_{i}+\Phi \varepsilon _{i})$ it is easily seen that 
$\sigma _{y}^{2}(1-\rho _{xy_{-1}}^{2})/\sigma ^{2}=(\sigma _{y}^{2}-\sigma
_{xy_{-1}}^{\prime }\sigma _{x}^{-2}\sigma _{xy_{-1}})/\sigma ^{2}\geq
(T-1)^{-1}tr(\Phi ^{\prime }Q\Phi ),$ with equality holding if $\rho =1$ or $%
\sigma _{v}^{2}=\beta =0$ (i.e. if $\Sigma _{zqz}=0$)$.$ We also have $\xi
^{\prime \prime }(\rho )-(T-1)^{-1}\times $ $tr(\Phi (\rho )^{\prime }Q\Phi
(\rho ))+2(\xi ^{\prime }(\rho ))^{2}\leq 0,$ with equality holding if $\rho
=1$ or $T=2.$ It follows that $\frac{\partial ^{2}\widetilde{l}^{c}(r)}{%
\partial r^{2}}|_{\rho }\leq 0,$ with equality holding if $\rho =1$ or if $%
T=2$ and $\sigma _{v}^{2}=\beta =0.$ Thus $\widetilde{l}^{c}(r)$ has a local
maximum at $\rho $ when $\rho \neq 1$ and, in case $T=2$, $\Sigma _{zqz}>0$.
Below we show that $\widetilde{l}^{c}(r)$ has a stationary point of
inflection at $\rho $ when $\rho =1$.\medskip

\textbf{\noindent }Derivation of (\ref{sml}): Given that the equality $%
y_{i}-ry_{i,-1}-X_{i}b=(\rho -r)y_{i,-1}+X_{i}(\beta -b)+\alpha _{i}\iota
+\varepsilon _{i}$ holds for any $r$ and $b,$ including for $r=\widehat{\rho 
}_{ML}$ and $b=\widehat{\beta }_{ML}$, we can rewrite $\widehat{\sigma }%
_{ML}^{2}$ as\vspace{-0.12in}%
\begin{multline}
\widehat{\sigma }_{ML}^{2}=(T-1)^{-1}N^{-1}\sum_{i=1}^{N}[((\rho -\widehat{%
\rho }_{ML})y_{i,-1}+X_{i}(\beta -\widehat{\beta }_{ML})+\varepsilon
_{i})^{\prime }\times  \label{sigmlhat} \\
Q((\rho -\widehat{\rho }_{ML})y_{i,-1}+X_{i}(\beta -\widehat{\beta }%
_{ML})+\varepsilon _{i})].
\end{multline}

Let $\beta _{ML}=$ p$\lim_{N\rightarrow \infty }\widehat{\beta }_{ML},$ $%
\sigma _{x}^{2}=$ p$\lim_{N\rightarrow \infty
}(T-1)^{-1}N^{-1}\tsum\nolimits_{i=1}^{N}X_{i}^{\prime }QX_{i},$ and $\sigma
_{xy_{-1}}^{2}=$ p$\lim_{N\rightarrow \infty
}(T-1)^{-1}N^{-1}\tsum\nolimits_{i=1}^{N}X_{i}^{\prime }Qy_{i,-1}.$ Then
combining p$\lim_{N\rightarrow \infty }N^{-1}\sum_{i=1}^{N}X_{i}^{\prime
}Q(y_{i}-\widehat{\rho }_{ML}y_{i,-1}-X_{i}\widehat{\beta }_{ML})=0$ from (%
\ref{likeqs}) with p$\lim_{N\rightarrow \infty
}N^{-1}\sum_{i=1}^{N}X_{i}^{\prime }Q(y_{i}-\rho y_{i,-1}-X_{i}\beta
)=p\lim_{N\rightarrow \infty }N^{-1}\sum_{i=1}^{N}X_{i}^{\prime
}Q\varepsilon _{i}=0$ gives%
\begin{equation}
\beta _{ML}-\beta =\sigma _{x}^{-2}\sigma _{xy_{-1}}(\rho -\rho _{ML}).
\label{plimbml}
\end{equation}

Using (\ref{sigmlhat}) and (\ref{plimbml}) and recalling that $\xi ^{\prime
}(\rho )=h(\rho )=-(T-1)^{-1}tr(Q\Phi )$, we obtain%
\begin{multline*}
\sigma _{ML}^{2}=\text{ p}\lim_{N\rightarrow \infty }\widehat{\sigma }%
_{ML}^{2}=(\rho -\rho _{ML})^{2}\sigma _{y}^{2}+2(\beta -\beta
_{ML})^{\prime }\sigma _{xy_{-1}}(\rho -\rho _{ML})+ \\
(\beta -\beta _{ML})^{\prime }\sigma _{x}^{2}(\beta -\beta _{ML})+2(\rho
-\rho _{ML})\sigma ^{2}(T-1)^{-1}tr(Q\Phi )+\sigma ^{2}= \\
(\rho -\rho _{ML})^{2}\sigma _{y}^{2}-\sigma _{xy_{-1}}^{\prime }\sigma
_{x}^{-2}\sigma _{xy_{-1}}(\rho -\rho _{ML})^{2}+2(\rho -\rho _{ML})\sigma
^{2}(T-1)^{-1}tr(Q\Phi )+\sigma ^{2}= \\
(\rho -\rho _{ML})^{2}\sigma _{y}^{2}(1-\rho _{xy_{-1}}^{2})-2(\rho -\rho
_{ML})\sigma ^{2}\xi ^{\prime }(\rho )+\sigma ^{2}.
\end{multline*}

Finally, using $\rho _{ML}-\rho =-\frac{\sigma ^{2}\xi ^{\prime }(\rho )}{%
\sigma _{y}^{2}(1-\rho _{xy_{-1}}^{2})},$ we find that%
\begin{equation*}
\sigma _{ML}^{2}/(\sigma _{y}^{2}(1-\rho _{xy_{-1}}^{2}))=-\left( \frac{%
\sigma ^{2}\xi ^{\prime }(\rho )}{\sigma _{y}^{2}(1-\rho _{xy_{-1}}^{2})}%
\right) ^{2}+\sigma ^{2}/(\sigma _{y}^{2}(1-\rho _{xy_{-1}}^{2})).
\end{equation*}

\textbf{\noindent Proof of the claim that }$\widetilde{l}^{c}(r)$\textbf{\
has an inflection point at }$\rho $\textbf{\ when }$\rho =1:$

We have already seen that $\frac{\partial \widetilde{l}^{c}(r)}{\partial r}%
|_{\rho =1}=\frac{\partial ^{2}\widetilde{l}^{c}(r)}{\partial r^{2}}|_{\rho
=1}=0.$ In addition, we have%
\begin{eqnarray*}
\frac{\partial ^{3}\widetilde{l}^{c}(r)}{\partial r^{3}} &=&(T-1)\xi
^{\prime \prime \prime }(r)+\frac{6(T-1)(r-\rho _{ML})(\sigma
_{ML}^{2}/(\sigma _{y}^{2}(1-\rho _{xy_{-1}}^{2})))}{(\sigma
_{ML}^{2}/(\sigma _{y}^{2}(1-\rho _{xy_{-1}}^{2}))+(r-\rho _{ML})^{2})^{3}}
\\
&&-\frac{2(T-1)(r-\rho _{ML})^{3}}{(\sigma _{ML}^{2}/(\sigma _{y}^{2}(1-\rho
_{xy_{-1}}^{2}))+(r-\rho _{ML})^{2})^{3}},
\end{eqnarray*}%
$\xi ^{\prime \prime \prime }(1)=\frac{(T-2)(T-3)}{12},$ $f(1)=\frac{T+1}{6}%
, $ $\xi ^{\prime }(1)=\frac{1}{2},$ $\lim_{\rho \rightarrow 1}\rho
_{xy_{-1}}^{2}=0,$ $\lim_{\rho \rightarrow 1}(\rho _{ML}-\rho )=-\frac{\xi
^{\prime }(1)}{f(1)}=-\frac{3}{T+1}$ and $\lim_{\rho \rightarrow 1}(\sigma
_{ML}^{2}/(\sigma _{y}^{2}(1-\rho _{xy_{-1}}^{2})))=-\left( \frac{\xi
^{\prime }(1)}{f(1)}\right) ^{2}+\frac{1}{f(1)}=3\frac{(2T-1)}{\left(
T+1\right) ^{2}}.$ It follows that $\frac{\partial ^{3}\widetilde{l}^{c}(r)}{%
\partial r^{3}}|_{\rho =1}=(T-1)\xi ^{\prime \prime \prime }(1)+\frac{%
(T-1)^{2}}{2}\neq 0$ (in fact $>0$) for $T\geq 2.\quad \square \bigskip $

We now present two lemmata that help to establish uniqueness and consistency
of our MMLEs:

\begin{lemma}
Let Assumption 1 hold. Then (i)\ $\widetilde{l}_{N}(\theta )$ has either no
local optima or one local maximum, namely $\widehat{\theta }_{W}=\widehat{%
\theta }_{C}$, and one local minimum on the set $\widetilde{\Omega }$ w.p.1.
(ii)\ $\widetilde{l}_{N}^{c}(r)$ has either no local optima or one local
maximum, namely $\widehat{\rho }_{W}=\widehat{\rho }_{C}$, and one local
minimum on the interval $[-1,\infty )$ w.p.1. (iii)\ The equation $\frac{%
\partial \widetilde{l}_{N}^{c}(r)}{\partial r}=0$ has either no solution on $%
[-1,\infty )$ or two solutions on $[-1,\infty )$, namely $\widehat{\rho }%
_{1} $ and $\widehat{\rho }_{2}$ with $\widehat{\rho }_{1}<\widehat{\rho }%
_{2}$ and $\widehat{\rho }_{1}=\widehat{\rho }_{W}=\widehat{\rho }_{C}$,
w.p.1.
\end{lemma}

\begin{lemma}
Let Assumption 1 hold. First let $\rho \neq 1.$ Then (i)\ $\widetilde{l}%
(\theta )$ has one local maximum and one local minimum but no inflection
point on the set $\widetilde{\Omega }$. The local maximum is attained at $%
\theta _{0}$. (ii)\ $\widetilde{l}^{c}(r)$ has one local maximum and one
local minimum but no inflection point on the interval $[-1,\infty )$. The
local maximum of $\widetilde{l}^{c}(r)$ is attained at $\rho $. (iii)\ The
equation $\frac{\partial \widetilde{l}^{c}(r)}{\partial r}=0$ has two
solutions on $[-1,\infty )$: $\rho _{1}$ and $\rho _{2}$ with $\rho
_{1}<\rho _{2}$ and $\rho _{1}=\rho .$

Now let $\rho =1.$ Then (iv)\ $\widetilde{l}(\theta )$ has one stationary
point of inflection but no local optima on $\widetilde{\Omega }$. The
inflection point is attained at $\theta _{0}$. (v)\ $\widetilde{l}^{c}(r)$
has one stationary point of inflection but no local optima on $[-1,\infty )$%
. The inflection point of $\widetilde{l}^{c}(r)$ is attained at $\rho =1$.
(vi)\ The equation $\frac{\partial \widetilde{l}^{c}(r)}{\partial r}=0$ has
only one solution on $[-1,\infty )$: $\rho _{1}=1.$
\end{lemma}

We first prove the following lemma, which summarizes some useful properties
of $\xi ^{\prime }(\rho ):$

\begin{lemma}
Let $\rho \geq -1.$ When $T\geq 2,$ $\xi ^{\prime }(\rho )>0,$ $\xi ^{\prime
}(1)=\frac{1}{2};$

When $T=2,$ $\xi ^{\prime }(\rho )=\frac{1}{2};$

When $T=3,$ $\xi ^{\prime }(-1)=\frac{1}{6}$ and $\xi ^{\prime \prime }(\rho
)=\frac{1}{6};$

When $T\geq 4$ and $T$ is even, $\xi ^{\prime }(-1)=\frac{1}{2(T-1)},$ $\xi
^{\prime \prime }(-1)=0,$ $\xi ^{\prime \prime }(\rho )>0$ when $\rho >-1,$
and $\xi ^{\prime \prime \prime }(\rho )>0;$

When $T\geq 5$ and $T$ is odd, $\xi ^{\prime }(-1)=\frac{1}{2T},$ $\xi
^{\prime \prime }(\rho )>0,$ $\xi ^{\prime \prime \prime }(-1)=-\frac{T-3}{4T%
}<0,$ $\xi ^{\prime \prime \prime }(-1/2)=\frac{2^{4-T}(2^{T}-3T+1)}{%
27T\left( T-1\right) }>0,$ and $\exists \rho _{\ast }$ with $-1<\rho _{\ast
}<-1/2$ such that $\xi ^{\prime \prime \prime }(\rho _{\ast })=0,$ $\xi
^{\prime \prime \prime }(\rho )<0$ for $\rho <\rho _{\ast }$ and $\xi
^{\prime \prime \prime }(\rho )>0\ $for $\rho >\rho _{\ast }.$
\end{lemma}

\textbf{Proof of lemma 3:} for the proof of most properties see Dhaene and
Jochmans (2016). Their proof uses that $\xi ^{\prime }(\rho
)=[T(T-1)]^{-1}\sum_{t=1}^{T-1}(T-t)\rho ^{t-1}=\frac{T-1-T\rho +\rho ^{T}}{%
T(T-1)(1-\rho )^{2}}$ when $\rho \neq 1,$ and Descartes' rule of signs. The
remaining claims, i.e., $\xi ^{\prime }(\rho )=\frac{1}{2}$ when $\rho =1$
or $T=2,$ $\xi ^{\prime \prime }(\rho )=\frac{1}{6}$ when $T=3,$ $\xi
^{\prime }(-1)=\frac{1}{2(T-1)}$ when $T$ is even, and $\xi ^{\prime }(-1)=%
\frac{1}{2T}$ when $T$ is odd, are easily verified.\quad $\square $

Thus when $\rho \geq -1,$ we have:\newline
If $T=2,$ then $\xi ^{\prime }(\rho )$ is strictly positive and constant; 
\newline
If $T=3,$ then $\xi ^{\prime }(\rho )$ is strictly positive and increasing
linearly; \newline
If $T\geq 4$ and $T$ is even, then $\xi ^{\prime }(\rho )$ is strictly
positive, non-decreasing and strictly convex;\newline
If $T\geq 5$ and $T$ is odd, then $\xi ^{\prime }(\rho )$ is strictly
positive, strictly increasing and first strictly concave and then strictly
convex.$\bigskip $

\textbf{\noindent Proof of lemma 1:}

We can write (\ref{foc1b}) as%
\begin{equation}
\xi ^{\prime }(r)\{\widehat{\sigma }%
_{ML}^{2}/(s_{y}^{2}(1-r_{xy_{-1}}^{2}))+(r-\widehat{\rho }_{ML})^{2}\}=(r-%
\widehat{\rho }_{ML}).  \label{foc1a}
\end{equation}%
Let $\zeta _{N}(r)=\xi ^{\prime }(r)\{\widehat{\sigma }%
_{ML}^{2}/(s_{y}^{2}(1-r_{xy_{-1}}^{2}))+(r-\widehat{\rho }_{ML})^{2}\}.$
Then $\zeta _{N}^{\prime }(r)=\xi ^{\prime \prime }(r)\{\widehat{\sigma }%
_{ML}^{2}/(s_{y}^{2}(1-r_{xy_{-1}}^{2}))+(r-\widehat{\rho }_{ML})^{2}\}+2(r-%
\widehat{\rho }_{ML})\xi ^{\prime }(r)$ and $\zeta _{N}^{\prime \prime
}(r)=\xi ^{\prime \prime \prime }(r)\{\widehat{\sigma }%
_{ML}^{2}/(s_{y}^{2}(1-r_{xy_{-1}}^{2}))+(r-\widehat{\rho }_{ML})^{2}\}+4(r-%
\widehat{\rho }_{ML})\xi ^{\prime \prime }(r)+2\xi ^{\prime }(r).$

By lemma 3 $\zeta _{N}(r)>0$ when $r\geq -1$. Hence any solution $r$ of (\ref%
{foc1a}) should satisfy $r>\widehat{\rho }_{ML}$. When $r\geq \max (-1,%
\widehat{\rho }_{ML}),$ we also have by lemma 3 that $\zeta _{N}^{\prime
}(r)>0$ and if $T$ is even that $\zeta _{N}^{\prime \prime }(r)>0$ while if $%
T$ is odd we either have $\zeta _{N}^{\prime \prime }(r)>0$ for all $r\geq
\max (-1,\widehat{\rho }_{ML})$ or $\zeta _{N}^{\prime \prime }(r)<0$ for
all $r$ on $[\max (-1,\widehat{\rho }_{ML}),\rho _{\ast \ast })$ and $\zeta
_{N}^{\prime \prime }(r)>0$ for all $r$ on $(\rho _{\ast \ast },\infty )$
with $\rho _{\ast \ast }>\max (-1,\rho _{ML})$ and equal to the solution of $%
\zeta _{N}^{\prime \prime }(r)=0$. It follows that w.p.1. the graph of $%
\zeta _{N}(r)$ either does not intersect the line $r-\widehat{\rho }_{ML}$
(this may well happen when $\rho $ is close or equal to unity, see Lancaster
for an example) or intersects the line $r-\widehat{\rho }_{ML}$ twice, say
at $r=\widehat{\rho }_{1}$ and $r=\widehat{\rho }_{2}$ with $\max (-1,%
\widehat{\rho }_{ML})<\widehat{\rho }_{1}<\widehat{\rho }_{2}.$ Both
solutions of (\ref{foc1a}) would correspond to local optima w.p.1. That is,
the possibility that $r-\widehat{\rho }_{ML}$ is a tangent to $\zeta _{N}(r)$
at $\widehat{\rho }_{1}$ and/or $\widehat{\rho }_{2}$ is an event with
probability zero, so $\widehat{\rho }_{1}$ and $\widehat{\rho }_{2}$ would
not correspond to (an) inflection point(s) w.p.1. It is clear that when $%
\left\vert \rho \right\vert \leq 1,$ $\widetilde{l}_{N}^{c}(r)$ and $%
\widetilde{l}_{N}(\theta )$ attain at most one local maximum on the interval 
$[-1,\infty )$ and the set $\widetilde{\Omega }$, respectively. Moreover, if
(\ref{foc1a}) has any solutions, then $\widehat{\rho }_{C}$ is one of them
and $\widehat{\rho }_{C}=\widehat{\rho }_{W}\geq \max (-1,\widehat{\rho }%
_{ML})$. Given that $\widetilde{l}_{N}^{c}(r)$ has a global maximum at $%
r=\infty $ (because $\lim_{r\uparrow \infty }\widetilde{l}_{N}^{c}(r)=\infty 
$), we conclude that if (\ref{foc1a}) has any solutions, then it has two
solutions $\widehat{\rho }_{1}$ and $\widehat{\rho }_{2}$ with $\widehat{%
\rho }_{1}<\widehat{\rho }_{2}$, where $\widehat{\rho }_{1}=\widehat{\rho }%
_{C}=\widehat{\rho }_{W}$ corresponds to a local maximum of $\widetilde{l}%
_{N}^{c}(r)$ and $\widehat{\rho }_{2}$ corresponds to a local minimum of $%
\widetilde{l}_{N}^{c}(r),$ because $\widetilde{l}_{N}^{c}(r)$ cannot attain
a local maximum at $\widehat{\rho }_{2}$. Likewise, given that $%
\lim_{r\uparrow \infty }\widetilde{l}_{N}(\widehat{\theta }%
(r))=\lim_{r\uparrow \infty }\widetilde{l}_{N}^{c}(r)=\infty ,$ we conclude
that if (\ref{foc1a}) has solutions $\widehat{\rho }_{1}$ and $\widehat{\rho 
}_{2}$ with $\widehat{\rho }_{1}<\widehat{\rho }_{2}$, then $\widetilde{l}%
_{N}(\theta )$ has two local optima on the set $\widetilde{\Omega },$ say $%
\widehat{\theta }_{1}$ and $\widehat{\theta }_{2},$ where $\widehat{\theta }%
_{1}=\widehat{\theta }(\widehat{\rho }_{1})=\widehat{\theta }(\widehat{\rho }%
_{W})=\widehat{\theta }(\widehat{\rho }_{C})$ corresponds to a local maximum
of $\widetilde{l}_{N}(\theta )$ and $\widehat{\theta }_{2}=\widehat{\theta }(%
\widehat{\rho }_{2})$ corresponds to a local minimum of $\widetilde{l}%
_{N}(\theta )$, because $\widetilde{l}_{N}(\theta )$ cannot attain a local
maximum at $\widehat{\theta }(\widehat{\rho }_{2})$.\quad $\square \bigskip $

\textbf{\noindent Proof of lemma 2:}

We can write (\ref{foc2b}) as%
\begin{equation}
\xi ^{\prime }(r)\{\sigma _{ML}^{2}/(\sigma _{y}^{2}(1-\rho
_{xy_{-1}}^{2}))+(r-\rho _{ML})^{2}\}=(r-\rho _{ML}).  \label{foc2a}
\end{equation}%
Let $\zeta (r)=\xi ^{\prime }(r)\{\sigma _{ML}^{2}/(\sigma _{y}^{2}(1-\rho
_{xy_{-1}}^{2}))+(r-\rho _{ML})^{2}\}.$ Then $\zeta ^{\prime }(r)=\xi
^{\prime \prime }(r)\{\sigma _{ML}^{2}/(\sigma _{y}^{2}(1-\rho
_{xy_{-1}}^{2}))+(r-\rho _{ML})^{2}\}+2(r-\rho _{ML})\xi ^{\prime }(r)$ and $%
\zeta ^{\prime \prime }(r)=\xi ^{\prime \prime \prime }(r)\{\sigma
_{ML}^{2}/(\sigma _{y}^{2}(1-\rho _{xy_{-1}}^{2}))+(r-\rho
_{ML})^{2}\}+4(r-\rho _{ML})\xi ^{\prime \prime }(r)+2\xi ^{\prime }(r).$

By lemma 3 $\zeta (r)>0$ when $r\geq -1$. Hence any solution $r$ of (\ref%
{foc2a}) should satisfy $r>\rho _{ML}$. When $r\geq \max (-1,\rho _{ML}),$
we also have by lemma 3 that $\zeta ^{\prime }(r)>0$ and if $T$ is even that 
$\zeta ^{\prime \prime }(r)>0$ while if $T$ is odd we either have $\zeta
^{\prime \prime }(r)>0$ for all $r\geq \max (-1,\rho _{ML})$ or $\zeta
^{\prime \prime }(r)<0$ for all $r$ in $[\max (-1,\rho _{ML}),\rho _{\ast
\ast })$ and $\zeta ^{\prime \prime }(r)>0$ for all $r$ in $(\rho _{\ast
\ast },\infty )$ where $\rho _{\ast \ast }$ satisfies $\rho _{\ast \ast
}>\max (-1,\rho _{ML})$ and $\zeta ^{\prime \prime }(\rho _{\ast \ast })=0$.
It follows that the graph of $\zeta (r)$ intersects the line $r-\rho _{ML}$
at most twice, say at $r=\rho _{1}$ and $r=\rho _{2}$ with $\max (-1,\rho
_{ML})\leq \rho _{1}\leq \rho _{2}$. On the other hand we have $\rho \geq
\max (-1,\rho _{ML})$ and using (\ref{abia}), $h(\rho )=\xi ^{\prime }(\rho
) $ and (\ref{sml}) it is easily verified that $\rho $ is a solution of (\ref%
{foc2a}). We have already seen in the main text that when $\rho \neq 1,$ $%
\widetilde{l}^{c}(r)$ and $\widetilde{l}(\theta )$ attain a local maximum at 
$\rho $ and $\theta _{0}=$ p$\lim_{N\rightarrow \infty }\widehat{\theta }%
(\rho )=$ p$\lim_{N\rightarrow \infty }(\rho ,\widehat{\sigma }^{2}(\rho ,%
\widehat{\beta }(\rho )),\widehat{\beta }(\rho ))^{\prime }$, respectively.
Given that $\widetilde{l}^{c}(r)$ has a global maximum at $r=\infty $
(because $\lim_{r\uparrow \infty }\widetilde{l}^{c}(r)=\infty $)$,$ $%
\widetilde{l}^{c}(r)$ cannot attain a local maximum at $\rho _{2}$ so we
conclude that (\ref{foc2a}) has two solutions, $\rho _{1}$ and $\rho _{2}$,
where $\rho _{1}=\rho $ and $\rho _{2}$ corresponds to a local minimum of $%
\widetilde{l}^{c}(r)$. Similarly, given that $\lim_{r\uparrow \infty }$p$%
\lim_{N\rightarrow \infty }\widetilde{l}(\widehat{\theta }%
(r))=\lim_{r\uparrow \infty }\widetilde{l}^{c}(r)=\infty ,$ $\widetilde{l}%
(\theta )$ cannot attain a local maximum at p$\lim_{N\rightarrow \infty }%
\widehat{\theta }(\rho _{2})$ so we conclude that p$\lim_{N\rightarrow
\infty }\widehat{\theta }(\rho _{1})=$ p$\lim_{N\rightarrow \infty }\widehat{%
\theta }(\rho )=\theta _{0}$ and that p$\lim_{N\rightarrow \infty }\widehat{%
\theta }(\rho _{2})$ corresponds to a local minimum of $\widetilde{l}(\theta
)$.

When $\rho =1,$ we know that $\widetilde{l}(\theta )$ and $\widetilde{l}%
^{c}(r)$ have an inflection point at $\theta _{0}$ and $\rho $,
respectively. When $\rho =1,$ we also have that $\rho _{ML}=1-\frac{3}{T+1},$
so that $\max (-1,\rho _{ML})=\rho _{ML}$ and $\frac{\partial \widetilde{l}%
^{c}(r)}{\partial r}|_{\rho _{ML}}=(T-1)\xi ^{\prime }(\rho _{ML})>0$.
Because $\frac{\partial \widetilde{l}^{c}(r)}{\partial r}$ is continuous for 
$r\geq \rho _{ML}$, because $\rho _{1}$ is the smallest $r>\rho _{ML}$ such
that $\frac{\partial \widetilde{l}^{c}(r)}{\partial r}=0,$ and because $%
\frac{\partial \widetilde{l}^{c}(r)}{\partial r}|_{\rho _{ML}}>0,$ we have $%
\frac{\partial \widetilde{l}^{c}(r)}{\partial r}|_{\rho _{1}-}>0.$ We now
show that $\rho _{1}$ is an inflection point. Suppose instead that $\rho
_{1} $ were a maximum (note that $\rho _{1}$ cannot be a minimum because $%
\frac{\partial \widetilde{l}^{c}(r)}{\partial r}|_{\rho _{1}-}>0$). Then $%
\rho _{2} $ must have been a minimum (because $\frac{\partial \widetilde{l}%
^{c}(r)}{\partial r}|_{+\infty }>0$) and there would be no inflection point
in the interval $[\rho _{ML},\infty )$. This would contradict that $%
\widetilde{l}^{c}(r)$ has at least one inflection point larger than $\rho
_{ML},$ namely at $r=1$. Thus $\rho _{1}$ is an inflection point. Remains to
show that $\rho _{1}=\rho _{2}=1.$ Since $\frac{\partial \widetilde{l}^{c}(r)%
}{\partial r}|_{\rho _{1}-}>0$ and $\rho _{1}$ is inflection point and $%
\frac{\partial \widetilde{l}^{c}(r)}{\partial r}=0$ has at most two
solutions, we also have $\frac{\partial \widetilde{l}^{c}(r)}{\partial r}%
|_{\rho _{1}+}>0$. Because $\rho _{1}>\rho _{ML},$ because $%
(T-1)^{-1}(\sigma _{ML}^{2}/\sigma _{y}^{2}+(r-\rho _{ML})^{2})$ is strictly
positive, increasing and strictly convex when $r>\rho _{ML}$, and because $%
(T-1)^{-1}(\sigma _{ML}^{2}/\sigma _{y}^{2}+(r-\rho _{ML})^{2})\frac{%
\partial \widetilde{l}^{c}(r)}{\partial r}=\zeta (r)-(r-\rho _{ML}),$ we
have $(\zeta (r)-(r-\rho _{ML}))|_{\rho _{1}-}>0,$ $(\zeta (r)-(r-\rho
_{ML}))|_{\rho _{1}}=0$ and $(\zeta (r)-(r-\rho _{ML}))|_{\rho _{1}+}>0.$ It
follows that $\zeta ^{\prime }(\rho _{1})=1$. Because $\rho _{ML}>-1/2,$ we
have both $\xi ^{\prime \prime \prime }(r)>0$ and $\zeta ^{\prime \prime
}(r)>0$ for all $r>\rho _{ML}.$ This implies that $\zeta ^{\prime }(r)>1$
for $r>\rho _{1}$ and hence that there exists no $r>\rho _{1}$ such that $%
\zeta (r)-(r-\rho _{ML})=0$. It follows that $\rho _{1}=\rho _{2}=\rho =1.$
Similarly, we obtain that p$\lim_{N\rightarrow \infty }\widehat{\theta }%
(\rho _{1})=$\thinspace p$\lim_{N\rightarrow \infty }\widehat{\theta }(\rho
_{2})=\theta _{0}$ $\quad \square $

\begin{lemma}
Let $\widetilde{\Theta }_{N}^{c}$ be the set of roots of $\frac{\partial 
\widetilde{l}_{N}^{c}}{\partial r}=0$ corresponding to local maxima of $%
\widetilde{l}_{N}^{c}$ on the interval $[-1,\infty ).$ Let Assumption 1 hold
and $\rho =1.$ Then $\lim_{N\rightarrow \infty }\Pr (\widetilde{\Theta }%
_{N}^{c}=\varnothing )>0.$
\end{lemma}

\textbf{Proof: }Let $\kappa (r)=\xi ^{\prime }(r)\{\widehat{\sigma }%
_{ML}^{2}/(s_{y}^{2}(1-r_{xy_{-1}}^{2}))+(r-\widehat{\rho }_{ML})^{2}\}-(r-%
\widehat{\rho }_{ML}).$ To save space we only prove the lemma for the model
without covariates so that $r_{xy_{-1}}^{2}=0.$

Note that $\frac{\partial \widetilde{l}_{N}^{c}}{\partial r}%
=0\Leftrightarrow \kappa (r)=0.$ When $\rho =1,$%
\begin{eqnarray*}
s_{y}^{2} &=&(T-1)^{-1}N^{-1}\tsum\nolimits_{i=1}^{N}\varepsilon
_{i}^{\prime }\Phi ^{\prime }Q\Phi \varepsilon _{i}\equiv (T-1)^{-1}A_{N}, \\
\widehat{\rho }_{ML}-1 &=&(\tsum\nolimits_{i=1}^{N}\varepsilon _{i}^{\prime
}\Phi ^{\prime }Q\Phi \varepsilon
_{i})^{-1}\tsum\nolimits_{i=1}^{N}\varepsilon _{i}^{\prime }\Phi ^{\prime
}Q\varepsilon _{i}\equiv A_{N}^{-1}B_{N},\text{\quad and} \\
\widehat{\sigma }_{ML}^{2}
&=&(T-1)^{-1}N^{-1}\tsum\nolimits_{i=1}^{N}\varepsilon _{i}^{\prime }(\Phi
(1-\widehat{\rho }_{ML})+I)^{\prime }Q(\Phi (1-\widehat{\rho }%
_{ML})+I)\varepsilon _{i}= \\
&&(T-1)^{-1}(C_{N}-A_{N}^{-1}B_{N}^{2})
\end{eqnarray*}
where $C_{N}\equiv N^{-1}\tsum\nolimits_{i=1}^{N}\varepsilon _{i}^{\prime
}Q\varepsilon _{i}.$

When $r$ is close to one, $\xi ^{\prime }(r)\approx \xi ^{\prime }(1)+\xi
^{\prime \prime }(1)(r-1)+\frac{1}{2}\xi ^{\prime \prime \prime
}(1)(r-1)^{2}.$ Note that $\xi ^{\prime }(1)=\frac{1}{2},$\ $\xi ^{\prime
\prime }(1)=\frac{1}{6}(T-2)$ and $\xi ^{\prime \prime \prime }(1)=\frac{1}{%
12}(T-2)(T-3).$

Let $z=r-1.$\ When $r$ is close to one, $\kappa (r)\approx (\frac{1}{2}+%
\frac{1}{6}(T-2)z+\frac{1}{24}%
(T-2)(T-3)z^{2})(A_{N}^{-1}C_{N}-2A_{N}^{-1}B_{N}z+z^{2})-z+A_{N}^{-1}B_{N}%
\approx \frac{1}{2}A_{N}^{-1}C_{N}+A_{N}^{-1}B_{N}+(\frac{1}{6}%
(T-2)A_{N}^{-1}C_{N}-A_{N}^{-1}B_{N}-1)z+(\frac{1}{2}+\frac{1}{24}%
(T-2)(T-3)A_{N}^{-1}C_{N}-\frac{1}{3}(T-2)A_{N}^{-1}B_{N})z^{2}\equiv 
\widetilde{\kappa }(z)$ where in the last step we have dropped the $z^{3}$%
-term and the $z^{4}$-term which are negligible when $r$ is close enough to
one. Note that when $T=2,$ the approximations can be replaced by exact
equalities.

Note that $\widetilde{\kappa }(z)=0\Leftrightarrow 6A_{N}\widetilde{\kappa }%
(z)=0.$ Solving $6A_{N}\widetilde{\kappa }(z)=0$ gives: $z_{1,2}=%
\{(6A_{N}+6B_{N}-(T-2)C_{N})\pm \sqrt{D_{N}}\}/(6A_{N}+\frac{1}{2}%
(T-2)(T-3)C_{N}-4(T-2)B_{N})$ where $D_{N}\equiv
36(A_{N}^{2}+B_{N}^{2}-A_{N}C_{N})+(T-2)^{2}C_{N}^{2}+12(T-2)(B_{N}C_{N}-A_{N}C_{N}+4B_{N}^{2})-(3C_{N}+6B_{N})(T-2)(T-3)C_{N}. 
$

It is easily seen that p$\lim_{N\rightarrow \infty }A_{N}=\frac{1}{6}%
(T+1)\sigma ^{2},$ p$\lim_{N\rightarrow \infty }B_{N}=-\frac{1}{2}\sigma
^{2} $ and\linebreak p$\lim_{N\rightarrow \infty }C_{N}=\sigma ^{2}.$ It
follows that p$\lim_{N\rightarrow \infty }(6A_{N}+\frac{1}{2}%
(T-2)(T-3)C_{N}-4(T-2)B_{N})=3(T-1)\sigma ^{2}+\frac{1}{2}(T-2)(T-3)\sigma
^{2}>0,$ p$\lim_{N\rightarrow \infty }(6A_{N}+6B_{N}-(T-2)C_{N})=0$ and p$%
\lim_{N\rightarrow \infty }D_{N}=0$ so that p$\lim_{N\rightarrow \infty
}z_{1,2}=0,$ as predicted by lemma 2. However, in finite samples of any size
we can have $D_{N}<0$, so that $\widetilde{\kappa }(z)=0$ does not have a
real solution. $\Pr (D_{N}<0)$ does not tend to zero when $N\rightarrow
\infty .$ We conclude that $\lim_{N\rightarrow \infty }\Pr (\widetilde{%
\Theta }_{N}^{c}=\varnothing )>0.$\quad $\square \bigskip $

\textbf{\noindent Proof of theorem 1:\medskip }

Lemma 1 implies that $\widehat{\theta }_{W}$ and $\widehat{\theta }_{C}$ are
uniquely defined when they exist. When $\widetilde{l}_{N}^{c}(r)$ and $%
\widetilde{l}_{N}(\theta )$ have local maxima on the interval $[-1,\infty )$
and the set $\widetilde{\Omega }$, respectively, this follows from lemma 1.
We will now turn to the other claims of the theorem. To prove the
consistency claims, we will verify the conditions of theorem 2.1 in Newey
and McFadden (NMcF, 1994). To simplify matters and following NMcF, we will
simply assume that the parameter space for $\theta $ is a very large compact
subset of $\widetilde{\Omega }$, viz. $\overline{\Omega }=\overline{\Omega }%
_{\rho }\times \overline{\Omega }_{\sigma ^{2}}\times \overline{\Omega }%
_{\beta }$ where $\overline{\Omega }_{\rho }=[-1,\rho _{u}]$ and $\overline{%
\Omega }_{\sigma ^{2}}=[1/\sigma _{u},\sigma _{u}]$ for some very large $%
\rho _{u},\sigma _{u}\in 
\mathbb{R}
^{+}$ and $\overline{\Omega }_{\beta }$ is a very large compact subset of $%
\mathbb{R}
^{K}.$

We will first prove the claims for $\widehat{\theta }_{C}$. Using that $%
\frac{1}{(T-1)}\left\vert \frac{\partial \widetilde{l}_{N}^{c}(r)}{\partial r%
}-\frac{\partial \widetilde{l}^{c}(r)}{\partial r}\right\vert =\medskip $%
\linebreak $\left\vert \frac{(r-\widehat{\rho }%
_{ML})s_{y}^{2}(1-r_{xy_{-1}}^{2})\{\sigma _{ML}^{2}+(r-\rho
_{ML})^{2}\sigma _{y}^{2}(1-\rho _{xy_{-1}}^{2})\}-(r-\rho _{ML})\sigma
_{y}^{2}(1-\rho _{xy_{-1}}^{2})\{\widehat{\sigma }_{ML}^{2}+(r-\widehat{\rho 
}_{ML})^{2}s_{y}^{2}(1-r_{xy_{-1}}^{2})\}}{(\widehat{\sigma }_{ML}^{2}+(r-%
\widehat{\rho }_{ML})^{2}s_{y}^{2}(1-r_{xy_{-1}}^{2}))(\sigma
_{ML}^{2}+(r-\rho _{ML})^{2}\sigma _{y}^{2}(1-\rho _{xy_{-1}}^{2}))}%
\right\vert \leq \medskip $\linebreak $\left\vert \frac{(r-\widehat{\rho }%
_{ML})s_{y}^{2}(1-r_{xy_{-1}}^{2})\{\sigma _{ML}^{2}+(r-\rho
_{ML})^{2}\sigma _{y}^{2}(1-\rho _{xy_{-1}}^{2})\}-(r-\rho _{ML})\sigma
_{y}^{2}(1-\rho _{xy_{-1}}^{2})\{\widehat{\sigma }_{ML}^{2}+(r-\widehat{\rho 
}_{ML})^{2}s_{y}^{2}(1-r_{xy_{-1}}^{2})\}}{\widehat{\sigma }_{ML}^{2}\sigma
_{ML}^{2}}\right\vert \equiv \medskip \linebreak \left\vert U(r)\right\vert $
and noting that the terms in the numerator of $U(r)$ are polynomials in $r$
and that $\sigma _{ML}^{2}=$ p$\lim_{N\rightarrow \infty }\widehat{\sigma }%
_{ML}^{2}>0$, it follows from p$\lim_{N\rightarrow \infty }U(r)=0$ $\forall
r\in \overline{\Omega }_{\rho }$ that p$\lim_{N\rightarrow \infty
}\sup_{r\in \overline{\Omega }_{\rho }}\left\vert U(r)\right\vert =0$ and
hence p$\lim_{N\rightarrow \infty }\sup_{r\in \overline{\Omega }_{\rho
}}\left\vert \frac{\partial \widetilde{l}_{N}^{c}(r)}{\partial r}-\frac{%
\partial \widetilde{l}^{c}(r)}{\partial r}\right\vert =0.\smallskip $%
\linebreak In a similar way it can be shown that p$\lim_{N\rightarrow \infty
}\sup_{r\in \overline{\Omega }_{\rho }}\left\vert \left( \frac{\partial 
\widetilde{l}_{N}^{c}(r)}{\partial r}\right) ^{2}-\left( \frac{\partial 
\widetilde{l}^{c}(r)}{\partial r}\right) ^{2}\right\vert =0$ and$\smallskip $%
\linebreak p$\lim_{N\rightarrow \infty }\sup_{r\in \overline{\Omega }_{\rho
}}\left\vert \frac{\partial ^{2}\widetilde{l}_{N}^{c}(r)}{\partial r^{2}}-%
\frac{\partial ^{2}\widetilde{l}^{c}(r)}{\partial r^{2}}\right\vert =0.$
Also the polynomials $\left( \frac{\partial \widetilde{l}^{c}(r)}{\partial r}%
\right) ^{2}$ and $\frac{\partial ^{2}\widetilde{l}^{c}(r)}{\partial r^{2}}$
are continuous on $\overline{\Omega }_{\rho }.$ Next we need to distinguish
between two cases, $\rho \neq 1$ and $\rho =1:$

When $\rho \neq 1,$ $\lim_{N\rightarrow \infty }\Pr (\widetilde{\Theta }%
_{N}^{c}=\varnothing )=0,$ where $\widetilde{\Theta }_{N}^{c}$ is the set of
roots of $\frac{\partial \widetilde{l}_{N}^{c}}{\partial r}=0$ corresponding
to local maxima of $\widetilde{l}_{N}^{c}$ on the interval $[-1,\infty ).$
Furthermore, it follows from lemma 2 above and theorem 2.1 in NMcF that $%
\widehat{\rho }_{C}$ converges in probability to $\rho ,$ which corresponds
to a unique local maximum of $\widetilde{l}^{c}(r)$ on $\overline{\Omega }%
_{\rho }$. It then follows straightforwardly that $\widehat{\theta }_{C}$ ($%
=(\widehat{\rho }_{C},\widehat{\sigma }^{2}(\widehat{\rho }_{C},\widehat{%
\beta }(\widehat{\rho }_{C})),\widehat{\beta }(\widehat{\rho }_{C}))^{\prime
}$ ) exists w.p.a.1. and is consistent.

When $\rho =1,$ $\lim_{N\rightarrow \infty }\Pr (\widetilde{\Theta }%
_{N}^{c}=\varnothing )>0$ by lemma 4, notwithstanding that $\frac{\partial 
\widetilde{l}^{c}(r)}{\partial r}|_{\rho }=0.$ However, because $\frac{%
\partial \widetilde{l}^{c}(r)}{\partial r}|_{\rho }=0,$ $\frac{\partial ^{2}%
\widetilde{l}^{c}(r)}{\partial r^{2}}|_{\rho }=0$ and $\frac{\partial ^{3}%
\widetilde{l}^{c}(r)}{\partial r^{3}}|_{\rho }>0$ ($\frac{\partial 
\widetilde{l}^{c}(r)}{\partial r}|_{\rho -}>0$ and $\frac{\partial 
\widetilde{l}^{c}(r)}{\partial r}|_{\rho +}>0,$ cf. proof of lemma 2), we
have that $\widehat{\rho }_{C}$ exists w.p.a.1. and by lemma 2 above and
theorem 2.1 in NMcF that $\widehat{\rho }_{C}$ converges in probability to $%
\rho ,$ which is a unique solution of $\frac{\partial \widetilde{l}^{c}(r)}{%
\partial r}|_{\rho }=0$ on $\overline{\Omega }_{\rho }$. It follows that
also when $\rho =1$, $\widehat{\theta }_{C}$ exists w.p.a.1. and is
consistent.

We now proceed to prove the claims for $\widehat{\theta }_{W}.$ To prove
consistency of $\widehat{\theta }_{W}$ we will make use of theorem 2.6 in
NMcF to verify the conditions of their theorem 2.1.

Let $\widetilde{l}_{N,i}(\theta )=(T-1)\xi (r)-0.5(T-1)\log
s^{2}-0.5s^{-2}(y_{i}-ry_{i,-1}-X_{i}b)^{\prime }Q(y_{i}-ry_{i,-1}-X_{i}b).$
In the notation of NMcF $\partial \widetilde{l}_{N,i}(\theta )/\partial
\theta =g(z_{i},\theta ).$ We assume that $\theta _{0}\in \overline{\Omega }%
, $ which is compact.

It is easily checked that $\partial \widetilde{l}_{N,i}(\theta )/\partial
\theta $ is continuous at each $\theta _{0}\in \overline{\Omega }$ w.p.1.
Furthermore, $E(\sup_{\theta \in \overline{\Omega }}(g(z_{i},\theta
)^{\prime }g(z_{i},\theta )))<\infty .$ To complete the proof, we again need
to distinguish between two cases, $\rho \neq 1$ and $\rho =1:$

When $\rho \neq 1,$ $\lim_{N\rightarrow \infty }\Pr (\widetilde{\Theta }%
_{N}=\varnothing )=0,$ where $\widetilde{\Theta }_{N}$ is the set of roots
of $\frac{\partial \widetilde{l}_{N}}{\partial \theta }=0$ corre- sponding
to local maxima of $\widetilde{l}_{N}$ on the set $\widetilde{\Omega }.$
Furthermore, it follows from lemma 2 above and theorem 2.6 in NMcF that $%
\widehat{\theta }_{W}$ converges in probability to $\theta _{0},$ which
corresponds to a unique local maximum of $\widetilde{l}(\theta )$ on $%
\overline{\Omega }$. Thus $\widehat{\theta }_{W}$ exists w.p.a.1. and is
consistent.

When $\rho =1,$ $\lim_{N\rightarrow \infty }\Pr (\widetilde{\Theta }%
_{N}=\varnothing )>0$ by lemma 4, notwithstanding that $\frac{\partial 
\widetilde{l}(r)}{\partial \theta }|_{\theta _{0}}=0.$ However, because $%
\frac{\partial \widetilde{l}(\theta )}{\partial \theta }|_{\theta _{0}}=0,$ $%
x^{\prime }\left( \frac{\partial ^{2}\widetilde{l}(\theta )}{\partial \theta
\partial \theta ^{\prime }}|_{\theta _{0}}\right) x\leq 0$ $\forall x\in 
\mathbb{R}
^{2+K},$ $\det \left( \frac{\partial ^{2}\widetilde{l}(\theta )}{\partial
\theta \partial \theta ^{\prime }}|_{\theta _{0}}\right) =0,$ $\frac{%
\partial ^{2}\widetilde{l}^{c}(r)}{\partial r^{2}}|_{\rho }=0$ and $\frac{%
\partial ^{3}\widetilde{l}^{c}(r)}{\partial r^{3}}|_{\rho }>0$ ($\frac{%
\partial \widetilde{l}^{c}(r)}{\partial r}|_{\rho -}>0$ and $\frac{\partial 
\widetilde{l}^{c}(r)}{\partial r}|_{\rho +}>0,$ cf. proof of lemma 2), we
have that $\widehat{\theta }_{W}$ exists w.p.a.1. and by lemma 2 above and
theorem 2.6 in NMcF that $\widehat{\theta }_{W}$ converges in probability to 
$\theta _{0},$ which is a unique solution of $\frac{\partial \widetilde{l}%
^{c}(\theta )}{\partial \theta }|_{\theta _{0}}=0$ on $\overline{\Omega }$.
Thus also when $\rho =1$, $\widehat{\theta }_{W}$ exists w.p.a.1. and is
consistent.

The proofs of the claims for $\widehat{\theta }_{F}$ are similar.\quad $%
\square \bigskip $

\textbf{\noindent Derivation of the minimum rate of convergence of }$%
\widehat{\rho }_{C}$\textbf{: }

We first state some preliminary results. Let $\breve{\sigma}^{2}=\widehat{%
\sigma }^{2}(1,\widehat{\beta }(1))$. Note that $\widehat{\beta }%
(1)=(\sum_{i=1}^{N}X_{i}^{\prime }QX_{i})^{-1}\sum_{i=1}^{N}X_{i}^{\prime
}Q\varepsilon _{i}$ and $\breve{\sigma}^{2}=\frac{1}{(T-1)N}%
\sum_{i=1}^{N}(\varepsilon _{i}-X_{i}\widehat{\beta }(1))^{\prime
}Q(\varepsilon _{i}-X_{i}\widehat{\beta }(1)).$ Let $\frac{\partial ^{j}%
\widetilde{l}_{N}^{c}(1)}{\partial r^{j}}=\frac{\partial ^{j}\widetilde{l}%
_{N}^{c}(r)}{\partial r^{j}}|_{r=1}$ and $\Phi =\Phi (1)$. Then%
\begin{eqnarray}
\frac{\partial \widetilde{l}_{N}^{c}(1)}{\partial r} &=&\breve{\sigma}%
^{-2}[(T-1)\xi ^{\prime }(1)(\breve{\sigma}^{2}-\sigma ^{2})+  \label{lcn} \\
&&(T-1)\xi ^{\prime }(1)\sigma ^{2}+N^{-1}\sum_{i=1}^{N}(\varepsilon
_{i}-X_{i}\widehat{\beta }(1))^{\prime }Q\Phi \varepsilon _{i}]  \notag
\end{eqnarray}%
and 
\begin{eqnarray*}
\frac{\partial ^{2}\widetilde{l}_{N}^{c}(1)}{\partial r^{2}} &=&(T-1)\xi
^{\prime \prime }(1)-\breve{\sigma}^{-2}N^{-1}\sum_{i=1}^{N}(\Phi
\varepsilon _{i}+X_{i}\frac{\partial \widehat{\beta }(1)}{\partial r}%
)^{\prime }Q\Phi \varepsilon _{i}- \\
&&\breve{\sigma}^{-4}\frac{\partial \widehat{\sigma }^{2}(1)}{\partial r}%
N^{-1}\sum_{i=1}^{N}(\varepsilon _{i}-X_{i}\widehat{\beta }(1))^{\prime
}Q\Phi \varepsilon _{i},
\end{eqnarray*}%
where $\frac{\partial \widehat{\beta }(1)}{\partial r}=-(%
\sum_{i=1}^{N}X_{i}^{\prime }QX_{i})^{-1}\sum_{i=1}^{N}X_{i}^{\prime }Q\Phi
\varepsilon _{i}$ and $\frac{\partial \widehat{\sigma }^{2}(1)}{\partial r}%
=-2(T-1)^{-1}N^{-1}\sum_{i=1}^{N}(\varepsilon _{i}-X_{i}\widehat{\beta }%
(1))^{\prime }Q(\Phi \varepsilon _{i}+X_{i}\frac{\partial \widehat{\beta }(1)%
}{\partial r}).$

Clearly $N^{1/2}\left( \frac{\partial \widetilde{l}_{N}^{c}(1)}{\partial r}%
\right) =O_{p}(1).$ Recall that $\xi ^{\prime \prime }(1)-(T-1)^{-1}tr(\Phi
^{\prime }Q\Phi )+2(\xi ^{\prime }(1))^{2}=0.$ Therefore we also have $%
N^{1/2}\left( \frac{\partial ^{2}\widetilde{l}_{N}^{c}(1)}{\partial r^{2}}%
\right) =O_{p}(1).$ Finally, we have 
\begin{eqnarray*}
\text{p}\lim\nolimits_{N\rightarrow \infty }\frac{\partial ^{3}\widetilde{l}%
_{N}^{c}(1)}{\partial r^{3}} &=&\frac{\partial ^{3}\widetilde{l}^{c}(1)}{%
\partial r^{3}}=\frac{T(T-1)(T+1)}{12}>0, \\
\xi ^{\prime \prime \prime \prime }(1) &=&\frac{1}{20}(T-2)(T-3)(T-4), \\
\text{p}\lim\nolimits_{N\rightarrow \infty }\frac{\partial ^{4}\widetilde{l}%
_{N}^{c}(1)}{\partial r^{4}} &=&\frac{\partial ^{4}\widetilde{l}^{c}(1)}{%
\partial r^{4}}=\left( T-1\right) \xi ^{\prime \prime \prime \prime }(1)+%
\frac{1}{6}(T-1)(T^{2}-10T+7)\neq 0, \\
\xi ^{\prime \prime \prime \prime \prime }(1) &=&\frac{1}{30}%
(T-2)(T-3)(T-4)(T-5)\text{\quad and} \\
\text{p}\lim\nolimits_{N\rightarrow \infty }\frac{\partial ^{5}\widetilde{l}%
_{N}^{c}(1)}{\partial r^{5}} &=&\frac{\partial ^{5}\widetilde{l}^{c}(1)}{%
\partial r^{5}}=\left( T-1\right) \xi ^{\prime \prime \prime \prime \prime
}(1)-\frac{1}{3}\left( T-1\right) (5T^{2}-20T+11)\neq 0.
\end{eqnarray*}

We now derive the minimum rate of convergence of $\widehat{\rho }_{C}$.
W.p.a.1 $\widehat{\rho }_{C}$ is a solution of the f.o.c. $\frac{\partial
^{2}\widetilde{l}_{N}^{c}(r)}{\partial r^{2}}\frac{\partial \widetilde{l}%
_{N}^{c}(r)}{\partial r}=0.$ Let $G_{N}^{c}(r)=N^{3/4}\frac{\partial ^{2}%
\widetilde{l}_{N}^{c}(r)}{\partial r^{2}}\frac{\partial \widetilde{l}%
_{N}^{c}(r)}{\partial r}.$ Forming a Taylor expansion of $G_{N}^{c}(\widehat{%
\rho }_{C})$ around $r=1$ gives that $\widehat{\rho }_{C}$ must solve%
\begin{equation*}
0=G_{N}^{c}(1)+\sum_{j=1}^{3}\frac{1}{j!}\frac{\partial ^{j}G_{N}^{c}(1)}{%
\partial r^{j}}(r-1)^{j}+P_{1,N}(N^{1/4}(r-1)),
\end{equation*}%
where $P_{1,N}(N^{1/4}(r-1))$ is a polynomial in $N^{1/4}(r-1)$ with
coefficients that are $o_{p}(1)$. That is, $\widehat{\rho }_{C}$ must solve 
\begin{eqnarray*}
0 &=&N^{-1/4}N\frac{\partial ^{2}\widetilde{l}_{N}^{c}(1)}{\partial r^{2}}%
\frac{\partial \widetilde{l}_{N}^{c}(1)}{\partial r}+ \\
&&N^{1/2}\left( \frac{\partial ^{3}\widetilde{l}_{N}^{c}(1)}{\partial r^{3}}%
\frac{\partial \widetilde{l}_{N}^{c}(1)}{\partial r}+\left( \frac{\partial
^{2}\widetilde{l}_{N}^{c}(1)}{\partial r^{2}}\right) ^{2}\right)
N^{1/4}(r-1)+ \\
&&\frac{1}{2}N^{-1/4}N^{1/2}\left( \frac{\partial ^{4}\widetilde{l}%
_{N}^{c}(1)}{\partial r^{4}}\frac{\partial \widetilde{l}_{N}^{c}(1)}{%
\partial r}+3\frac{\partial ^{3}\widetilde{l}_{N}^{c}(1)}{\partial r^{3}}%
\frac{\partial ^{2}\widetilde{l}_{N}^{c}(1)}{\partial ^{2}r}\right)
N^{1/2}(r-1)^{2}+ \\
&&\frac{1}{3!}\left( \frac{\partial ^{5}\widetilde{l}_{N}^{c}(1)}{\partial
r^{5}}\frac{\partial \widetilde{l}_{N}^{c}(1)}{\partial r}+4\frac{\partial
^{4}\widetilde{l}_{N}^{c}(1)}{\partial r^{4}}\frac{\partial ^{2}\widetilde{l}%
_{N}^{c}(1)}{\partial r^{2}}+3\left( \frac{\partial ^{3}\widetilde{l}%
_{N}^{c}(1)}{\partial r^{3}}\right) ^{2}\right) N^{3/4}(r-1)^{3}+ \\
&&P_{1,N}(N^{1/4}(r-1))
\end{eqnarray*}%
or equivalently%
\begin{eqnarray}
0 &=&N^{1/2}\left( \frac{\partial ^{3}\widetilde{l}_{N}^{c}(1)}{\partial
r^{3}}\frac{\partial \widetilde{l}_{N}^{c}(1)}{\partial r}\right)
N^{1/4}(r-1)+  \label{rexp} \\
&&\frac{1}{2}\left( \frac{\partial ^{3}\widetilde{l}_{N}^{c}(1)}{\partial
^{3}r}\right) ^{2}N^{3/4}(r-1)^{3}+P_{2,N}(N^{1/4}(r-1)),  \notag
\end{eqnarray}%
where $P_{2,N}(N^{1/4}(r-1))$ is another polynomial in $N^{1/4}(r-1)$ with
coefficients that are $o_{p}(1)$. It follows that $N^{1/4}(\widehat{\rho }%
_{C}-1)=O_{p}(1),$ i.e., the rate of convergence of $\widehat{\rho }_{C}$ is
at least $N^{1/4}$.\textbf{\medskip }

\textbf{\noindent Proof of theorem 2 and corollary 1:\medskip }

We will show in the proof below that $Z_{1,N}\overset{d}{\rightarrow }%
Z_{1}\sim N(0,48T^{-2}((T-1)(T+1))^{-1})$ and that there exists a sequence $%
\{U_{N}\}$ with $U_{N}=O_{p}(N^{-1/2})$ such that if $Z_{1,N}+U_{N}>0,$ then 
$M_{N}^{c}(r),$ which is given in (\ref{mnc}), has two local minima attained
at values $\widetilde{\rho }$ such that $N^{1/2}(\widetilde{\rho }%
-1)^{2}=Z_{1,N}+o_{p}(1),$ whereas if $Z_{1,N}+U_{N}<0,$ then $M_{N}^{c}(r)$%
\ has one local minimum attained at $r=\widehat{\rho }$ with $N^{1/2}(%
\widehat{\rho }-1)^{2}=o_{p}(1)$. Furthermore, if $Z_{1,N}+U_{N}>0,$ then
the sign of $N^{1/4}(\widehat{\rho }-1)$ is determined by the remainder $%
R_{1,N}^{c}(N^{1/4}(\widehat{\rho }-1))$ in (\ref{mnc}). We first examine
this remainder: 
\begin{equation*}
N^{1/4}R_{1,N}^{c}(N^{1/4}(\widehat{\rho }-1))=N^{1/4}(\widehat{\rho }%
-1)R_{2,N}^{c}(N^{1/2}(\widehat{\rho }-1)^{2})+R_{3,N}^{c}(N^{1/4}(\widehat{%
\rho }-1))\text{ }
\end{equation*}%
where%
\begin{eqnarray*}
&&R_{2,N}^{c}(N^{1/2}(\widehat{\rho }-1)^{2})=2N\frac{\partial ^{2}%
\widetilde{l}_{N}^{c}(1)}{\partial r^{2}}\frac{\partial \widetilde{l}%
_{N}^{c}(1)}{\partial r}+N^{1/2}\frac{\partial ^{3}\widetilde{l}_{N}^{c}(1)}{%
\partial r^{3}}\frac{\partial ^{2}\widetilde{l}_{N}^{c}(1)}{\partial r^{2}}%
N^{1/2}(\widehat{\rho }-1)^{2}+ \\
&&\qquad \frac{1}{3}N^{1/2}\frac{\partial ^{4}\widetilde{l}_{N}^{c}(1)}{%
\partial r^{4}}\frac{\partial \widetilde{l}_{N}^{c}(1)}{\partial r}N^{1/2}(%
\widehat{\rho }-1)^{2}+\frac{1}{6}\frac{\partial ^{4}\widetilde{l}_{N}^{c}(1)%
}{\partial r^{4}}\frac{\partial ^{3}\widetilde{l}_{N}^{c}(1)}{\partial ^{3}r}%
N(\widehat{\rho }-1)^{4}\quad \text{and}
\end{eqnarray*}%
\begin{equation*}
R_{3,N}^{c}(N^{1/4}(\widehat{\rho }-1))=o_{p}(1).
\end{equation*}%
However, if $N^{1/2}(\widehat{\rho }-1)^{2}=Z_{1,N}+o_{p}(1),$ then $%
R_{2,N}^{c}(N^{1/2}(\widehat{\rho }-1)^{2})=o_{p}(1).$ Therefore we need to
consider $R_{3,N}^{c}(N^{1/4}(\widehat{\rho }-1)).$ We have 
\begin{eqnarray*}
&&N^{1/2}R_{3,N}^{c}(N^{1/4}(\widehat{\rho }-1))=N^{1/4}R_{4,N}^{c}(N^{1/2}(%
\widehat{\rho }-1)^{2})+ \\
&&\qquad N^{5/4}(\widehat{\rho }-1)^{5}R_{5,N}^{c}(N^{1/2}(\widehat{\rho }%
-1)^{2})+R_{6,N}^{c}(N^{1/4}(\widehat{\rho }-1))
\end{eqnarray*}%
where 
\begin{eqnarray*}
&&R_{4,N}^{c}(N^{1/2}(\widehat{\rho }-1)^{2})=(N^{1/2}\frac{\partial ^{2}%
\widetilde{l}_{N}^{c}(1)}{\partial r^{2}})^{2}N^{1/2}(\widehat{\rho }%
-1)^{2}+(\frac{8}{4!}N^{1/2}\frac{\partial ^{4}\widetilde{l}_{N}^{c}(1)}{%
\partial r^{4}}\frac{\partial ^{2}\widetilde{l}_{N}^{c}(1)}{\partial r^{2}}+
\\
&&\qquad \frac{2}{4!}N^{1/2}\frac{\partial ^{5}\widetilde{l}_{N}^{c}(1)}{%
\partial r^{5}}\frac{\partial \widetilde{l}_{N}^{c}(1)}{\partial r})N(%
\widehat{\rho }-1)^{4}+(\frac{30}{6!}\frac{\partial ^{5}\widetilde{l}%
_{N}^{c}(1)}{\partial r^{5}}\frac{\partial ^{3}\widetilde{l}_{N}^{c}(1)}{%
\partial r^{3}}+ \\
&&\qquad \frac{20}{6!}(\frac{\partial ^{4}\widetilde{l}_{N}^{c}(1)}{\partial
r^{4}})^{2})N^{3/2}(\widehat{\rho }-1)^{6}+o_{p}(1), \\
&& \\
&&R_{5,N}^{c}(N^{1/2}(\widehat{\rho }-1)^{2})=\frac{2}{5!}N^{1/2}\frac{%
\partial ^{6}\widetilde{l}_{N}^{c}(1)}{\partial r^{6}}\frac{\partial 
\widetilde{l}_{N}^{c}(1)}{\partial r}+\frac{10}{5!}N^{1/2}\frac{\partial ^{5}%
\widetilde{l}_{N}^{c}(1)}{\partial r^{5}}\frac{\partial ^{2}\widetilde{l}%
_{N}^{c}(1)}{\partial r^{2}}+ \\
&&\qquad \frac{1}{5!}\frac{\partial ^{6}\widetilde{l}_{N}^{c}(1)}{\partial
r^{6}}\frac{\partial ^{3}\widetilde{l}_{N}^{c}(1)}{\partial r^{3}}N^{1/2}(%
\widehat{\rho }-1)^{2}+\frac{10}{6!}\frac{\partial ^{5}\widetilde{l}%
_{N}^{c}(1)}{\partial r^{5}}\frac{\partial ^{4}\widetilde{l}_{N}^{c}(1)}{%
\partial r^{4}}N^{1/2}(\widehat{\rho }-1)^{2}\quad \text{and} \\
&& \\
&&R_{6,N}^{c}(N^{1/4}(\widehat{\rho }-1))=o_{p}(1).
\end{eqnarray*}%
If $N^{1/2}(\widehat{\rho }-1)^{2}=-2(\frac{\partial ^{3}\widetilde{l}%
_{N}^{c}(1)}{\partial r^{3}})^{-1}(N^{1/2}\frac{\partial \widetilde{l}%
_{N}^{c}(1)}{\partial r})+o_{p}(1)$ then\vspace{-0.12in} 
\begin{eqnarray*}
&&R_{4,N}^{c}(N^{1/2}(\widehat{\rho }-1)^{2})=(N^{1/2}\frac{\partial ^{2}%
\widetilde{l}_{N}^{c}(1)}{\partial r^{2}})^{2}N^{1/2}(\widehat{\rho }-1)^{2}+
\\
&&\qquad \frac{1}{3}N^{1/2}\frac{\partial ^{4}\widetilde{l}_{N}^{c}(1)}{%
\partial r^{4}}\frac{\partial ^{2}\widetilde{l}_{N}^{c}(1)}{\partial r^{2}}N(%
\widehat{\rho }-1)^{4}+\frac{1}{36}(\frac{\partial ^{4}\widetilde{l}%
_{N}^{c}(1)}{\partial r^{4}})^{2}N^{3/2}(\widehat{\rho }-1)^{6}+o_{p}(1).
\end{eqnarray*}%
It follows that if $Z_{1,N}+U_{N}>0,$ then the value of $M_{N}^{c}(r)$ is in
fact minimized at 
\begin{equation*}
N^{1/2}(\widehat{\rho }-1)^{2}=Z_{1,N}+N^{-1/2}R_{7,N}^{c}+o_{p}(N^{-1/2})
\end{equation*}%
where 
\begin{eqnarray*}
R_{7,N}^{c} &\equiv &(\frac{\partial ^{3}\widetilde{l}_{N}^{c}(1)}{\partial
r^{3}})^{-2}(-2(N^{1/2}\frac{\partial ^{2}\widetilde{l}_{N}^{c}(1)}{\partial
r^{2}})^{2}- \\
&&\frac{4}{3}(\frac{\partial ^{4}\widetilde{l}_{N}^{c}(1)}{\partial r^{4}}%
)(N^{1/2}\frac{\partial ^{2}\widetilde{l}_{N}^{c}(1)}{\partial r^{2}}%
)Z_{1,N}-\frac{1}{6}(\frac{\partial ^{4}\widetilde{l}_{N}^{c}(1)}{\partial
r^{4}})^{2}Z_{1,N}^{2}).
\end{eqnarray*}%
Furthermore%
\begin{eqnarray*}
R_{2,N}^{c}(N^{1/2}(\widehat{\rho }-1)^{2}) &=&(N^{1/2}\frac{\partial ^{3}%
\widetilde{l}_{N}^{c}(1)}{\partial r^{3}}\frac{\partial ^{2}\widetilde{l}%
_{N}^{c}(1)}{\partial r^{2}}+ \\
&&\frac{1}{6}\frac{\partial ^{4}\widetilde{l}_{N}^{c}(1)}{\partial r^{4}}%
\frac{\partial ^{3}\widetilde{l}_{N}^{c}(1)}{\partial r^{3}}N^{1/2}(\widehat{%
\rho }-1)^{2})\times N^{-1/2}R_{7,N}^{c}+o_{p}(N^{-1/2}).
\end{eqnarray*}%
We also have 
\begin{eqnarray*}
R_{5,N}^{c}(N^{1/2}(\widehat{\rho }-1)^{2}) &=&\frac{10}{5!}N^{1/2}\frac{%
\partial ^{5}\widetilde{l}_{N}^{c}(1)}{\partial r^{5}}\frac{\partial ^{2}%
\widetilde{l}_{N}^{c}(1)}{\partial r^{2}}+ \\
&&\frac{10}{6!}\frac{\partial ^{5}\widetilde{l}_{N}^{c}(1)}{\partial r^{5}}%
\frac{\partial ^{4}\widetilde{l}_{N}^{c}(1)}{\partial r^{4}}N^{1/2}(\widehat{%
\rho }-1)^{2}+o_{p}(1).
\end{eqnarray*}

We conclude that if $Z_{1,N}+U_{N}>0,$ then $N^{1/2}(\widehat{\rho }%
-1)^{2}=Z_{1,N}+O_{p}(N^{-1/2}),$ 
\begin{eqnarray*}
&&R_{1,N}^{c}(N^{1/4}(\widehat{\rho }-1))= \\
&&\qquad N^{-1/2}R_{4,N}^{c}(N^{1/2}(\widehat{\rho }-1)^{2})+N^{1/4}(%
\widehat{\rho }-1)N^{-3/4}R_{N}^{c}(N^{1/2}(\widehat{\rho }%
-1)^{2})+o_{p}(N^{-3/4})\text{\quad }
\end{eqnarray*}%
where%
\begin{equation*}
R_{N}^{c}(N^{1/2}(\widehat{\rho }-1)^{2})\equiv N^{1/2}R_{2,N}^{c}(N^{1/2}(%
\widehat{\rho }-1)^{2})+Z_{1,N}^{2}R_{5,N}^{c}(N^{1/2}(\widehat{\rho }%
-1)^{2})+o_{p}(1),
\end{equation*}%
and%
\begin{equation*}
sgn(N^{1/4}(\widehat{\rho }-1))=sgn(-R_{N}^{c}(N^{1/2}(\widehat{\rho }%
-1)^{2})).
\end{equation*}%
It also follows that $U_{N}=O_{p}(N^{-1/2}).$ If $Z_{1,N}+U_{N}<0,$ then $%
M_{N}^{c}(r)$\ has one local minimum and its value is minimized at $N^{1/2}(%
\widehat{\rho }-1)^{2}=o_{p}(1).$

Next we derive the limiting distributions of $N^{1/2}\left( \frac{\partial 
\widetilde{l}_{N}^{c}(1)}{\partial r}\right) $ and $N^{1/2}\left( \frac{%
\partial ^{2}\widetilde{l}_{N}^{c}(1)}{\partial r^{2}}\right) $. Using that
under normality of $\varepsilon _{i}$ for any constant $T\times T$ matrices $%
M_{1}$ and $M_{2},$ $E(\varepsilon _{i}^{\prime }M_{1}\varepsilon
_{i}\varepsilon _{i}^{\prime }M_{2}\varepsilon _{i})=\sigma
^{4}(tr(M_{1})tr(M_{2})+tr(M_{1}M_{2}+M_{1}^{\prime }M_{2}))$, we find that 
\begin{align*}
N^{1/2}(\breve{\sigma}^{2}-\sigma ^{2})/\sigma ^{2}\overset{d}{\rightarrow }%
V_{1}& \sim N(0,2/(T-1)), \\
N^{1/2}(N^{-1}\sum_{i=1}^{N}\varepsilon _{i}^{\prime }\Phi ^{\prime }Q\Phi
\varepsilon _{i}-\sigma ^{2}tr(\Phi ^{\prime }Q\Phi ))/\sigma ^{2}\overset{d}%
{\rightarrow }V_{2}& \sim N(0,2tr(\Phi ^{\prime }Q\Phi \Phi ^{\prime }Q\Phi
)), \\
N^{1/2}(N^{-1}\sum_{i=1}^{N}\varepsilon _{i}^{\prime }Q\Phi \varepsilon
_{i}-\sigma ^{2}tr(Q\Phi ))/\sigma ^{2}\overset{d}{\rightarrow }V_{3}& \sim
N(0,(tr(Q\Phi Q\Phi )+tr(\Phi ^{\prime }Q\Phi ))),
\end{align*}%
and%
\begin{align*}
E(V_{1}V_{2})& =2(T-1)^{-1}tr(Q\Phi ^{\prime }Q\Phi )=\frac{1}{6}(T+1), \\
E(V_{1}V_{3})& =2(T-1)^{-1}tr(Q\Phi )=-1,\text{\quad and} \\
E(V_{2}V_{3})& =2tr(\Phi ^{\prime }Q\Phi Q\Phi )=-\frac{1}{6}(T-1)(T+1).
\end{align*}%
It is now easily seen that 
\begin{eqnarray*}
N^{1/2}\left( \frac{\partial \widetilde{l}_{N}^{c}(1)}{\partial r}\right) 
\overset{d}{\rightarrow }V_{4} &\equiv &(T-1)\xi ^{\prime }(1)V_{1}+V_{3}=%
\frac{1}{2}(T-1)V_{1}+V_{3}, \\
N^{1/2}\left( \frac{\partial ^{2}\widetilde{l}_{N}^{c}(1)}{\partial r^{2}}%
\right) \overset{d}{\rightarrow }V_{5} &\equiv &(2(T-1)\xi ^{\prime \prime
}(1)-tr(\Phi ^{\prime }Q\Phi ))V_{1}-V_{2}-4\xi ^{\prime }(1)V_{3}= \\
&&\frac{1}{6}(T-1)(T-5)V_{1}-V_{2}-2V_{3}\quad \text{and} \\
Z_{1,N}\overset{d}{\rightarrow }Z_{1} &\equiv &\left( -\frac{1}{2}\frac{%
\partial ^{3}\widetilde{l}^{c}(1)}{\partial r^{3}}\right)
^{-1}V_{4}=-24\left( T(T-1)(T+1)\right) ^{-1}V_{4},
\end{eqnarray*}%
where we have used that $\xi ^{\prime }(1)=\frac{1}{2},$ $\xi ^{\prime
\prime }(1)=\frac{1}{6}(T-2),$ $tr(\Phi ^{\prime }Q\Phi )=\frac{1}{6}%
(T-1)(T+1)$ and $\frac{\partial ^{3}\widetilde{l}^{c}(1)}{\partial r^{3}}=%
\frac{T(T-1)(T+1)}{12}.$ Clearly 
\begin{equation*}
Z_{2,N}=N^{1/2}(\breve{\sigma}^{2}-\sigma ^{2})\overset{d}{\rightarrow }%
Z_{2}\equiv \sigma ^{2}V_{1}.
\end{equation*}%
Noting that 
\begin{eqnarray*}
N^{1/2}(\widehat{\sigma }_{n}^{2}-\sigma ^{2}) &=&\frac{1}{\widehat{\rho }}%
\left\{ Z_{2,N}+2N^{-1/2}(1-\widehat{\rho })(T-1)^{-1}\sum_{i=1}^{N}%
\varepsilon _{i}^{\prime }Q\Phi \varepsilon _{i}+N^{1/2}(1-\widehat{\rho }%
)\sigma ^{2}+\right. \\
&&\left. N^{-1/2}(\widehat{\rho }-1)^{2}(T-1)^{-1}\times
\sum_{i=1}^{N}\varepsilon _{i}^{\prime }\Phi ^{\prime }Q\Phi \varepsilon
_{i}\right\} +o_{p}(1)\quad \text{and} \\
K_{+} &=&\sigma ^{2}tr(\Phi ^{\prime }Q\Phi )/(T-1),
\end{eqnarray*}%
we find that 
\begin{equation*}
N^{1/2}(\widehat{\sigma }_{n}^{2}-\sigma ^{2})\overset{d}{\rightarrow }%
(Z_{2}+K_{+}Z_{1})\times \mathbf{1}\{Z_{1}>0\}+Z_{2}\times \mathbf{1}%
\{Z_{1}\leq 0\}.
\end{equation*}%
Furthermore, using the delta method, we obtain 
\begin{equation*}
N^{1/4}(\widehat{\sigma }_{C}^{2}-\sigma ^{2})-N^{1/4}(\widehat{\rho }%
_{C}-1)\sigma ^{2}=o_{p}(1).
\end{equation*}%
Finally, it is easily seen that 
\begin{equation*}
Z_{3,N}\overset{d}{\rightarrow }Z_{3}\sim N(0,\sigma ^{2}(\Sigma
_{xqx})^{-1}),
\end{equation*}%
that $E(Z_{1}Z_{2})=E(Z_{1}Z_{3})=E(Z_{2}Z_{3})=0$, and that $%
N^{1/2}R_{2,N}^{c}(Z_{1,N}+U_{N})\overset{d}{\rightarrow }R_{2}^{c}$, $%
R_{5,N}^{c}(Z_{1,N}+U_{N})\overset{d}{\rightarrow }R_{5}^{c}$ and $%
R_{N}^{c}(Z_{1,N}+U_{N})\overset{d}{\rightarrow }R^{c}$ for some $R_{2}^{c},$
$R_{5}^{c}$ and $R^{c}$.

Next let 
\begin{equation*}
\tilde{R}_{2}^{c}\equiv \left( \frac{\partial ^{3}\widetilde{l}_{N}^{c}(1)}{%
\partial r^{3}}\right) ^{-1}\left( -2V_{5}^{3}-\frac{5}{3}\frac{\partial ^{4}%
\widetilde{l}^{c}(1)}{\partial r^{4}}V_{5}^{2}Z_{1}-\frac{7}{18}(\frac{%
\partial ^{4}\widetilde{l}^{c}(1)}{\partial r^{4}})^{2}V_{5}Z_{1}^{2}-\frac{1%
}{36}(\frac{\partial ^{4}\widetilde{l}^{c}(1)}{\partial r^{4}}%
)^{3}Z_{1}^{3}\right)
\end{equation*}%
and 
\begin{equation*}
\tilde{R}_{5}^{c}\equiv \frac{1}{12}\frac{\partial ^{5}\widetilde{l}^{c}(1)}{%
\partial r^{5}}(V_{5}+\frac{1}{6}\frac{\partial ^{4}\widetilde{l}^{c}(1)}{%
\partial r^{4}}Z_{1}).
\end{equation*}%
When $Z_{1}>0,$ $R_{2}^{c}=\tilde{R}_{2}^{c},$ $R_{5}^{c}=\tilde{R}_{5}^{c}$
and 
\begin{equation}
R^{c}=\tilde{R}_{2}^{c}+Z_{1}^{2}\tilde{R}_{5}^{c}.  \label{rc}
\end{equation}

Noting that $tr(Q\Phi Q\Phi )=-\frac{1}{12}(T-1)(T-5)$ and $tr(\Phi ^{\prime
}Q\Phi \Phi ^{\prime }Q\Phi )=\frac{1}{180}(2T^{4}+5T^{2}-7),$ we have 
\begin{eqnarray*}
V_{4} &\sim &N(0,\frac{1}{12}(T-1)(T+1)), \\
V_{5} &\sim &N(0,\frac{1}{90}(T-1)(T+1)(2T^{2}+7))\quad \text{and} \\
&&E(V_{4}V_{5})=-\frac{1}{12}(T-1)(T+1).
\end{eqnarray*}%
We can decompose $V_{5}$ as $V_{5}=-V_{4}+V_{0}$ so that $E(V_{0}V_{4})=0$
and 
\begin{equation*}
V_{0}\sim N(0,\frac{1}{180}(T-1)(T+1)(4T^{2}-1)).
\end{equation*}%
Let 
\begin{eqnarray*}
\kappa &=&-2\left( \frac{\partial ^{3}\widetilde{l}_{N}^{c}(1)}{\partial
r^{3}}\right) ^{-1}\frac{\partial ^{4}\widetilde{l}^{c}(1)}{\partial r^{4}},
\\
\tilde{\kappa}_{0}(T) &=&2-\frac{5}{3}\kappa +\frac{7}{18}\kappa ^{2}-\frac{1%
}{36}\kappa ^{3}+\frac{1}{18}\left( \frac{\partial ^{3}\widetilde{l}%
_{N}^{c}(1)}{\partial r^{3}}\right) ^{-1}\frac{\partial ^{5}\widetilde{l}%
^{c}(1)}{\partial r^{5}}(\kappa -6)\quad \text{and} \\
\tilde{\kappa}_{2}(T) &=&6-\frac{5}{3}\kappa .
\end{eqnarray*}%
Then we have 
\begin{equation*}
\tilde{R}_{2}^{c}+Z_{1}^{2}\tilde{R}_{5}^{c}=(\frac{\partial ^{3}\widetilde{l%
}_{N}^{c}(1)}{\partial r^{3}})^{-1}(\tilde{\kappa}_{0}(T)V_{4}^{3}+\tilde{%
\kappa}_{1}(T)V_{4}^{2}V_{0}+\tilde{\kappa}_{2}(T)V_{4}V_{0}^{2}+\tilde{%
\kappa}_{3}(T)V_{0}^{3})
\end{equation*}%
for some $\tilde{\kappa}_{1}(T)$ and $\tilde{\kappa}_{3}(T).$ It is easily
verified that $\tilde{\kappa}_{0}(T)>0$ and $\tilde{\kappa}_{2}(T)>0$ for $%
T\geq 4,$ $\tilde{\kappa}_{0}(2)=0,$ $\tilde{\kappa}_{0}(3)<0,$ $\tilde{%
\kappa}_{2}(2)<0$ and $\tilde{\kappa}_{2}(3)<0.$ Furthermore, because $V_{0}$
is a Gaussian r.v. with mean zero, the conditional p.d.f. of $\tilde{\kappa}%
_{1}(T)V_{4}^{2}V_{0}+\tilde{\kappa}_{3}(T)V_{0}^{3}$ given $V_{4}$ (or
equivalently, given $Z_{1}$) is symmetric around zero. Also, $V_{0}^{2}\geq
0.$ Noting that $B^{c}=\mathbf{1}(R^{c}>0),$ it is now easily seen that $%
E((-1)^{B^{c}}Z_{1}^{1/2}|Z_{1},Z_{1}>0)>0$ when $T\geq 4$, while $%
E((-1)^{B^{c}}Z_{1}^{1/2}|Z_{1},Z_{1}>0)<0$ when $T=2$ or $T=3.$ We can
conclude that $E((-1)^{B^{c}}Z_{1}^{1/2}|Z_{1}>0)>0$ when $T\geq 4$, while $%
E((-1)^{B^{c}}Z_{1}^{1/2}|Z_{1}>0)<0$ when $T=2$ or $T=3.\quad \square
\bigskip $

\textbf{\noindent Derivation of the rate of convergence of }$\widehat{\rho }%
_{F}$\textbf{\ and the limiting distribution of }$\widehat{\theta }_{F}$%
\textbf{\ when }$\rho =1$\textbf{:}

Let\vspace{-0.12in}%
\begin{eqnarray*}
\Psi _{N,n}(\theta _{n}) &\equiv &(\Psi _{\rho ,N,n}(r),\Psi _{\sigma
_{n}^{2},N,n}(\theta _{n}),\Psi _{\beta ,N,n}^{\prime }(\theta
_{n}))^{\prime }= \\
&&(\frac{\partial \widetilde{l}_{N}^{c}(r)}{\partial r},s_{n}^{2}r\frac{%
\partial \widetilde{l}_{n,N}(\theta _{n})}{\partial s_{n}^{2}},s_{n}^{2}r%
\frac{\partial \widetilde{l}_{n,N}(\theta _{n})}{\partial b^{\prime }}%
)^{\prime }\quad \text{and}
\end{eqnarray*}%
\begin{equation*}
M_{N}(\theta _{n})=N\left( \Psi _{N,n}^{\prime }(\theta _{n})\right)
W_{N}\left( \Psi _{N,n}(\theta _{n})\right) .
\end{equation*}%
Note that $\Psi _{N,n}(\theta _{n})$ is a reparametrized and rescaled
version of $(\frac{\partial \widetilde{l}_{N}^{c}(r)}{\partial r},\frac{%
\partial \widetilde{l}_{N}(\theta )}{\partial s^{2}},\frac{\partial 
\widetilde{l}_{N}(\theta )}{\partial b^{\prime }})^{\prime }$ and that p$%
\lim\nolimits_{N\rightarrow \infty }\partial \Psi _{N,n}(\theta _{\ast
})/\partial r=0.$ Let $G_{N}$ be an $(2+K)\times (2+K)$ matrix with\vspace{%
-0.12in}%
\begin{gather*}
G_{N,1,1}=\frac{1}{2}\frac{\partial ^{3}\widetilde{l}_{N}^{c}(1)}{\partial
r^{3}},\quad G_{N,2,2}=\frac{\partial \Psi _{\sigma _{n}^{2},N,n}(\theta
_{\ast })}{\partial s_{n}^{2}},\quad G_{N,2,3}=\frac{\partial \Psi _{\sigma
_{n}^{2},N,n}(\theta _{\ast })}{\partial b}, \\
G_{N,3,3}=\frac{\partial \Psi _{\beta ,N,n}(\theta _{\ast })}{\partial
b^{\prime }},\quad G_{N,2,1}=\frac{1}{2}\frac{\partial ^{2}\Psi _{\sigma
_{n}^{2},N,n}(\theta _{\ast })}{\partial r^{2}}
\end{gather*}%
and the other elements of $G_{N}$ equal to zero. Note that p$%
\lim\nolimits_{N\rightarrow \infty }G_{N}=G$ has full rank. Similarly to the
analysis for $M_{N}^{c}(r)$, we consider a Taylor expansion of $M_{N}(\theta
_{n})$ around $\theta _{n}=\theta _{\ast }.$ Let $\widehat{\theta }=\widehat{%
\theta }_{F},$ $\widehat{\theta }_{n}=\widehat{\theta }_{n,F}$ and $\widehat{%
\omega }_{n}=((\widehat{\rho }-1)^{2},(\widehat{\sigma }_{n}^{2}-\sigma
^{2}),\widehat{\beta }^{\prime })^{\prime }.$ Substituting $\widehat{\theta }%
_{n}$ for $\theta _{n}$ we obtain\vspace{-0.12in}%
\begin{eqnarray}
M_{N}(\widehat{\theta }_{n}) &=&N\left( \Psi _{N,n}^{\prime }(\theta _{\ast
})\right) W_{N}\left( \Psi _{N,n}(\theta _{\ast })\right) +2N^{1/2}\Psi
_{N,n}^{\prime }(\theta _{\ast })W_{N}G_{N}N^{1/2}\widehat{\omega }_{n}+ 
\notag \\
&&N^{1/2}\widehat{\omega }_{n}^{\prime }G_{N}^{\prime }W_{N}G_{N}N^{1/2}%
\widehat{\omega }_{n}+R_{1,N}(N^{1/4}(\widehat{\rho }-1)),  \label{obf1}
\end{eqnarray}%
where $R_{1,N}(N^{1/4}(\widehat{\rho }-1))=o_{p}(1).\vspace{0.06in}$

Let 
\begin{equation*}
W_{N}=\left[ 
\begin{array}{cc}
W_{N,1,1} & \underline{W}_{N,1,2} \\ 
\underline{W}_{N,2,1} & \underline{W}_{N,2,2}%
\end{array}%
\right] \;\text{and\ }G_{N}=\left[ 
\begin{array}{cc}
G_{N,1,1} & \underline{G}_{N,1,2} \\ 
\underline{G}_{N,2,1} & \underline{G}_{N,2,2}%
\end{array}%
\right] ,
\end{equation*}%
where $\underline{W}_{N,2,1}=\underline{W}_{N,1,2}^{\prime },$ $\underline{G}%
_{N,2,1}=(G_{N,2,1},0^{\prime })^{\prime }$ and $\underline{G}%
_{N,1,2}^{\prime }=0$ are $(K+1)-$vectors, and let 
\begin{equation*}
\Psi _{N,n}(\theta _{\ast })=(\Psi _{\rho ,N,n}(1),\underline{\Psi }%
_{N,n}^{\prime }(\theta _{\ast }))^{\prime }\quad \text{and\quad }\widehat{%
\omega }_{n}=((\widehat{\rho }-1)^{2},\widehat{\underline{\omega }}%
_{n}^{\prime })^{\prime }.
\end{equation*}%
Then we have the following result:

\begin{lemma}
There exists a sequence $\{\widetilde{U}_{N}\}$ with $\widetilde{U}%
_{N}=o_{p}(1)$ such that if $Z_{1,N}+\widetilde{U}_{N}>0,$ then the value of 
$M_{N}(\widehat{\theta }_{n})$ in (\ref{obf1}) is minimized at $N^{1/2}(%
\widehat{\rho }-1)^{2}=Z_{1,N}+o_{p}(1)$ and $N^{1/2}\widehat{\underline{%
\omega }}_{n}=N^{1/2}\widehat{\underline{\omega }}_{+}$ where $N^{1/2}%
\widehat{\underline{\omega }}_{+}=\underline{G}_{N,2,2}^{-1}\underline{G}%
_{N,2,1}G_{N,1,1}^{-1}N^{1/2}\Psi _{\rho ,N,n}(1)-\underline{G}%
_{N,2,2}^{-1}N^{1/2}\underline{\Psi }_{N,n}(\theta _{\ast })+o_{p}(1)$
(i.e., at $N^{1/2}\widehat{\omega }_{n}=-G_{N}^{-1}N^{1/2}\Psi _{N,n}(\theta
_{\ast })+o_{p}(1)$ ), whereas if $Z_{1,N}+\widetilde{U}_{N}<0,$ the value
of $M_{N}(\widehat{\theta }_{n})$ is minimized at $N^{1/2}(\widehat{\rho }%
-1)^{2}=o_{p}(1)$ and $N^{1/2}\widehat{\underline{\omega }}_{n}=N^{1/2}%
\widehat{\underline{\omega }}_{-}$ where $N^{1/2}\widehat{\underline{\omega }%
}_{-}\equiv N^{1/2}\widehat{\underline{\omega }}_{+}+K_{-,N}Z_{1,N}$\ with $%
K_{-,N}\equiv \underline{G}_{N,2,2}^{-1}\underline{W}_{N,2,2}^{-1}\underline{%
W}_{N,2,1}G_{N,1,1}+\underline{G}_{N,2,2}^{-1}\underline{G}_{N,2,1}$ (i.e.,
at $N^{1/2}\widehat{\omega }_{n}=(o_{p}(1),N^{1/2}\widehat{\underline{\omega 
}}_{-}^{\prime })^{\prime }$ ).
\end{lemma}

\textbf{\noindent Proof of lemma 5: }Minimizing $M_{N}(\widehat{\theta }%
_{n}) $ given in (\ref{obf1}) w.r.t. $N^{1/2}\widehat{\omega }_{n}$ is
equivalent to minimizing 
\begin{equation*}
\widetilde{M}_{N}(\widehat{\theta }_{n})\equiv 2(G_{N}^{-1}N^{1/2}\Psi
_{N,n}(\theta _{\ast }))^{\prime }\widetilde{W}_{N}N^{1/2}\widehat{\omega }%
_{n}+N^{1/2}\widehat{\omega }_{n}^{\prime }\widetilde{W}_{N}N^{1/2}\widehat{%
\omega }_{n}+R_{1,N}(N^{1/4}(\widehat{\rho }-1))
\end{equation*}%
w.r.t. $N^{1/2}\widehat{\omega }_{n},$ where $\widetilde{W}%
_{N}=G_{N}^{\prime }W_{N}G_{N}.\vspace{0.06in}$ Since $W_{N}$ is PD and $%
G_{N}$ has full rank, $\widetilde{W}_{N}$ is also PD. Partition $\widetilde{W%
}_{N}$ as $\left[ 
\begin{array}{cc}
\widetilde{W}_{N,1,1} & \widetilde{W}_{N,1,2} \\ 
\widetilde{W}_{N,2,1} & \widetilde{W}_{N,2,2}%
\end{array}%
\right] $ $\vspace{0.06in}$where $\widetilde{W}_{N,1,1}$ is a scalar and $%
\widetilde{W}_{N,2,1}=\widetilde{W}_{N,1,2}^{\prime }$ is a $(K+1)-$vector.
Given the value of $N^{1/2}(\widehat{\rho }-1)^{2},$ $\widetilde{M}_{N}(%
\widehat{\theta }_{n})$ is minimized at $N^{1/2}\widehat{\underline{\omega }}%
_{n}=N^{1/2}\widehat{\underline{\omega }}_{+}-\widetilde{W}_{N,2,2}^{-1}%
\widetilde{W}_{N,2,1}(N^{1/2}(\widehat{\rho }-1)^{2}-Z_{1,N})$. Substituting
this expression for $N^{1/2}\widehat{\underline{\omega }}_{n}$ in $%
\widetilde{M}_{N}(\widehat{\theta }_{n})$ and noting that $%
G_{N,1,1}^{-1}N^{1/2}\Psi _{\rho ,N,n}(1)=-Z_{1,N}$ gives%
\begin{equation}
\widetilde{M}_{N}(\widehat{\theta }_{n})=(-2Z_{1,N}N^{1/2}(\widehat{\rho }%
-1)^{2}+N(\widehat{\rho }-1)^{4})(\widetilde{W}_{N,1,1}-\widetilde{W}%
_{N,2,1}^{\prime }\widetilde{W}_{N,2,2}^{-1}\widetilde{W}_{N,2,1})+%
\widetilde{R}_{N}+o_{p}(1),  \label{mtilap}
\end{equation}%
where $\widetilde{R}_{N}$ does not depend on $\widehat{\theta }_{n}.$ Noting
that p$\lim\nolimits_{N\rightarrow \infty }(\widetilde{W}_{N,1,1}-\widetilde{%
W}_{N,2,1}^{\prime }\widetilde{W}_{N,2,2}^{-1}\widetilde{W}_{N,2,1})>0$
(because p$\lim\nolimits_{N\rightarrow \infty }\widetilde{W}_{N}$ is PD) and 
$\widetilde{W}_{N,2,2}^{-1}\widetilde{W}_{N,2,1}=\underline{G}_{N,2,2}^{-1}%
\underline{W}_{N,2,2}^{-1}\underline{W}_{N,2,1}G_{N,1,1}+$ $\underline{G}%
_{N,2,2}^{-1}\underline{G}_{N,2,1}$ $=$ $K_{-,N},$ the claims in the lemma
follow straightforwardly.$\quad \square \medskip $\newpage

\textbf{\noindent Proof of theorem 3:}

According to lemma 5, if $Z_{1,N}+\widetilde{U}_{N}>0,$ then the value of $%
M_{N}(\widehat{\theta }_{n})$ in (\ref{obf1}) is minimized at $N^{1/2}(%
\widehat{\rho }-1)^{2}=Z_{1,N}+o_{p}(1)$ and $N^{1/2}\widehat{\underline{%
\omega }}_{n}=N^{1/2}\widehat{\underline{\omega }}_{+}$. The sign of $%
N^{1/4}(\widehat{\rho }-1)$ is such that it minimizes the value of $%
R_{1,N}(N^{1/4}(\widehat{\rho }-1))$ in (\ref{obf1}) where\vspace*{-0.07in}%
\begin{eqnarray}
&&N^{1/4}R_{1,N}(N^{1/4}(\widehat{\rho }-1))=2N^{1/2}(\Psi _{N,n}(\theta
_{\ast })+\frac{\partial \Psi _{N,n}(\theta _{\ast })}{\partial \underline{w}%
_{n}^{\prime }}\widehat{\underline{\omega }}_{n})^{\prime }W_{N}\times 
\notag \\
&&N^{1/2}(\frac{\partial \Psi _{N,n}(\theta _{\ast })}{\partial r}+\frac{%
\partial ^{2}\Psi _{N,n}(\theta _{\ast })}{\partial r\partial \underline{w}%
_{n}^{\prime }}\widehat{\underline{\omega }}_{n})N^{1/4}(\widehat{\rho }-1)+
\label{rem1} \\
&&(\frac{2}{3!}N^{1/2}(\Psi _{N,n}(\theta _{\ast })+\frac{\partial \Psi
_{N,n}(\theta _{\ast })}{\partial \underline{w}_{n}^{\prime }}\widehat{%
\underline{\omega }}_{n})^{\prime }W_{N}\frac{\partial ^{3}\Psi
_{N,n}(\theta _{\ast })}{\partial r^{3}}+  \notag \\
&&\quad N^{1/2}(\frac{\partial \Psi _{N,n}(\theta _{\ast })}{\partial r}+%
\frac{\partial ^{2}\Psi _{N,n}(\theta _{\ast })}{\partial r\partial 
\underline{w}_{n}^{\prime }}\widehat{\underline{\omega }}_{n})^{\prime }W_{N}%
\frac{\partial ^{2}\Psi _{N,n}(\theta _{\ast })}{\partial r^{2}})N^{3/4}(%
\widehat{\rho }-1)^{3}+  \notag \\
&&\frac{1}{3!}\frac{\partial ^{3}\Psi _{N,n}^{\prime }(\theta _{\ast })}{%
\partial r^{3}}W_{N}\frac{\partial ^{2}\Psi _{N,n}(\theta _{\ast })}{%
\partial r^{2}}N^{5/4}(\widehat{\rho }-1)^{5}+o_{p}(1).  \notag
\end{eqnarray}%
We can write (\ref{rem1}) as $N^{1/4}R_{1,N}(N^{1/4}(\widehat{\rho }%
-1))=N^{1/4}(\widehat{\rho }-1)R_{2,N}(N^{1/4}(\widehat{\rho }%
-1))+R_{3,N}(N^{1/4}(\widehat{\rho }-1))$ where $R_{3,N}(N^{1/4}(\widehat{%
\rho }-1))=o_{p}(1)$. Noting that\vspace*{-0.07in}%
\begin{eqnarray*}
&&N^{1/2}\Psi _{\sigma _{n}^{2},N,n}(\theta _{\ast })\overset{d}{\rightarrow 
}\frac{(T-1)}{2}V_{1},\quad N^{1/2}\Psi _{\beta ,N,n}(\theta _{\ast })%
\overset{d}{\rightarrow }V_{6}\sim N(0,\sigma ^{2}\Sigma _{xqx}), \\
&&\text{p}\lim\nolimits_{N\rightarrow \infty }\frac{\partial \Psi
_{N,n}(\theta _{\ast })}{\partial \underline{w}_{n}^{\prime }}=\text{p}%
\lim\nolimits_{N\rightarrow \infty }(0,\frac{\partial \Psi _{\sigma
_{n}^{2},N,n}(\theta _{\ast })}{\partial \underline{w}_{n}},\frac{\partial
\Psi _{\beta ,N,n}^{\prime }(\theta _{\ast })}{\partial \underline{w}_{n}}%
)^{\prime }, \\
&&\text{p}\lim\nolimits_{N\rightarrow \infty }\frac{\partial \Psi _{\sigma
_{n}^{2},N,n}(\theta _{\ast })}{\partial s_{n}^{2}}=-\frac{(T-1)}{2\sigma
^{2}},\quad \text{p}\lim\nolimits_{N\rightarrow \infty }\frac{\partial \Psi
_{\beta ,N,n}^{\prime }(\theta _{\ast })}{\partial b}=-\Sigma _{xqx}, \\
&&\text{p}\lim\nolimits_{N\rightarrow \infty }\frac{\partial \Psi _{\sigma
_{n}^{2},N,n}(\theta _{\ast })}{\partial b}=\text{p}\lim\nolimits_{N%
\rightarrow \infty }\frac{\partial \Psi _{\beta ,N,n}^{\prime }(\theta
_{\ast })}{\partial s_{n}^{2}}=0, \\
&&\text{p}\lim\nolimits_{N\rightarrow \infty }\frac{\partial ^{j+1}\Psi
_{N,n}(\theta _{\ast })}{\partial r^{j}\partial \underline{w}_{n}^{\prime }}=%
\text{p}\lim\nolimits_{N\rightarrow \infty }(0,\frac{\partial ^{j+1}\Psi
_{\sigma _{n}^{2},N,n}(\theta _{\ast })}{\partial r^{j}\partial \underline{w}%
_{n}},0)^{\prime },\text{\quad }j=1,2, \\
&&\text{p}\lim\nolimits_{N\rightarrow \infty }\frac{\partial ^{2}\Psi
_{\sigma _{n}^{2},N,n}(\theta _{\ast })}{\partial r\partial s_{n}^{2}}=-%
\frac{(T-1)}{2\sigma ^{2}},\quad \text{p}\lim\nolimits_{N\rightarrow \infty }%
\frac{\partial ^{2}\Psi _{\sigma _{n}^{2},N,n}(\theta _{\ast })}{\partial
r\partial \beta }=0, \\
&&\text{p}\lim\nolimits_{N\rightarrow \infty }\frac{\partial ^{3}\Psi
_{\sigma _{n}^{2},N,n}(\theta _{\ast })}{\partial r^{2}\partial s_{n}^{2}}=-%
\frac{(T^{2}-1)}{6\sigma ^{2}},\quad \text{p}\lim\nolimits_{N\rightarrow
\infty }\frac{\partial ^{2}\Psi _{\sigma _{n}^{2},N,n}(\theta _{\ast })}{%
\partial r\partial \beta }=0, \\
&&N^{1/2}\frac{\partial \Psi _{\sigma _{n}^{2},N,n}(\theta _{\ast })}{%
\partial r}\overset{d}{\rightarrow }-V_{3},\quad N^{1/2}\frac{\partial \Psi
_{\beta ,N,n}(\theta _{\ast })}{\partial r}\overset{d}{\rightarrow }%
V_{7}\sim N(0,\sigma ^{2}\Sigma _{xq\phi \phi ^{\prime }qx}), \\
&&\text{p}\lim\nolimits_{N\rightarrow \infty }\frac{\partial ^{4}\widetilde{l%
}_{N}^{c}(1)}{\partial r^{4}}=\frac{\partial ^{4}\widetilde{l}^{c}(1)}{%
\partial r^{4}}=\frac{1}{60}(T-1)(T+1)(3T^{2}-20T-2), \\
&&\text{p}\lim\nolimits_{N\rightarrow \infty }\frac{\partial ^{3}\Psi
_{\sigma _{n}^{2},N,n}(\theta _{\ast })}{\partial r^{3}}=0,\quad \text{p}%
\lim\nolimits_{N\rightarrow \infty }\frac{\partial ^{3}\Psi _{\beta
,N,n}(\theta _{\ast })}{\partial r^{3}}=0, \\
&&\text{p}\lim\nolimits_{N\rightarrow \infty }\frac{\partial ^{2}\Psi
_{\sigma _{n}^{2},N,n}(\theta _{\ast })}{\partial r^{2}}=tr(\Phi ^{\prime
}Q\Phi )\quad \text{and}\quad \text{p}\lim\nolimits_{N\rightarrow \infty }%
\frac{\partial ^{2}\Psi _{\beta ,N,n}(\theta _{\ast })}{\partial r^{2}}=0,
\end{eqnarray*}%
and recalling that $N^{1/2}\frac{\partial \widetilde{l_{N}}^{c}(1)}{\partial
r}\overset{d}{\rightarrow }V_{4},$ $N^{1/2}\frac{\partial ^{2}\widetilde{%
l_{N}}^{c}(1)}{\partial r^{2}}\overset{d}{\rightarrow }V_{5}$ and p$%
\lim\nolimits_{N\rightarrow \infty }\frac{\partial ^{3}\widetilde{l}%
_{N}^{c}(1)}{\partial r^{3}}=\frac{\partial ^{3}\widetilde{l}^{c}(1)}{%
\partial r^{3}}=\frac{1}{12}T(T-1)(T+1),$ in other words, noting that 
\begin{eqnarray*}
&&N^{1/2}\Psi _{N,n}(\theta _{\ast })\overset{d}{\rightarrow }\Psi
_{n}(\theta _{\ast }),\quad N^{1/2}\frac{\partial \Psi _{N,n}(\theta _{\ast
})}{\partial r}\overset{d}{\rightarrow }\Psi _{n,\rho }(\theta _{\ast }), \\
&&p\lim\nolimits_{N\rightarrow \infty }\frac{\partial \Psi _{N,n}(\theta
_{\ast })}{\partial \underline{w}_{n}^{\prime }}=\Psi _{n,\omega }(\theta
_{\ast }), \\
&&p\lim\nolimits_{N\rightarrow \infty }\frac{\partial ^{j+1}\Psi
_{N,n}(\theta _{\ast })}{\partial r^{j}\partial \underline{w}_{n}^{\prime }}%
=\Psi _{n,\rho ^{j}\omega }(\theta _{\ast }),\text{\quad }j=1,2, \\
&&p\lim\nolimits_{N\rightarrow \infty }\frac{\partial ^{2}\Psi _{N,n}(\theta
_{\ast })}{\partial r^{2}}=\Psi _{n,\rho ^{2}}(\theta _{\ast })\quad \text{%
and} \\
&&p\lim\nolimits_{N\rightarrow \infty }\frac{\partial ^{3}\Psi _{N,n}(\theta
_{\ast })}{\partial r^{3}}=\Psi _{n,\rho ^{3}}(\theta _{\ast }),
\end{eqnarray*}%
it follows that p$\lim_{N\rightarrow \infty }G_{N}=G$ and that if $Z_{1,N}+%
\widetilde{U}_{N}>0,$ then $N^{1/2}(\widehat{\rho }-1)^{2}\overset{d}{%
\rightarrow }Z_{1}$, $N^{1/2}\widehat{\underline{\omega }}_{n}=N^{1/2}%
\widehat{\underline{\omega }}_{+}\overset{d}{\rightarrow }\underline{\omega }%
_{+}=-\underline{G}_{2,2}^{-1}\underline{G}_{2,1}Z_{1}-\underline{G}%
_{2,2}^{-1}\underline{\Psi }_{n}(\theta _{\ast })=-(\Psi _{n,\omega
}^{\prime }(\theta _{\ast })W\Psi _{n,\omega }(\theta _{\ast }))^{-1}\times $
$\Psi _{n,\omega }^{\prime }(\theta _{\ast })W(\Psi _{n}(\theta _{\ast })+%
\frac{1}{2}\Psi _{n,\rho ^{2}}(\theta _{\ast })Z_{1})$ and $R_{2,N}(N^{1/4}(%
\widehat{\rho }-1))\overset{d}{\rightarrow }R_{2},$ which after lengthy but
simple calculations can be shown to obey:\vspace{-0.08in} 
\begin{equation}
R_{2}=(Z_{1}\frac{\partial ^{3}\widetilde{l}^{c}(1)}{\partial r^{3}}%
+2V_{4})(W_{1,1}-\underline{W}_{1,2}\underline{W}_{2,2}^{-1}\underline{W}%
_{2,1})(\frac{1}{6}\frac{\partial ^{4}\widetilde{l}^{c}(1)}{\partial r^{4}}%
Z_{1}+V_{5}).  \label{rtwee}
\end{equation}%
Let $\Psi _{n,\omega }=\Psi _{n,\omega }(\theta _{\ast })$ and $\mathcal{M}%
=-(\Psi _{n,\omega }^{\prime }W\Psi _{n,\omega })^{-1}\Psi _{n,\omega
}^{\prime }W.$ To derive (\ref{rtwee}) we have used that $W(I+\Psi
_{n,\omega }\mathcal{M})=diag(W_{1,1}-\underline{W}_{1,2}\underline{W}%
_{2,2}^{-1}\underline{W}_{2,1},0,...,0)$ and $W(I+\Psi _{n,\omega }\mathcal{M%
})\Psi _{n,\rho \omega }(\theta _{\ast })=0.$

Recall that $Z_{1}=-(\frac{1}{2}\frac{\partial ^{3}\widetilde{l}^{c}(1)}{%
\partial r^{3}})^{-1}V_{4}.$ Hence $R_{2,N}=o_{p}(1)$. This result also
follows directly from (\ref{rem1}) and $N^{1/2}(\Psi _{N,n}(\theta _{\ast })+%
\frac{\partial \Psi _{N,n}(\theta _{\ast })}{\partial \underline{w}%
_{n}^{\prime }}\widehat{\underline{\omega }}_{n}+\frac{1}{2}\frac{\partial
^{2}\Psi _{N,n}(\theta _{\ast })}{\partial r^{2}}(\widehat{\rho }%
-1)^{2})=o_{p}(1)$ when $Z_{1,N}+\widetilde{U}_{N}>$\linebreak $0$. The
latter result holds because in the just identified case $\Psi _{N,n}(%
\widehat{\theta }_{n})=0$ when $Z_{1,N}+\widetilde{U}_{N}>0$. Just as in the
case of $N^{1/4}R_{1,N}^{c}(N^{1/4}(\widehat{\rho }-1))$ in the proof of
theorem 2, we need\linebreak to consider a higher order expansion of the
remainder term $N^{1/4}R_{1,N}(N^{1/4}(\widehat{\rho }-1))$ in order to find
the limiting distribution of the sign of $N^{1/4}(\widehat{\rho }-1)$ when $%
Z_{1}>0,$ i.e., the dis-\linebreak tribution of $B$ given $Z_{1}>0.$ This
means we need to consider $R_{3,N}(N^{1/4}(\widehat{\rho }-1)).$ We have 
\begin{eqnarray*}
&&N^{1/2}R_{3,N}(N^{1/4}(\widehat{\rho }-1))=N^{1/4}R_{4,N}(N^{1/2}(\widehat{%
\rho }-1)^{2})+ \\
&&\qquad N^{1/4}(\widehat{\rho }-1)R_{5,N}(N^{1/2}(\widehat{\rho }%
-1)^{2})+R_{6,N}(N^{1/4}(\widehat{\rho }-1))
\end{eqnarray*}%
where $R_{6,N}(N^{1/4}(\widehat{\rho }-1))=o_{p}(1),$ 
\begin{equation*}
R_{4,N}(N^{1/2}(\widehat{\rho }-1)^{2})=\hspace{4in}
\end{equation*}%
\begin{eqnarray*}
&&(N(\frac{\partial \Psi _{N,n}(\theta _{\ast })}{\partial r}+\frac{\partial
^{2}\Psi _{N,n}(\theta _{\ast })}{\partial r\partial \underline{w}%
_{n}^{\prime }}\widehat{\underline{\omega }}_{n})^{\prime }W_{N}(\frac{%
\partial \Psi _{N,n}(\theta _{\ast })}{\partial r}+\frac{\partial ^{2}\Psi
_{N,n}(\theta _{\ast })}{\partial r\partial \underline{w}_{n}^{\prime }}%
\widehat{\underline{\omega }}_{n})+ \\
&&\quad \frac{1}{2!}N(\kappa _{1,N},\text{ }\kappa _{2,N},\text{ }\ldots ,%
\text{ }\kappa _{K+2,N})^{\prime }W_{N}\frac{\partial ^{2}\Psi _{N,n}(\theta
_{\ast })}{\partial r^{2}}+ \\
&&\quad N(\Psi _{N,n}(\theta _{\ast })+\frac{\partial \Psi _{N,n}(\theta
_{\ast })}{\partial \underline{w}_{n}^{\prime }}\widehat{\underline{\omega }}%
_{n})^{\prime }W_{N}\frac{\partial ^{3}\Psi _{N,n}(\theta _{\ast })}{%
\partial r^{2}\partial \underline{w}_{n}^{\prime }}\widehat{\underline{%
\omega }}_{n})N^{1/2}(\widehat{\rho }-1)^{2}+ \\
&&(\frac{2}{3!}N^{1/2}(\frac{\partial \Psi _{N,n}(\theta _{\ast })}{\partial
r}+\frac{\partial ^{2}\Psi _{N,n}(\theta _{\ast })}{\partial r\partial 
\underline{w}_{n}^{\prime }}\widehat{\underline{\omega }}_{n})^{\prime }W_{N}%
\frac{\partial ^{3}\Psi _{N,n}(\theta _{\ast })}{\partial r^{3}}+ \\
&&\quad \frac{1}{2!}N^{1/2}\frac{\partial ^{2}\Psi _{N,n}^{\prime }(\theta
_{\ast })}{\partial r^{2}}W_{N}\frac{\partial ^{3}\Psi _{N,n}(\theta _{\ast
})}{\partial r^{2}\partial \underline{w}_{n}^{\prime }}\widehat{\underline{%
\omega }}_{n}+ \\
&&\quad \frac{2}{4!}N^{1/2}(\Psi _{N,n}(\theta _{\ast })+\frac{\partial \Psi
_{N,n}(\theta _{\ast })}{\partial \underline{w}_{n}^{\prime }}\widehat{%
\underline{\omega }}_{n})^{\prime }W_{N}\frac{\partial ^{4}\Psi
_{N,n}(\theta _{\ast })}{\partial r^{4}})N(\widehat{\rho }-1)^{4}+ \\
&&(\frac{1}{4!}\frac{\partial ^{2}\Psi _{N,n}^{\prime }(\theta _{\ast })}{%
\partial r^{2}}W_{N}\frac{\partial ^{4}\Psi _{N,n}(\theta _{\ast })}{%
\partial r^{4}}+\frac{1}{3!3!}\frac{\partial ^{3}\Psi _{N,n}^{\prime
}(\theta _{\ast })}{\partial r^{3}}W_{N}\frac{\partial ^{3}\Psi
_{N,n}(\theta _{\ast })}{\partial r^{3}})N^{3/2}(\widehat{\rho }%
-1)^{6},\quad \text{and} \\
&& \\
&&R_{5,N}(N^{1/2}(\widehat{\rho }-1)^{2})=N^{3/2}(\kappa _{1,N},\text{ }%
\kappa _{2,N},\text{ }\ldots ,\text{ }\kappa _{K+2,N})^{\prime }W_{N}(\frac{%
\partial \Psi _{N,n}(\theta _{\ast })}{\partial r}+\frac{\partial ^{2}\Psi
_{N,n}(\theta _{\ast })}{\partial r\partial \underline{w}_{n}^{\prime }}%
\widehat{\underline{\omega }}_{n})+ \\
&&(\frac{1}{3!}N(\kappa _{1,N},\text{ }\kappa _{2,N},\text{ }\ldots ,\text{ }%
\kappa _{K+2,N})^{\prime }W_{N}\frac{\partial ^{3}\Psi _{N,n}(\theta _{\ast
})}{\partial r^{3}}+ \\
&&\quad \frac{2}{3!}N(\Psi _{N,n}(\theta _{\ast })+\frac{\partial \Psi
_{N,n}(\theta _{\ast })}{\partial \underline{w}_{n}^{\prime }}\widehat{%
\underline{\omega }}_{n})^{\prime }W_{N}\frac{\partial ^{4}\Psi
_{N,n}(\theta _{\ast })}{\partial r^{3}\partial \underline{w}_{n}^{\prime }}%
\widehat{\underline{\omega }}_{n}+ \\
&&\quad N(\frac{\partial \Psi _{N,n}(\theta _{\ast })}{\partial r}+\frac{%
\partial ^{2}\Psi _{N,n}(\theta _{\ast })}{\partial r\partial \underline{w}%
_{n}^{\prime }}\widehat{\underline{\omega }}_{n})^{\prime }W_{N}\frac{%
\partial ^{3}\Psi _{N,n}(\theta _{\ast })}{\partial r^{2}\partial \underline{%
w}_{n}^{\prime }}\widehat{\underline{\omega }}_{n})N^{1/2}(\widehat{\rho }%
-1)^{2}+ \\
&&(\frac{2}{5!}N^{1/2}(\Psi _{N,n}(\theta _{\ast })+\frac{\partial \Psi
_{N,n}(\theta _{\ast })}{\partial \underline{w}_{n}^{\prime }}\widehat{%
\underline{\omega }}_{n})^{\prime }W_{N}\frac{\partial ^{5}\Psi
_{N,n}(\theta _{\ast })}{\partial r^{5}}+\frac{1}{3!}N^{1/2}\frac{\partial
^{2}\Psi _{N,n}^{\prime }(\theta _{\ast })}{\partial r^{2}}W_{N}\frac{%
\partial ^{4}\Psi _{N,n}(\theta _{\ast })}{\partial r^{3}\partial \underline{%
w}_{n}^{\prime }}\widehat{\underline{\omega }}_{n}+ \\
&&\quad \frac{2}{4!}N^{1/2}(\frac{\partial \Psi _{N,n}(\theta _{\ast })}{%
\partial r}+\frac{\partial ^{2}\Psi _{N,n}(\theta _{\ast })}{\partial
r\partial \underline{w}_{n}^{\prime }}\widehat{\underline{\omega }}%
_{n})^{\prime }W_{N}\frac{\partial ^{4}\Psi _{N,n}(\theta _{\ast })}{%
\partial r^{4}}+\frac{1}{3!}N^{1/2}\frac{\partial ^{3}\Psi _{N,n}^{\prime
}(\theta _{\ast })}{\partial r^{3}}W_{N}\frac{\partial ^{3}\Psi
_{N,n}(\theta _{\ast })}{\partial r^{2}\partial \underline{w}_{n}^{\prime }}%
\widehat{\underline{\omega }}_{n})\times \qquad \\
&&N(\widehat{\rho }-1)^{4}+(\frac{1}{5!}\frac{\partial ^{2}\Psi
_{N,n}^{\prime }(\theta _{\ast })}{\partial r^{2}}W_{N}\frac{\partial
^{5}\Psi _{N,n}(\theta _{\ast })}{\partial r^{5}}+\frac{2}{3!4!}\frac{%
\partial ^{3}\Psi _{N,n}^{\prime }(\theta _{\ast })}{\partial r^{3}}W_{N}%
\frac{\partial ^{4}\Psi _{N,n}(\theta _{\ast })}{\partial r^{4}})N^{3/2}(%
\widehat{\rho }-1)^{6}
\end{eqnarray*}%
with $\kappa _{k,N}=\widehat{\underline{\omega }}_{n}^{\prime }K_{k,N}%
\widehat{\underline{\omega }}_{n}$ and $K_{k,N}=\frac{\partial ^{2}\Psi
_{k,N,n}(\theta _{\ast })}{\partial w_{n}\partial \underline{w}_{n}^{\prime }%
}$ for $k=1,2,...,K+2,$ where $\Psi _{k,N,n}(\theta )$ is the kth element of 
$\Psi _{N,n}(\theta ).$ It is easily seen that p$\lim\nolimits_{N\rightarrow
\infty }K_{k,N}=\mathbf{0}$ for $k=1,3,4,...,K+2,$ and that p$%
\lim\nolimits_{N\rightarrow \infty }K_{2,N}=diag(\sigma ^{-4}(T-1),$ $\sigma
^{-2}\Sigma _{xqx}).$ Using that when $Z_{1,N}+\widetilde{U}_{N}>0,$ $%
N^{1/2}(\Psi _{N,n}(\theta _{\ast })+\frac{\partial \Psi _{N,n}(\theta
_{\ast })}{\partial \underline{w}_{n}^{\prime }}\widehat{\underline{\omega }}%
_{n}+\frac{1}{2}\frac{\partial ^{2}\Psi _{N,n}(\theta _{\ast })}{\partial
r^{2}}(\widehat{\rho }-1)^{2})=(\widehat{\rho }%
-1)F_{1,N}+N^{-1/2}F_{2,N}+o_{p}(N^{-1/2})=o_{p}(1)$ where $F_{1,N}=$ $(0,$ $%
0,$ $N^{-1/2}\sum\nolimits_{i=1}^{N}(X_{i}^{\prime }Q_{i}\Phi \varepsilon
_{i})^{\prime })^{\prime }=O_{p}(1)$ and $F_{2,N}=(\frac{1}{2}\frac{\partial
^{3}\widetilde{l}_{N}^{c}(1)}{\partial r^{3}}R_{7,N},$ $\sigma
^{-2}\sum\nolimits_{i=1}^{N}(\varepsilon _{i}^{\prime
}QX_{i})(\sum\nolimits_{i=1}^{N}(X_{i}^{\prime
}QX_{i}))^{-1}\sum\nolimits_{i=1}^{N}(X_{i}^{\prime }Q_{i}\varepsilon _{i}),$
$\mathbf{0}^{\prime })^{\prime }=O_{p}(1)$ with $R_{7,N}=O_{p}(1)$ defined
below, we obtain the following expressions for $R_{4,N}(N^{1/2}(\widehat{%
\rho }-1)^{2}),$ $R_{5,N}(N^{1/2}(\widehat{\rho }-1)^{2})$ and $%
R_{2,N}(N^{1/4}(\widehat{\rho }-1)):$\pagebreak 
\begin{eqnarray}
&&R_{4,N}(N^{1/2}(\widehat{\rho }-1)^{2})=N((\frac{\partial \Psi
_{N,n}(\theta _{\ast })}{\partial r}+\frac{\partial ^{2}\Psi _{N,n}(\theta
_{\ast })}{\partial r\partial \underline{w}_{n}^{\prime }}\widehat{%
\underline{\omega }}_{n})^{\prime }W_{N}(\frac{\partial \Psi _{N,n}(\theta
_{\ast })}{\partial r}+\frac{\partial ^{2}\Psi _{N,n}(\theta _{\ast })}{%
\partial r\partial \underline{w}_{n}^{\prime }}\widehat{\underline{\omega }}%
_{n})+  \notag \\
&&\quad \frac{1}{2!}(\kappa _{1,N},\text{ }\kappa _{2,N},\text{ }\ldots ,%
\text{ }\kappa _{K+2,N})^{\prime }W_{N}\frac{\partial ^{2}\Psi _{N,n}(\theta
_{\ast })}{\partial r^{2}})N^{1/2}(\widehat{\rho }-1)^{2}+  \notag \\
&&\frac{2}{3!}N^{1/2}(\frac{\partial \Psi _{N,n}(\theta _{\ast })}{\partial r%
}+\frac{\partial ^{2}\Psi _{N,n}(\theta _{\ast })}{\partial r\partial 
\underline{w}_{n}^{\prime }}\widehat{\underline{\omega }}_{n})^{\prime }W_{N}%
\frac{\partial ^{3}\Psi _{N,n}(\theta _{\ast })}{\partial r^{3}}N(\widehat{%
\rho }-1)^{4}+  \notag \\
&&\frac{1}{3!3!}\frac{\partial ^{3}\Psi _{N,n}^{\prime }(\theta _{\ast })}{%
\partial r^{3}}W_{N}\frac{\partial ^{3}\Psi _{N,n}(\theta _{\ast })}{%
\partial r^{3}}N^{3/2}(\widehat{\rho }-1)^{6}+o_{p}(1),\quad  \notag \\
&&  \notag \\
&&R_{5,N}(N^{1/2}(\widehat{\rho }-1)^{2})=N^{3/2}(\kappa _{1,N},\text{ }%
\kappa _{2,N},\text{ }\ldots ,\text{ }\kappa _{K+2,N})^{\prime }W_{N}(\frac{%
\partial \Psi _{N,n}(\theta _{\ast })}{\partial r}+\frac{\partial ^{2}\Psi
_{N,n}(\theta _{\ast })}{\partial r\partial \underline{w}_{n}^{\prime }}%
\widehat{\underline{\omega }}_{n})+  \notag \\
&&(\frac{1}{3!}N(\kappa _{1,N},\text{ }\kappa _{2,N},\text{ }\ldots ,\text{ }%
\kappa _{K+2,N})^{\prime }W_{N}\frac{\partial ^{3}\Psi _{N,n}(\theta _{\ast
})}{\partial r^{3}}+  \notag \\
&&\quad N(\frac{\partial \Psi _{N,n}(\theta _{\ast })}{\partial r}+\frac{%
\partial ^{2}\Psi _{N,n}(\theta _{\ast })}{\partial r\partial \underline{w}%
_{n}^{\prime }}\widehat{\underline{\omega }}_{n})^{\prime }W_{N}\frac{%
\partial ^{3}\Psi _{N,n}(\theta _{\ast })}{\partial r^{2}\partial \underline{%
w}_{n}^{\prime }}\widehat{\underline{\omega }}_{n})N^{1/2}(\widehat{\rho }%
-1)^{2}+  \notag \\
&&(\frac{2}{4!}N^{1/2}(\frac{\partial \Psi _{N,n}(\theta _{\ast })}{\partial
r}+\frac{\partial ^{2}\Psi _{N,n}(\theta _{\ast })}{\partial r\partial 
\underline{w}_{n}^{\prime }}\widehat{\underline{\omega }}_{n})^{\prime }W_{N}%
\frac{\partial ^{4}\Psi _{N,n}(\theta _{\ast })}{\partial r^{4}}+  \notag \\
&&\quad \frac{1}{3!}N^{1/2}\frac{\partial ^{3}\Psi _{N,n}^{\prime }(\theta
_{\ast })}{\partial r^{3}}W_{N}\frac{\partial ^{3}\Psi _{N,n}(\theta _{\ast
})}{\partial r^{2}\partial \underline{w}_{n}^{\prime }}\widehat{\underline{%
\omega }}_{n})N(\widehat{\rho }-1)^{4}+  \notag \\
&&(\frac{2}{3!4!}\frac{\partial ^{3}\Psi _{N,n}^{\prime }(\theta _{\ast })}{%
\partial r^{3}}W_{N}\frac{\partial ^{4}\Psi _{N,n}(\theta _{\ast })}{%
\partial r^{4}})N^{3/2}(\widehat{\rho }-1)^{6}+o_{p}(1),\quad \text{and}
\label{rvijf} \\
&&  \notag \\
&&R_{2,N}(N^{1/4}(\widehat{\rho }-1))=N^{-1/4}N^{1/4}(\widehat{\rho }-1)%
\underline{R}_{2,N}(N^{1/2}(\widehat{\rho }-1)^{2},F_{1,N})+o_{p}(N^{-1/4}) 
\notag
\end{eqnarray}%
with\vspace{-0.1in}%
\begin{eqnarray}
\underline{R}_{2,N}(N^{1/2}(\widehat{\rho }-1)^{2},F_{N}) &=&2(N^{1/2}(\frac{%
\partial \Psi _{N,n}(\theta _{\ast })}{\partial r}+\frac{\partial ^{2}\Psi
_{N,n}(\theta _{\ast })}{\partial r\partial \underline{w}_{n}^{\prime }}%
\widehat{\underline{\omega }}_{n})^{\prime }W_{N}F_{N}+  \notag \\
&&\frac{1}{6}\frac{\partial ^{3}\Psi _{N,n}^{\prime }(\theta _{\ast })}{%
\partial r^{3}}W_{N}F_{N}N^{1/2}(\widehat{\rho }-1)^{2})+o_{p}(1).
\label{r2n}
\end{eqnarray}%
It follows that if $Z_{1,N}+\widetilde{U}_{N}>0,$ then the value of $M_{N}(%
\widehat{\theta }_{n})$ in (\ref{obf1}) is in fact minimized at $N^{1/2}%
\widehat{\underline{\omega }}_{n}=N^{1/2}\widehat{\underline{\omega }}_{+}$
and $N^{1/2}(\widehat{\rho }-1)^{2}=Z_{1,N}+N^{-1/2}R_{7,N}+o_{p}(N^{-1/2})$
where\vspace{-0.1in} 
\begin{eqnarray}
R_{7,N} &\equiv &-\frac{1}{2}\widetilde{Q}_{N}^{-1}(N((\frac{\partial \Psi
_{N,n}(\theta _{\ast })}{\partial r}+\frac{\partial ^{2}\Psi _{N,n}(\theta
_{\ast })}{\partial r\partial \underline{w}_{n}^{\prime }}\widehat{%
\underline{\omega }}_{n})^{\prime }W_{N}(\frac{\partial \Psi _{N,n}(\theta
_{\ast })}{\partial r}+\frac{\partial ^{2}\Psi _{N,n}(\theta _{\ast })}{%
\partial r\partial \underline{w}_{n}^{\prime }}\widehat{\underline{\omega }}%
_{n})+  \notag \\
&&\frac{1}{2}(\kappa _{1,N},\text{ }\kappa _{2,N},\text{ }\ldots ,\text{ }%
\kappa _{K+2,N})^{\prime }W_{N}\frac{\partial ^{2}\Psi _{N,n}(\theta _{\ast
})}{\partial r^{2}})+\underline{R}_{2,N}(Z_{1,N},F_{1,N})+  \notag \\
&&\frac{2}{3}N^{1/2}(\frac{\partial \Psi _{N,n}(\theta _{\ast })}{\partial r}%
+\frac{\partial ^{2}\Psi _{N,n}(\theta _{\ast })}{\partial r\partial 
\underline{w}_{n}^{\prime }}\widehat{\underline{\omega }}_{n})^{\prime }W_{N}%
\frac{\partial ^{3}\Psi _{N,n}(\theta _{\ast })}{\partial r^{3}}Z_{1,N}+ 
\notag \\
&&\frac{1}{12}\frac{\partial ^{3}\Psi _{N,n}^{\prime }(\theta _{\ast })}{%
\partial r^{3}}W_{N}\frac{\partial ^{3}\Psi _{N,n}(\theta _{\ast })}{%
\partial r^{3}}Z_{1,N}^{2})  \label{r7n}
\end{eqnarray}%
with%
\begin{equation*}
\widetilde{Q}_{N}=\widetilde{W}_{N,1,1}-\widetilde{W}_{N,2,1}^{\prime }%
\widetilde{W}_{N,2,2}^{-1}\widetilde{W}_{N,2,1}.
\end{equation*}

We conclude that if $Z_{1,N}+\widetilde{U}_{N}>0,$ then $N^{1/2}(\widehat{%
\rho }-1)^{2}=Z_{1,N}+O_{p}(N^{-1/2}),$ 
\begin{eqnarray*}
&&R_{1,N}(N^{1/4}(\widehat{\rho }-1))=N^{-1/2}R_{4,N}(N^{1/2}(\widehat{\rho }%
-1)^{2})+(\widehat{\rho }-1)^{2}\underline{R}_{2,N}(N^{1/2}(\widehat{\rho }%
-1)^{2},F_{1,N})+ \\
&&\qquad N^{1/4}(\widehat{\rho }-1)N^{-3/4}R_{N}(N^{1/2}(\widehat{\rho }%
-1)^{2})+o_{p}(N^{-3/4})\text{\quad }
\end{eqnarray*}%
where%
\begin{equation*}
R_{N}(N^{1/2}(\widehat{\rho }-1)^{2})\equiv \underline{R}_{2,N}(N^{1/2}(%
\widehat{\rho }-1)^{2},F_{2,N})+\underline{R}_{4,N}(N^{1/2}(\widehat{\rho }%
-1)^{2},F_{1,N})+R_{5,N}(N^{1/2}(\widehat{\rho }-1)^{2})
\end{equation*}%
with%
\begin{eqnarray}
\underline{R}_{4,N}(N^{1/2}(\widehat{\rho }-1)^{2},F_{N}) &=&N^{1/2}(\frac{%
\partial ^{3}\Psi _{N,n}(\theta _{\ast })}{\partial r^{2}\partial \underline{%
w}_{n}^{\prime }}\widehat{\underline{\omega }}_{n})^{\prime
}W_{N}F_{N}N^{1/2}(\widehat{\rho }-1)^{2}+  \notag \\
&&\frac{1}{12}\frac{\partial ^{4}\Psi _{N,n}^{\prime }(\theta _{\ast })}{%
\partial r^{4}}W_{N}F_{N}N(\widehat{\rho }-1)^{4}+o_{p}(1).
\end{eqnarray}%
Furthermore%
\begin{equation*}
sgn(N^{1/4}(\widehat{\rho }-1))=sgn(-R_{N}(N^{1/2}(\widehat{\rho }-1)^{2})).
\end{equation*}%
It also follows that $\widetilde{U}_{N}=O_{p}(N^{-1/2}).$ If $Z_{1,N}+%
\widetilde{U}_{N}<0,$ then $M_{N}(\widehat{\theta }_{n})$\ has one local
minimum and its value is minimized at $N^{1/4}(\widehat{\rho }-1)=o_{p}(1).$

It is easily seen that $\underline{R}_{2,N}(Z_{1,N}+\widetilde{U}%
_{N},F_{2,N})\overset{d}{\rightarrow }\underline{R}_{2}$, $R_{5,N}(Z_{1,N}+%
\widetilde{U}_{N})\overset{d}{\rightarrow }R_{5}$ and $R_{N}(Z_{1,N}+%
\widetilde{U}_{N})\overset{d}{\rightarrow }R$ for some $\underline{R}_{2},$ $%
R_{5}$ and $R$. Likewise, when $Z_{1,N}+\widetilde{U}_{N}>0,$ $\underline{R}%
_{2,N}(Z_{1,N}+\widetilde{U}_{N})\overset{d}{\rightarrow }\underline{\tilde{R%
}}_{2},$ $R_{5,N}(Z_{1,N}+\widetilde{U}_{N})\overset{d}{\rightarrow }\tilde{R%
}_{5}$ and $R_{N}(Z_{1,N}+\widetilde{U}_{N})\overset{d}{\rightarrow }\tilde{R%
}$ for some $\underline{\tilde{R}}_{2},$ $\tilde{R}_{5}$ and $\tilde{R}.$ We
obtain formulae for $\underline{\tilde{R}}_{2}$ and $\tilde{R}_{5}$ from the
expressions for $\underline{R}_{2,N}(N^{1/2}(\widehat{\rho }%
-1)^{2},F_{2,N}), $ $R_{7,N}$ and $R_{5,N}(N^{1/2}(\widehat{\rho }-1)^{2})$
in (\ref{r2n}), (\ref{r7n}) and (\ref{rvijf}), respectively, by replacing
appropriately scaled versions of the derivatives of $\Psi _{N,n}(\theta )$
at $\theta _{\ast }$ by their stochastic limits, $N^{1/2}\widehat{\underline{%
\omega }}_{n}$ by $\underline{\omega }_{+}$, and $Z_{1,N}$ or $N^{1/2}(%
\widehat{\rho }-1)^{2}$ by $Z_{1}.$

Now, when $Z_{1}>0,$ $\underline{R}_{2}=\underline{\tilde{R}}_{2},$ $R_{5}=%
\tilde{R}_{5}$ and $R=\tilde{R}=\underline{\tilde{R}}_{2}+\tilde{R}_{5}.$ We
also have p$\lim_{N\rightarrow \infty }K_{-,N}=K_{-}.$ We conclude that 
\begin{equation*}
\left[ 
\begin{array}{c}
N^{1/4}(\widehat{\rho }_{F}-1) \\ 
N^{1/2}\widehat{\underline{\omega }}_{n}%
\end{array}%
\right] \overset{d}{\rightarrow }\left[ 
\begin{array}{c}
(-1)^{B}Z_{1}^{1/2} \\ 
\underline{\omega }_{+}%
\end{array}%
\right] \mathbf{1}\{Z_{1}>0\}+\left[ 
\begin{array}{c}
0 \\ 
\underline{\omega }_{+}+K_{-}Z_{1}%
\end{array}%
\right] \mathbf{1}\{Z_{1}\leq 0\}
\end{equation*}%
where $B=\mathbf{1}(R>0).$ Regarding $\Sigma _{\omega }$ we note that $%
E(V_{6}V_{4})=E(V_{6}V_{1})=0.\quad \square \bigskip \bigskip $\pagebreak

\textbf{\noindent Regularity conditions:}$\smallskip $

Let $f(y_{i}^{+};\theta )$ denote the density of the random vector $%
y_{i}^{+} $ so that $l_{i}(\theta )=\ln f(y_{i}^{+};\theta ).$ Let $p=\dim
(\theta ).$ We can write $\theta $ as $(\theta _{1},...,\theta _{p})^{\prime
}.$ For any $1\times p$ vector $a=(a_{1},...,a_{p}),$ let $%
l_{i}^{(a)}(\theta )$ denote $\partial ^{a.}l_{i}(\theta )/\partial
^{a_{1}}\theta _{1}\partial ^{a_{2}}\theta _{2}...\partial ^{a_{p}}\theta
_{p}$ where $a.=\tsum\nolimits_{k=1}^{p}a_{k}.$ Let $l^{(a)}(\theta
)=\tsum\nolimits_{i=1}^{N}l_{i}^{(a)}(\theta )$ and define $%
f^{(a)}(y_{i}^{+};\theta )$ similarly. Let w.p.1 denote "with probability
1". Let $S_{k}(\theta )=\partial l(\theta )/\partial \theta _{k},$ $1\leq
k\leq p,$ and $S_{k}=S_{k}(\theta _{\ast }).$ Let $S_{1}^{(q+j)},$ $j=0,1,$
denote $\partial ^{q+j}l(\theta )/\partial ^{q+j}\theta _{1}|_{\theta _{\ast
}}.$ Let $q\in 
\mathbb{N}
$ be such that conditions (B1$^{\prime }$)-(B3$^{\prime }$) given in K2013
hold, i.e., $\theta _{1}$ is $q$-th order identified. Then the regularity
conditions for the case of i.h.d. data are given by (cf. Bottai, 2003):

\begin{description}
\item[(A1$^{\prime }$)] $\theta _{\ast }\in \Theta $, a compact subset of $%
\mathbb{R}
^{p}$ that contains an open neighbourhood $\mathcal{N}$ of $\theta _{\ast }$.%
\vspace{-0.06in}

\item[(A2$^{\prime }$)] Distinct values of $\theta $ in $\Theta $ correspond
to distinct probability distributions.\vspace{-0.06in}

\item[(A3$^{\prime }$)] W.p.1 (under $\theta _{\ast }$), $f(y_{i}^{+};\theta
)>0$ for all $\theta \in \mathcal{N}$.\vspace{-0.06in}

\item[(A4$^{\prime }$)] There exist $B_{i}(y_{i}^{+})$ such that $\left\vert
l_{i}(\theta )\right\vert \leq B_{i}(y_{i}^{+})$ for all $\theta \in \Theta $
and $E\left\vert B_{i}(y_{i}^{+})\right\vert ^{1+\xi }<\infty $ for some $%
\xi >0$ and all $i\in \{1,...,N\}.$\vspace{-0.06in}

\item[(A5$^{\prime }$)] For all $\theta $ in $\mathcal{N}$ and all $a$ with $%
1\leq a.\leq 2q+1$, the derivatives $f^{(a)}(y_{i}^{+};\theta )$ and $%
l_{i}^{(a)}(\theta )$ exist w.p.1 and there exist $B_{i}(y_{i}^{+})$ such
that $\left\vert l_{i}^{(a)}(\theta )\right\vert \leq B_{i}(y_{i}^{+})$ for
all $\theta \in \mathcal{N}$ and $E\left\vert B_{i}(y_{i}^{+})\right\vert
^{1+\xi }<\infty $ for some $\xi $ and all $i\in \{1,...,N\}.$ Furthermore
(for all $a$ with $1\leq a.\leq 2q+1$), $\int \sup_{\theta \in \mathcal{N}%
}\left\vert f^{(a)}(y_{i}^{+};\theta )\right\vert dy_{i}<\infty $ , $\int
\sup_{\theta \in \mathcal{N}}[\{f^{(a)}(y_{i}^{+};\theta
)\}^{2}/f(y_{i}^{+};\theta )]dy_{i}<\infty ,$ and $\int \sup_{\theta \in 
\mathcal{N}}\{\left\vert l_{i}^{(a)}(\theta )\right\vert ^{j}f^{(a^{\prime
})}(y_{i}^{+};\theta )\}dy_{i}<\infty $ for $j=1,2$ and all $a^{\prime }$
with $0\leq a^{\prime }.\leq 2q+1,$ where $dy_{i}$ is short for $%
dy_{i,1}dy_{i,2},...,dy_{i,T}$\vspace{-0.06in}

\item[(A6a$^{\prime }$)] For all $a$ with $1\leq a.\leq 2q+1,$ there exist $%
\xi (a)>2$ such that $\sup_{\theta \in \mathcal{N}}E_{\theta }\left\vert
l_{i}^{(a)}(\theta ^{\prime })\right\vert ^{\xi (a)}$ $<\infty $ for all $%
\theta ^{\prime }\in \mathcal{N}$ and all $i\in \{1,...,N\}.$\vspace{-0.06in}

\item[(A6b$^{\prime }$)] When $a.=2q+1$ there exists $\varpi >0$ and some
function $h(.)$ satisfying $E\left\vert h(y_{i}^{+})\right\vert ^{\xi }$ $%
<\infty $ for some $\xi >2$ and all $i\in \{1,...,N\}$, such that for any $%
\theta $ and $\theta ^{\prime }$ in $\mathcal{N}$, w.p.1, $\left\vert
l^{(a)}(\theta )-l^{(a)}(\theta ^{\prime })\right\vert \leq \left\Vert
\theta -\theta ^{\prime }\right\Vert ^{\varpi }h(y_{i}^{+}),$ where $%
\left\Vert \theta \right\Vert =\left( \tsum\nolimits_{k=1}^{p}\theta
_{k}^{2}\right) ^{1/2}.$\vspace{-0.06in}

\item[(A7$^{\prime }$)] Conditions (B1$^{\prime }$)-(B3$^{\prime }$) given
in K2013 hold and w.p.1, $S_{2},...,S_{p}$ are linearly independent.
Furthermore, for each $\theta \neq \theta _{\ast }$, $S_{1}(\theta )\neq 0$
with positive probability.\vspace{-0.06in}
\end{description}

Conditions (A1)-(A7) alluded to in theorem 4 are equal to (A1$^{\prime }$%
)-(A7$^{\prime }$), respectively, with $\theta $ and $l_{i}(\theta )$
replaced by $\theta _{n}=(r_{n},d_{n}^{\prime })^{\prime }$ and $\widetilde{l%
}_{n,i}(\theta _{n}),$ respectively, and with $f(y_{i}^{+};\theta )$
denoting the density of the random vector $y_{i}^{+}$ used in the likelihood
function for the FEMLE for $\underline{\theta }_{n}$ (so $\widetilde{l}%
_{n,i}(\theta _{n})\neq \ln f(y_{i}^{+};\theta )$), $p=2+K$, $q=2$ and $%
\theta _{\ast }=(\rho _{\ast },\delta _{\ast }^{\prime })^{\prime
}=(1,\delta _{\ast }^{\prime })^{\prime }$. Conditions (A1)-(A7) are
satisfied when the data are i.i.d. and normal but in general have to be
checked.$\smallskip \smallskip $

\textbf{\noindent Proof of theorem 4:}

We first prove that if (A1)-(A7) hold, the restricted estimator $\widetilde{%
\mathcal{\theta }}_{n}=\widetilde{\mathcal{\theta }}_{n,N}$ that satisfies $A%
\widetilde{\mathcal{\theta }}_{n,N}=a_{N}$, where $A_{1,.}=(1$ $\mathbf{0}%
^{\prime }),$ is root-$N$ consistent under the parameter sequence $\mathcal{%
\theta }_{0,n,N}\linebreak $with $\mathcal{\theta }_{0,n,N}\rightarrow 
\mathcal{\theta }_{\ast }$ and $A\mathcal{\theta }_{0,n,N}=a_{N}$ (so that $A%
\mathcal{\theta }_{\ast }=a=\lim_{N\rightarrow \infty }a_{N}$). Consistency
of $\widetilde{\mathcal{\theta }}_{n,N}$ follows from Theorem 2.1 in NMcF,
cf. the proof of Theorem 1. The proof is similar to that of consistency of $%
\widehat{\theta }_{n}$ given in K2013wp. Let $\mathcal{\ddot{\theta}}%
_{n,N}=(a_{N},\mathcal{\ddot{\delta}}_{n,N}^{\prime })^{\prime }$ be the
estimator that maximizes $\widetilde{l}_{n,N}(\theta _{n,N})$ subject to $%
r_{n,N}=a_{N}$. One can show that $\mathcal{\ddot{\delta}}_{n,N}-\mathcal{%
\delta }_{n,N}=O_{p}(N^{-1/2})$ by expanding the likelihood equations for $%
\delta $ corresponding to $\widetilde{l}_{n,N}(\theta _{n,N})$ with $%
r_{n,N}=a_{N}$. Following Davidson and MacKinnon (1993, pp. 276-277), one
can similarly show that $\widetilde{\mathcal{\theta }}_{n,N}-\mathcal{\theta 
}_{0,n,N}=O_{p}(N^{-1/2})$. In a similar way we can prove that the
restricted FE(Q)MLE for $\sigma _{v,n}^{2}$ is consistent under the
parameter sequence $\underline{\mathcal{\theta }}_{0,N,n}$ with $\underline{%
\mathcal{\theta }}_{0,N,n}\rightarrow \underline{\mathcal{\theta }}_{\ast
}\in \underline{\Theta }$ and $A\mathcal{\theta }_{0,N,n}=a_{N},$ where $%
\underline{\Theta }$ is a compact parameter set, see Kruiniger (2025a).

Let $\overline{S}_{N,i}(\widetilde{\underline{\mathcal{\theta }}}_{n,N};%
\mathcal{\theta }_{0,n,N})=(A\overline{H}^{-1}(\widetilde{\underline{%
\mathcal{\theta }}}_{n,N})\overline{\mathcal{J}}(\widetilde{\mathcal{\theta }%
}_{n,N})\overline{H}^{-1}(\widetilde{\underline{\mathcal{\theta }}}%
_{n,N})A^{\prime })^{-1/2}A\overline{H}^{-1}(\widetilde{\underline{\mathcal{%
\theta }}}_{n,N})\frac{\partial \widetilde{l}_{n,i}(\widetilde{\mathcal{%
\theta }}_{n,N})}{\partial \theta _{n}}\linebreak $when $\widetilde{\mathcal{%
\theta }}_{n,N}\neq \theta _{\ast },$ $\mathcal{\theta }_{0,n,N}\neq \theta
_{\ast }$ (for all$\,\sigma ^{2}>0$), $A\mathcal{\theta }_{0,n,N}=a_{N},$
and $\mathcal{\theta }_{0,n,N}\in \Theta .$ For any $\{\widetilde{\mathcal{%
\theta }}_{n,N}\}$ such that $\widetilde{\mathcal{\theta }}_{n,N}\rightarrow 
\mathcal{\theta }_{\ast },$ the sequence $\overline{QLM}(\widetilde{%
\underline{\mathcal{\theta }}}_{n,N};\mathcal{\theta }_{0,n,N})\equiv
(N^{-1/2}\tsum_{i=1}^{N}\overline{S}_{N,i}(\widetilde{\underline{\mathcal{%
\theta }}}_{n,N};\mathcal{\theta }_{0,n,N}))^{\prime }\times $ $%
(N^{-1/2}\tsum_{i=1}^{N}\overline{S}_{N,i}(\widetilde{\underline{\mathcal{%
\theta }}}_{n,N};\mathcal{\theta }_{0,n,N}))$ converges when $\widetilde{%
\mathcal{\theta }}_{n,N}(=\widetilde{\mathcal{\theta }}_{n})\rightarrow 
\mathcal{\theta }_{\ast }$ (for any $\sigma ^{2}>0$) and $\mathcal{\theta }%
_{0,n,N}\rightarrow \mathcal{\theta }_{\ast }$ (for any $\sigma ^{2}>0$),
while $A\mathcal{\theta }_{0,n,N}=a_{N}$. Its limit can be obtained by
applying de l'H\^{o}pital's rule in the event that p$\lim_{N\rightarrow
\infty }\det (A\times adj(\overline{H}(\widetilde{\underline{\mathcal{\theta 
}}}_{n,N}))\overline{\mathcal{J}}(\widetilde{\mathcal{\theta }}_{n,N})adj(%
\overline{H}(\widetilde{\underline{\mathcal{\theta }}}_{n,N}))A^{\prime })=0$%
.

Following Davidson and MacKinnon (1993, pp. 276-277), we obtain $N^{1/2}(%
\widetilde{\mathcal{\theta }}_{n}-\mathcal{\theta }_{0,n,N})\overset{asy}{=}%
\linebreak -\overline{H}^{-1}(I-A^{\prime }(A\overline{H}^{-1}A^{\prime
})^{-1}A\overline{H}^{-1})N^{1/2}\frac{\partial \widetilde{l}_{n,N}(\mathcal{%
\theta }_{0,n,N})}{\partial \theta _{n}}$ where $\overline{H}\mathcal{=}%
\overline{H}(\underline{\mathcal{\theta }}_{0,n,N})$ and $\mathcal{\theta }%
_{0,n,N}\neq \theta _{\ast }.$ We also have $N^{1/2}\frac{\partial 
\widetilde{l}_{n,N}(\widetilde{\mathcal{\theta }}_{n})}{\partial \theta _{n}}%
\overset{asy}{=}N^{1/2}\frac{\partial \widetilde{l}_{n,N}(\mathcal{\theta }%
_{0,n,N})}{\partial \theta _{n}}+\overline{H}N^{1/2}(\widetilde{\mathcal{%
\theta }}_{n}-\mathcal{\theta }_{0,n,N}).$ Hence $N^{1/2}\frac{\partial 
\widetilde{l}_{n,N}(\widetilde{\mathcal{\theta }}_{n})}{\partial \theta _{n}}%
\overset{asy}{=}\linebreak A^{\prime }(A\overline{H}^{-1}A^{\prime })^{-1}A%
\overline{H}^{-1}N^{1/2}\frac{\partial \widetilde{l}_{n,N}(\mathcal{\theta }%
_{0,n,N})}{\partial \theta _{n}}$ and $(A\overline{H}^{-1}(\widetilde{%
\underline{\mathcal{\theta }}}_{n})\overline{\mathcal{J}}(\widetilde{%
\mathcal{\theta }}_{n})\overline{H}^{-1}(\widetilde{\underline{\mathcal{%
\theta }}}_{n})A^{\prime })^{-1/2}A\overline{H}^{-1}(\widetilde{\underline{%
\mathcal{\theta }}}_{n})\times $\linebreak $N^{1/2}\frac{\partial \widetilde{%
l}_{n,N}(\widetilde{\mathcal{\theta }}_{n})}{\partial \theta _{n}}\overset{%
asy}{=}(A\overline{H}^{-1}(\underline{\mathcal{\theta }}_{0,n,N})E_{%
\underline{\mathcal{\theta }}_{0,n,N}}\left( \overline{\mathcal{J}}(\mathcal{%
\theta }_{0,n,N})\right) \overline{H}^{-1}(\underline{\mathcal{\theta }}%
_{0,n,N})A^{\prime })^{-1/2}A\overline{H}^{-1}(\underline{\mathcal{\theta }}%
_{0,n,N})\times \linebreak N^{1/2}\frac{\partial \widetilde{l}_{n,N}(%
\mathcal{\theta }_{0,n,N})}{\partial \theta _{n}}$ under the parameter
sequence $\underline{\mathcal{\theta }}_{0,N,n}$ with $\underline{\mathcal{%
\theta }}_{0,N,n}\rightarrow \underline{\mathcal{\theta }}_{\ast }$, $A%
\mathcal{\theta }_{0,n,N}=a_{N}$ and $\mathcal{\theta }_{0,n,N}\neq \theta
_{\ast },$ where we have used that p$\lim_{N\rightarrow \infty }[\overline{%
\mathcal{J}}(\widetilde{\mathcal{\theta }}_{n})-E_{\underline{\mathcal{%
\theta }}_{0,n,N}}\left( \overline{\mathcal{J}}(\mathcal{\theta }%
_{0,n,N})\right) ]=0$ under $\underline{\mathcal{\theta }}%
_{0,N,n}\rightarrow \underline{\mathcal{\theta }}_{\ast }$ and the fact that 
$\overline{\mathcal{J}}(\widetilde{\mathcal{\theta }}_{n})$ and $E_{%
\underline{\mathcal{\theta }}_{0,n,N}}\left( \overline{\mathcal{J}}(\mathcal{%
\theta }_{0,n,N})\right) $ are positive definite when $\widetilde{\mathcal{%
\theta }}_{n,N}\neq \theta _{\ast }$ and $\mathcal{\theta }_{0,n,N}\neq
\theta _{\ast }$.

Let $S_{N,i}(\underline{\mathcal{\theta }}_{n})=A\overline{H}^{-1}(%
\underline{\mathcal{\theta }}_{n})\frac{\partial \widetilde{l}_{n,N,i}(%
\mathcal{\theta }_{n})}{\partial \theta _{n}}$, $S_{N,i}=S_{N,i}(\underline{%
\mathcal{\theta }}_{0,N,n})$ and $\underline{\mathcal{\theta }}%
_{0,N,n}\rightarrow \underline{\mathcal{\theta }}_{\ast }\in \underline{%
\Theta }$ with $A\mathcal{\theta }_{0,N,n}=a_{N}$ and $\mathcal{\theta }%
_{0,n,N}\neq \theta _{\ast }.$ Under conditions (A5)-(A6) and similarly to
Bottai (2003), $E_{\underline{\mathcal{\theta }}_{0,N,n}}(S_{N,i})=0,$ $Var_{%
\underline{\mathcal{\theta }}_{0,N,n}}(S_{N,i})=A\overline{H}^{-1}(%
\underline{\mathcal{\theta }}_{0,N,n})E_{\underline{\mathcal{\theta }}%
_{0,N,n}}(\mathcal{J}_{i}(\mathcal{\theta }_{0,N,n}))\overline{H}^{-1}(%
\underline{\mathcal{\theta }}_{0,N,n})A^{\prime }$ and$\linebreak
\sup_{i}\sup_{\underline{\mathcal{\theta }}_{n}\in \mathcal{N}}E_{\underline{%
\mathcal{\theta }}_{0,N,n}}(\left\vert \lambda ^{\prime }S_{N,i}\right\vert
^{\varsigma })<\infty $ for some $\varsigma >2$ and for all $\lambda \in 
\mathbb{R}
^{J}$ where $\mathcal{N}\subset \underline{\Theta }$ is an open
neighbourhood around $\underline{\mathcal{\theta }}_{\ast }$.\thinspace We
also have $N^{-1}\tsum_{i=1}^{N}Var_{\underline{\mathcal{\theta }}%
_{0,N,n}}(\lambda ^{\prime }S_{N,i})>0$ uniformly in $N$ for all $\lambda
\in 
\mathbb{R}
^{J}\backslash \{\mathbf{0}\}.$ Thus the Lyapunov conditions are satisfied
and by (a multivariate version of) Lindeberg's CLT for triangular arrays, $%
(\tsum_{i=1}^{N}Var_{\underline{\mathcal{\theta }}_{0,N,n}}(S_{N,i}))^{-1/2}%
\times \linebreak \tsum_{i=1}^{N}S_{N,i}$ converges under the parameter
sequence $\underline{\mathcal{\theta }}_{0,N,n}$ to $N(0,I_{J}).$ It follows
that $(A\overline{H}^{-1}(\widetilde{\underline{\mathcal{\theta }}}_{n})%
\overline{\mathcal{J}}(\widetilde{\mathcal{\theta }}_{n})\overline{H}^{-1}(%
\widetilde{\underline{\mathcal{\theta }}}_{n})A^{\prime })^{-1/2}A\overline{H%
}^{-1}(\widetilde{\underline{\mathcal{\theta }}}_{n})N^{1/2}\frac{\partial 
\widetilde{l}_{n,N}(\widetilde{\mathcal{\theta }}_{n})}{\partial \theta _{n}}%
\overset{d}{\rightarrow }N(0,I_{J})$ and $QLM(\widetilde{\underline{\mathcal{%
\theta }}}_{n})\overset{d}{\rightarrow }\chi ^{2}(J)$ under the parameter
sequence $\underline{\mathcal{\theta }}_{0,N,n}$ with $\underline{\mathcal{%
\theta }}_{0,N,n}\rightarrow \underline{\mathcal{\theta }}_{\ast }$, $A%
\mathcal{\theta }_{0,N,n}=a_{N}$ and $\mathcal{\theta }_{0,n,N}\neq \theta
_{\ast }.$

Next let $\widetilde{S}_{N,i}(\underline{\mathcal{\theta }}_{n})=\widetilde{A%
}\widetilde{\mathcal{H}}^{-1}(\underline{\mathcal{\theta }}_{n})\widetilde{S}%
_{i}(\mathcal{\theta }_{n})$, and $\widetilde{S}_{N,i}=\widetilde{S}_{N,i}(%
\underline{\mathcal{\theta }}_{\ast }).$ Under condition (A7) and similarly
to Bottai (2003), we can show that $(\tsum_{i=1}^{N}Var_{\underline{\mathcal{%
\theta }}_{0,N,n}}(\widetilde{S}_{N,i}))^{-1/2}\times \linebreak
\tsum_{i=1}^{N}\widetilde{S}_{N,i}\overset{d}{\rightarrow }N(0,I_{J})$ and $%
QLM(\underline{\mathcal{\theta }}_{\ast })\overset{d}{\rightarrow }\chi
^{2}(J)$ when $\underline{\mathcal{\theta }}_{0,N,n}=\underline{\mathcal{%
\theta }}_{\ast }.$ We conclude that$\linebreak \lim_{N\rightarrow \infty
}\sup_{\underline{\theta }_{0,n}\in \mathcal{N}}\left\vert \Pr_{\underline{%
\theta }_{0,n}}\{QLM(\mathcal{\theta }_{0,n})>\chi _{J,\alpha }^{2}\}-\alpha
\right\vert =0.$\quad $\square \bigskip $

\begin{table}[hbp] \centering%
\caption{Estimators of $\rho$; Design S; 5000
replications.\label{key}} 
\begin{tabular}{||c|c|c|c|c|c|c|c|c||}
\hline\hline
{\small N=100} & {\small T=4} & NM & \multicolumn{2}{|c|}{MMLC} & 
\multicolumn{2}{|c|}{FEML} & \multicolumn{2}{|c||}{REML} \\ \hline
${\small \sigma }_{\mu }^{2}$ & $\rho $ &  & bias & RMSE & bias & RMSE & bias
& RMSE \\ \hline
1 & \multicolumn{1}{|l|}{0.50} & .075 & .019 & .126 & .023 & .140 & .017 & 
.125 \\ 
1 & \multicolumn{1}{|l|}{0.80} & .396 & -.010 & .132 & .010 & .147 & .038 & 
.156 \\ 
1 & \multicolumn{1}{|l|}{0.90} & .468 & -.040 & .132 & -.012 & .135 & .038 & 
.149 \\ 
1 & \multicolumn{1}{|l|}{0.95} & .471 & -.065 & .139 & -.009 & .138 & .042 & 
.146 \\ 
1 & \multicolumn{1}{|l|}{0.98} & .485 & -.076 & .144 & .009 & .138 & .038 & 
.140 \\ 
1 & \multicolumn{1}{|l|}{1.00} & .481 & -.084 & .148 & .026 & .135 & .035 & 
.136 \\ \hline
0 & \multicolumn{1}{|l|}{0.50} & .079 & .017 & .125 & .021 & .138 & .005 & 
.098 \\ 
0 & \multicolumn{1}{|l|}{0.80} & .385 & -.012 & .131 & .008 & .146 & .019 & 
.139 \\ 
0 & \multicolumn{1}{|l|}{0.90} & .459 & -.042 & .132 & -.013 & .135 & .030 & 
.147 \\ 
0 & \multicolumn{1}{|l|}{0.95} & .474 & -.064 & .141 & -.010 & .139 & .037 & 
.145 \\ 
0 & \multicolumn{1}{|l|}{1.00} & .481 & -.086 & .150 & .026 & .135 & .026 & 
.135 \\ \hline
25 & \multicolumn{1}{|l|}{0.50} & .077 & .016 & .125 & .019 & .138 & .022 & 
.143 \\ 
25 & \multicolumn{1}{|l|}{0.80} & .400 & -.010 & .133 & .011 & .148 & .045 & 
.173 \\ 
25 & \multicolumn{1}{|l|}{0.90} & .461 & -.043 & .134 & -.015 & .136 & .040
& .159 \\ 
25 & \multicolumn{1}{|l|}{0.95} & .474 & -.064 & .140 & -.010 & .137 & .041
& .148 \\ 
25 & \multicolumn{1}{|l|}{1.00} & .479 & -.089 & .152 & .025 & .137 & .035 & 
.137 \\ \hline\hline
\end{tabular}
\linebreak NM: relative frequency that $\widehat{\rho }_{LAN}$ does not
exist (No Maximum).%
\end{table}%

\begin{table}[hbp] \centering%
\caption{Estimators of $\rho$; Design NS; 5000
replications.\label{key}} 
\begin{tabular}{||c|c|c|c|c|c|c|c|c||}
\hline\hline
{\small N=100} & {\small T=4} & NM & \multicolumn{2}{|c|}{MMLC} & 
\multicolumn{2}{|c|}{FEML} & \multicolumn{2}{|c||}{REML} \\ \hline
${\small \sigma }_{\mu }^{2}$ & $\rho $ &  & bias & RMSE & bias & RMSE & bias
& RMSE \\ \hline
1 & \multicolumn{1}{|l|}{0.50} & .326 & .010 & .143 & .014 & .141 & .015 & 
.142 \\ 
1 & \multicolumn{1}{|l|}{0.80} & .467 & -.069 & .148 & .028 & .161 & .028 & 
.163 \\ 
1 & \multicolumn{1}{|l|}{0.90} & .471 & -.085 & .153 & .010 & .146 & .021 & 
.149 \\ 
1 & \multicolumn{1}{|l|}{0.95} & .478 & -.085 & .150 & -.001 & .133 & .016 & 
.139 \\ 
1 & \multicolumn{1}{|l|}{0.98} & .483 & -.088 & .152 & .005 & .137 & .024 & 
.139 \\ 
1 & \multicolumn{1}{|l|}{1.00} & .481 & -.084 & .148 & .026 & .135 & .035 & 
.136 \\ \hline\hline
\end{tabular}%
\end{table}%

\begin{table}[htp] \centering%
\caption{Estimators of $\rho$; Design S; 5000
replications.\label{key}} 
\begin{tabular}{||c|c|c|c|c|c|c|c|c||}
\hline\hline
{\small N=100} & {\small T=9} & NM & \multicolumn{2}{|c|}{MMLC} & 
\multicolumn{2}{|c|}{FEML} & \multicolumn{2}{|c||}{REML} \\ \hline
${\small \sigma }_{\mu }^{2}$ & $\rho $ &  & bias & RMSE & bias & RMSE & bias
& RMSE \\ \hline
1 & \multicolumn{1}{|l|}{0.50} & .000 & .000 & .042 & .000 & .042 & .000 & 
.041 \\ 
1 & \multicolumn{1}{|l|}{0.80} & .130 & .006 & .064 & .008 & .069 & .005 & 
.061 \\ 
1 & \multicolumn{1}{|l|}{0.90} & .375 & -.004 & .060 & .007 & .070 & .011 & 
.067 \\ 
1 & \multicolumn{1}{|l|}{0.95} & .455 & -.020 & .061 & -.004 & .064 & .014 & 
.067 \\ 
1 & \multicolumn{1}{|l|}{0.98} & .489 & -.032 & .063 & .001 & .062 & .017 & 
.063 \\ 
1 & \multicolumn{1}{|l|}{1.00} & .490 & -.041 & .068 & .013 & .057 & .016 & 
.058 \\ \hline\hline
\end{tabular}%
\end{table}%

\begin{table}[hbp] \centering%
\caption{Estimators of $\rho$; Design S; 5000
replications.\label{key}} 
\begin{tabular}{||c|c|c|c|c|c|c|c|c||}
\hline\hline
{\small N=500} & {\small T=4} & NM & \multicolumn{2}{|c|}{MMLC} & 
\multicolumn{2}{|c|}{FEML} & \multicolumn{2}{|c||}{REML} \\ \hline
${\small \sigma }_{\mu }^{2}$ & $\rho $ &  & bias & RMSE & bias & RMSE & bias
& RMSE \\ \hline
1 & \multicolumn{1}{|l|}{0.50} & .001 & .003 & .052 & .002 & .048 & .002 & 
.046 \\ 
1 & \multicolumn{1}{|l|}{0.80} & .306 & .007 & .084 & .017 & .099 & .008 & 
.077 \\ 
1 & \multicolumn{1}{|l|}{0.90} & .442 & -.016 & .082 & -.003 & .089 & .015 & 
.090 \\ 
1 & \multicolumn{1}{|l|}{0.95} & .482 & -.035 & .085 & -.017 & .087 & .024 & 
.095 \\ 
1 & \multicolumn{1}{|l|}{0.98} & .498 & -.047 & .090 & -.014 & .089 & .029 & 
.093 \\ 
1 & \multicolumn{1}{|l|}{1.00} & .512 & -.054 & .092 & .018 & .087 & .024 & 
.088 \\ \hline
0 & \multicolumn{1}{|l|}{0.50} & .001 & .002 & .053 & .002 & .050 & .001 & 
.042 \\ 
0 & \multicolumn{1}{|l|}{0.80} & .293 & .007 & .085 & .019 & .101 & .004 & 
.068 \\ 
0 & \multicolumn{1}{|l|}{0.90} & .438 & -.018 & .084 & -.007 & .090 & .009 & 
.085 \\ 
0 & \multicolumn{1}{|l|}{0.95} & .473 & -.034 & .085 & -.016 & .085 & .020 & 
.093 \\ 
0 & \multicolumn{1}{|l|}{1.00} & .493 & -.056 & .094 & .018 & .088 & .018 & 
.088 \\ \hline
25 & \multicolumn{1}{|l|}{0.50} & .002 & .002 & .054 & .000 & .049 & .000 & 
.049 \\ 
25 & \multicolumn{1}{|l|}{0.80} & .314 & .009 & .085 & .017 & .097 & .058 & 
.135 \\ 
25 & \multicolumn{1}{|l|}{0.90} & .443 & -.016 & .082 & -.004 & .088 & .055
& .118 \\ 
25 & \multicolumn{1}{|l|}{0.95} & .471 & -.036 & .084 & -.016 & .085 & .043
& .102 \\ 
25 & \multicolumn{1}{|l|}{1.00} & .489 & -.058 & .096 & .018 & .090 & .023 & 
.090 \\ \hline\hline
\end{tabular}
\linebreak NM: relative frequency that $\widehat{\rho }_{LAN}$ does not
exist (No Maximum).%
\end{table}%

\begin{table}[hbp] \centering%
\caption{Estimators of $\rho$; Design NS; 5000
replications.\label{key}} 
\begin{tabular}{||c|c|c|c|c|c|c|c|c||}
\hline\hline
{\small N=500} & {\small T=4} & NM & \multicolumn{2}{|c|}{MMLC} & 
\multicolumn{2}{|c|}{FEML} & \multicolumn{2}{|c||}{REML} \\ \hline
${\small \sigma }_{\mu }^{2}$ & $\rho $ &  & bias & RMSE & bias & RMSE & bias
& RMSE \\ \hline
1 & \multicolumn{1}{|l|}{0.50} & .180 & .016 & .091 & .004 & .064 & .004 & 
.064 \\ 
1 & \multicolumn{1}{|l|}{0.80} & .489 & -.036 & .088 & .019 & .104 & .021 & 
.105 \\ 
1 & \multicolumn{1}{|l|}{0.90} & .502 & -.051 & .094 & .027 & .101 & .032 & 
.102 \\ 
1 & \multicolumn{1}{|l|}{0.95} & .484 & -.056 & .094 & .009 & .091 & .019 & 
.092 \\ 
1 & \multicolumn{1}{|l|}{0.98} & .481 & -.058 & .096 & -.004 & .088 & .011 & 
.090 \\ 
1 & \multicolumn{1}{|l|}{1.00} & .512 & -.054 & .092 & .018 & .087 & .024 & 
.088 \\ \hline\hline
\end{tabular}%
\end{table}%

\begin{table}[htp] \centering%
\caption{Estimators of $\rho$; Design S; 5000
replications.\label{key}} 
\begin{tabular}{||c|c|c|c|c|c|c|c|c||}
\hline\hline
{\small N=500} & {\small T=9} & NM & \multicolumn{2}{|c|}{MMLC} & 
\multicolumn{2}{|c|}{FEML} & \multicolumn{2}{|c||}{REML} \\ \hline
${\small \sigma }_{\mu }^{2}$ & $\rho $ &  & bias & RMSE & bias & RMSE & bias
& RMSE \\ \hline
1 & \multicolumn{1}{|l|}{0.50} & .000 & .001 & .020 & .001 & .020 & .000 & 
.017 \\ 
1 & \multicolumn{1}{|l|}{0.80} & .006 & .003 & .028 & .002 & .026 & .001 & 
.022 \\ 
1 & \multicolumn{1}{|l|}{0.90} & .272 & .005 & .039 & .009 & .046 & .003 & 
.033 \\ 
1 & \multicolumn{1}{|l|}{0.95} & .420 & -.007 & .036 & .001 & .041 & .007 & 
.039 \\ 
1 & \multicolumn{1}{|l|}{0.98} & .460 & -.019 & .039 & -.006 & .039 & .009 & 
.041 \\ 
1 & \multicolumn{1}{|l|}{1.00} & .488 & -.027 & .044 & .010 & .039 & .012 & 
.039 \\ \hline\hline
\end{tabular}%
\end{table}%

\begin{table}[hbp] \centering%
\caption{Empirical size of Quasi LM test based on Modified Likelihood; Nominal size is 0.05; T=9; 10000
replications.\label{key}}%
\begin{tabular}{||l|c|c|c|c|c|c||}
\hline\hline
model & \multicolumn{2}{|c|}{S-Normal} & \multicolumn{2}{|c|}{S-ChiSq} & 
\multicolumn{2}{|c||}{NS-Normal} \\ \hline
$\rho $ & $N=100$ & $N=500$ & $N=100$ & $N=500$ & $N=100$ & $N=500$ \\ \hline
0.50 & .0547 & .0474 & .0531 & .0491 & .0533 & .0483 \\ 
0.80 & .0540 & .0514 & .0551 & .0519 & .0575 & .0522 \\ 
0.90 & .0557 & .0500 & .0529 & .0510 & .0527 & .0539 \\ 
0.95 & .0495 & .0510 & .0496 & .0498 & .0501 & .0512 \\ 
0.98 & .0502 & .0508 & .0482 & .0512 & .0472 & .0443 \\ 
0.99 & .0512 & .0518 & .0528 & .0508 & .0496 & .0506 \\ \hline\hline
\end{tabular}
\end{table}%

\begin{table}[hbp] \centering%
\caption{Empirical power of Quasi LM test based on Modified Likelihood; $H_{0}:\rho=0.8$; Nominal size is 0.05; T=9; 5000
replications.\label{key}}%
\begin{tabular}{||l|c|c|c|c|c|c||}
\hline\hline
model & \multicolumn{2}{|c|}{S-Normal} & \multicolumn{2}{|c|}{S-ChiSq} & 
\multicolumn{2}{|c||}{NS-Normal} \\ \hline
true $\rho $ & $N=100$ & $N=500$ & $N=100$ & $N=500$ & $N=100$ & $N=500$ \\ 
\hline
0.50 & .999 & 1.000 & .966 & .999 & .993 & 1.000 \\ 
0.60 & .919 & 1.000 & .828 & .999 & .781 & 1.000 \\ 
0.70 & .375 & .926 & .399 & .878 & .274 & .783 \\ 
0.90 & .170 & .743 & .207 & .775 & .141 & .667 \\ 
0.95 & .421 & .989 & .404 & .956 & .391 & .982 \\ 
0.99 & .714 & 1.000 & .529 & .992 & .726 & 1.000 \\ \hline\hline
\end{tabular}
\end{table}%
\newpage

\end{document}